\documentclass[twocolumn]{aastex701}

\newcommand{\W}{$\lambda$}
\newcommand{\mstar}{$M_*$}
\newcommand{\msun}{M$_\odot$}
\newcommand{\zsun}{Z$_\odot$}
\newcommand{\te}{$T_{\mathrm{e}}$}
\newcommand{\den}{$n_{\mathrm{e}}$}
\newcommand{\densp}{$n_{\mathrm{e}}(\mathrm{S}^{+})$}
\newcommand{\temotp}{$T_{\mathrm{e}}(\mathrm{O}^{2+})$}
\newcommand{\temop}{$T_{\mathrm{e}}(\mathrm{O}^{+})$}
\newcommand{\temstp}{$T_{\mathrm{e}}(\mathrm{S}^{2+})$}
\newcommand{\temsp}{$T_{\mathrm{e}}(\mathrm{S}^{+})$}
\newcommand{\ebvgas}{$E(B-V)_\mathrm{gas}$}

\begin{document}

\title{The AURORA Survey: High-Redshift Empirical Metallicity Calibrations from Electron Temperature Measurements at $z=2-10$}

\shorttitle{AURORA High-$z$ Metallicity Calibrations}
\shortauthors{Sanders et al.}

\correspondingauthor{Ryan L. Sanders}


\author[0000-0003-4792-9119]{Ryan L. Sanders}\affiliation{Department of Physics and Astronomy, University of Kentucky, 505 Rose Street, Lexington, KY 40506, USA}\email[show]{ryan.sanders@uky.edu}

\author[0000-0003-3509-4855]{Alice E. Shapley}\affiliation{Department of Physics \& Astronomy, University of California, Los Angeles, 430 Portola Plaza, Los Angeles, CA 90095, USA}\email{aes@astro.ucla.edu}

\author[0000-0001-8426-1141]{Michael W. Topping}\affiliation{Steward Observatory, University of Arizona, 933 N Cherry Avenue, Tucson, AZ 85721, USA}\email{michaeltopping@arizona.edu}

\author[0000-0001-9687-4973]{Naveen A. Reddy}\affiliation{Department of Physics \& Astronomy, University of California, Riverside, 900 University Avenue, Riverside, CA 92521, USA}\email{naveenr@ucr.edu}

\author[0000-0002-4153-053X]{Danielle A. Berg}\affiliation{Department of Astronomy, The University of Texas at Austin, 2515 Speedway, Stop C1400, Austin, TX 78712, USA}\email{daberg@austin.utexas.edu}

\author[0000-0002-0101-336X]{Ali Ahmad Khostovan}
\affiliation{Department of Physics and Astronomy, University of Kentucky, 505 Rose Street, Lexington, KY 40506, USA}\email{ali.ahmad.khostovan@uky.edu}

\author[0000-0002-4989-2471]{Rychard J. Bouwens}\affiliation{Leiden Observatory, Leiden University, NL-2300 RA Leiden, Netherlands}\email{bouwens@strw.leidenuniv.nl}

\author[0000-0003-2680-005X]{Gabriel Brammer}\affiliation{Niels Bohr Institute, University of Copenhagen, Lyngbyvej 2, DK2100 Copenhagen \O, Denmark}\affiliation{Cosmic Dawn Center (DAWN), Copenhagen, Denmark}\email{gabriel.brammer@nbi.ku.dk}

\author[0000-0002-1482-5818]{Adam C. Carnall}\affiliation{Institute for Astronomy, University of Edinburgh, Royal Observatory, Edinburgh, EH9 3HJ, UK}\email{adamc@roe.ac.uk}

\author[0000-0002-3736-476X]{Fergus Cullen}\affiliation{Institute for Astronomy, University of Edinburgh, Royal Observatory, Edinburgh, EH9 3HJ, UK}\email{fc@roe.ac.uk}

\author[0000-0003-2842-9434]{Romeel Dav\'e}\affiliation{Institute for Astronomy, University of Edinburgh, Royal Observatory, Edinburgh, EH9 3HJ, UK}\email{Romeel.Dave@ed.ac.uk}

\author{James S. Dunlop}\affiliation{Institute for Astronomy, University of Edinburgh, Royal Observatory, Edinburgh, EH9 3HJ, UK}\email{jsd@roe.ac.uk}

\author[0000-0001-7782-7071]{Richard S. Ellis}\affiliation{Department of Physics \& Astronomy, University College London. Gower St., London WC1E 6BT, UK}\email{richard.ellis@ucl.ac.uk}

\author[0000-0003-4264-3381]{N. M. F\"orster Schreiber}\affiliation{Max-Planck-Institut f\"ur extraterrestrische Physik (MPE), Giessenbachstr.1, D-85748 Garching, Germany}\email{forster@mpe.mpg.de}

\author[0000-0002-0658-1243]{Steven R. Furlanetto}\affiliation{Department of Physics \& Astronomy, University of California, Los Angeles, 430 Portola Plaza, Los Angeles, CA 90095, USA}\email{sfurlane@astro.ucla.edu}

\author[0000-0002-3254-9044]{Karl Glazebrook}\affiliation{Centre for Astrophysics and Supercomputing, Swinburne University of Technology, P.O. Box 218, Hawthorn, VIC 3122, Australia}\email{kglazebrook@swin.edu.au}

\author[0000-0002-8096-2837]{Garth D. Illingworth}\affiliation{Department of Astronomy and Astrophysics, UCO/Lick Observatory, University of California, Santa Cruz, CA 95064, USA}\email{gdi@ucolick.org}

\author[0000-0001-5860-3419]{Tucker Jones}\affiliation{Department of Physics and Astronomy, University of California Davis, 1 Shields Avenue, Davis, CA 95616, USA}\email{tdjones@ucdavis.edu}

\author[0000-0002-7613-9872]{Mariska Kriek}\affiliation{Leiden Observatory, Leiden University, NL-2300 RA Leiden, Netherlands}\email{kriek@strw.leidenuniv.nl}

\author[0000-0003-4368-3326]{Derek J. McLeod}\affiliation{Institute for Astronomy, University of Edinburgh, Royal Observatory, Edinburgh, EH9 3HJ, UK}\email{derek.mcleod@ed.ac.uk}

\author{Ross J. McLure}\affiliation{Institute for Astronomy, University of Edinburgh, Royal Observatory, Edinburgh, EH9 3HJ, UK}\email{rjm@roe.ac.uk}

\author[0000-0002-7064-4309]{Desika Narayanan}\affiliation{Department of Astronomy, University of Florida, 211 Bryant Space Sciences Center, Gainesville, FL, USA}\email{desika.narayanan@ufl.edu}

\author[0000-0001-5851-6649]{Pascal A. Oesch}\affiliation{Department of Astronomy, University of Geneva, Chemin Pegasi 51, 1290 Versoix, Switzerland}\affiliation{Niels Bohr Institute, University of Copenhagen, Lyngbyvej 2, DK2100 Copenhagen \O, Denmark}\affiliation{Cosmic Dawn Center (DAWN), Copenhagen, Denmark}\email{pascal.oesch@unige.ch}

\author[0000-0003-4464-4505]{Anthony J. Pahl}\affiliation{The Observatories of the Carnegie Institution for Science, 813 Santa Barbara Street, Pasadena, CA 91101, USA}\email{apahl@carnegiescience.edu}

\author[0000-0002-5139-4359]{Max Pettini}\affiliation{Institute of Astronomy, Madingley Road, Cambridge CB3 OHA, UK}\email{pettini@ast.cam.ac.uk}

\author[0000-0001-7144-7182]{Daniel Schaerer}\affiliation{Department of Astronomy, University of Geneva, Chemin Pegasi 51, 1290 Versoix, Switzerland}\email{daniel.schaerer@unige.ch}

\author{Daniel P. Stark}\affiliation{Department of Astronomy, University of California, Berkeley, Berkeley, CA 94720, USA}\email{dpstark@berkeley.edu}

\author[0000-0002-4834-7260]{Charles C. Steidel}\affiliation{Cahill Center for Astronomy and Astrophysics, California Institute of Technology, MS 249-17, Pasadena, CA 91125, USA}\email{ccs@astro.caltech.edu}

\author[0000-0001-5940-338X]{Mengtao Tang}\affiliation{Steward Observatory, University of Arizona, 933 N Cherry Avenue, Tucson, AZ 85721, USA}\email{tangmtasua@arizona.edu}

\author[0000-0003-1249-6392]{Leonardo Clarke}\affiliation{Department of Physics \& Astronomy, University of California, Los Angeles, 430 Portola Plaza, Los Angeles, CA 90095, USA}\email{leoclarke@astro.ucla.edu}

\author[0000-0002-7622-0208]{Callum T. Donnan}
\affiliation{NSF's National Optical-Infrared Astronomy Research Laboratory, 950 N. Cherry Ave., Tucson, AZ 85719, USA}\email{callum.donnan@noirlab.edu}

\author{Emily Kehoe}\affiliation{Department of Physics \& Astronomy, University of California, Los Angeles, 430 Portola Plaza, Los Angeles, CA 90095, USA}\email{ekehoe@astro.ucla.edu}

\begin{abstract}
We present detections of auroral emission lines of [\ion{O}{3}], [\ion{O}{2}], [\ion{S}{3}], and [\ion{S}{2}] in deep {\it JWST}/NIRSpec spectroscopy for 41 star-forming galaxies at $z=1.4-7.2$ from the AURORA survey.
We combine these new observations with 98 star-forming galaxies at $z=1.3-10.6$ with detected auroral lines drawn from the literature to form a sample of 139 high-redshift galaxies with robust electron temperature and direct-method oxygen abundance determinations.
This sample notably covers a wider dynamic range in metallicity than previous work, spanning $0.02-0.9$~Z$_\odot$. 
We calibrate empirical relations between 19 emission-line ratios and oxygen abundance, providing a robust tool set to infer accurate gas-phase metallicities of high-redshift galaxies when auroral lines are not detected.
While calibrations based on lines of $\alpha$ elements (O, Ne, S, Ar) appear reliable, we find significant scatter in calibrations involving lines of N driven by a high dispersion in N/O at fixed O/H, suggesting that N-based line ratios are less reliable tracers of the oxygen abundance at high redshift.
These new high-redshift calibrations are notably offset from those based on typical $z\sim0$ galaxy and \ion{H}{2} region samples, and are better matched by samples of extreme local galaxies that are analogs of high-redshift sources.
The new metallicity calibrations presented in this work pave the way for robust studies of galaxy chemical evolution in the early Universe, leading to a better understanding of baryon cycling and galaxy formation from Cosmic Noon through the Epoch of Reionization.
\end{abstract}


\section{Introduction}\label{sec:intro}

Heavy elements (i.e., metals) play a vital role in the interstellar medium (ISM) of galaxies.
Many important physical processes are sensitive to the abundance of metals relative to hydrogen (i.e, metallicity), including ISM cooling, dust production, star formation, stellar evolution, and the emergent ionizing spectrum of massive stars.
Metallicity is also intimately linked to galaxy formation and growth, governed by the cycle of baryons into and out of galaxies \citep[e.g.,][]{maiolino2019,peroux2020}.
The metallicity of the ISM, as traced by the gas-phase oxygen abundance (O/H), is modulated by metal-poor gas accretion, metal-enriched gas outflows, and the regulation of the star formation rate (SFR) via feedback from supernovae or accreting supermassive black holes \citep[e.g.,][]{finlator2008,peeples2011,dave2012,lilly2013}.

Significant effort has gone into spectroscopic observing campaigns to characterize ISM metallicity in large samples of galaxies across a wide range of redshifts.
Such efforts have shown that ISM metallicity and stellar mass (\mstar) are positively correlated (the ``mass-metallicity relation'') in both the local Universe \citep[e.g.,][]{lequeux1979,tremonti2004,andrews2013,curti2020,yates2020} and at high redshifts up to $z>6$ \citep[e.g.,][]{erb2006,maiolino2008,zahid2011,troncoso2014,sanders2021,topping2021,nakajima2023,curti2024,chemerynska2024,sarkar2025}, with metallicity decreasing as redshift increases at fixed mass.
The evolving mass-metallicity relation, and its secondary dependence on SFR \citep[e.g.,][]{mannucci2010,laralopez2010,sanders2018}, provide key constraints on baryon cycling parameters across Cosmic history.

One of the most robust ways to determine the ISM metallicity of ionized gas in \ion{H}{2} regions is known as the ``direct'' or temperature-based method that requires constraints on the electron temperature (\te) of the ionized gas.
\te\ can be derived from the ratio of two collisionally-excited transitions of the same metal ion originating from different electron upper energy levels (e.g., [\ion{O}{3}]\W4364/[\ion{O}{3}]\W5008).
With known \te, observed intensity ratios of a collisionally-excited metal line to a \ion{H}{1} recombination line (e.g., [\ion{O}{3}]\W5008/H$\beta$) can be converted to a number density ratio of the two ionic species ($n(\mathrm{O}^{2+})/n(\mathrm{H}^{+})$) using the ratio of their \te-sensitive emissivities \citep[e.g.,][]{aller1984,osterbrock2006}.
The difficulty in widely applying the direct method for galaxy metallicity studies is due to the intrinsic faintness of the higher-level ``auroral'' emission lines (e.g., [\ion{O}{3}]\W4364, [\ion{O}{2}]\W\W7322,7332), which are typically hundreds of times fainter than H$\alpha$ or [\ion{O}{3}]\W5008 and thus require deep spectroscopy to detect.

An alternative approach is to utilize ``strong-line'' metallicity calibrations.
These relations between intensity ratios of the brightest rest-optical emission lines and metallicity can be calibrated based on either theoretical photoionization models \citep[e.g.,][]{kewley2002,dopita2016,kewley2019} or relatively small samples of galaxies or \ion{H}{2} regions with empirical direct-method metallicities from sufficiently deep spectroscopy \citep[e.g.,][]{pettini2004,marino2013,curti2020}.
Since the strong-line technique only requires detections of bright lines, it has enabled metallicity estimates for large samples of galaxies numbering in the hundreds of thousands at $z\sim0$ and thousands at $z>1$.
However, the accuracy of the resulting metallicities relies on the assumption that the physical nebular properties of the calibrating sample (or models) are well-matched to those of the sample to which the calibrations are applied.

It is now well known that the properties of the ionized ISM at $z>2$ are distinct from those present in typical $z\sim0$ objects, with more extreme conditions at high redshifts driven by increased electron densities and a harder ionizing spectrum due to $\alpha$-enhanced massive stars \citep[e.g.,][]{steidel2014,steidel2016,shapley2015,sanders2016den,sanders2020,strom2017,strom2018,shapley2019,topping2020a,topping2020b,cullen2021,isobe2023a,stanton2024,topping2025}.
Consequently, strong-line calibrations based on representative $z\sim0$ samples will not reliably yield accurate metallicities if applied to high-redshift samples.
Until the launch of the James Webb Space Telescope ({\it JWST}), it was not possible to detect auroral lines at $z>1$ in a large enough sample to produce high-redshift calibrations \citep[e.g.,][]{patricio2018,sanders2020}.
Strong-line calibrations for use at high redshifts were instead constructed from samples of low-redshift galaxies that were analogs of $z>2$ populations based on their enhanced SFRs or excitation properties \citep{bian2018,perezmontero2021,nakajima2022}, though the reliability of this analog technique has not been thoroughly validated.

The excellent sensitivity and wide wavelength coverage offered by the NIRSpec instrument onboard {\it JWST} have now enabled the detetection of auroral lines for large samples of galaxies at $z>1$, yielding direct-method metallicity determinations for galaxies up to $z\sim10$ \citep[e.g.,][]{schaerer2022,curti2023,jones2023,sanders2024,laseter2024,rogers2024,hsiao2024,morishita2024,langeroodi2024,welch2024,welch2025,arellano2025,chakraborty2025,cataldi2025,scholte2025}.
These observations have led to the construction of the first strong-line metallicity calibrations based on high redshift galaxies rather than local analogs, opening a new era of accurate chemical evolution studies in the early Universe.
Early work utilized auroral-detected samples of $\sim20-50$ galaxies, with most at moderately low metallicities (12+log(O/H$)\sim7.7-8.4$) where [\ion{O}{3}]\W4364 detectability is maximized
\citep{nakajima2023,laseter2024,sanders2024,scholte2025,chakraborty2025}.
Recently, \citet{cataldi2025} presented strong-line calibrations based on a sample of 112 galaxies spanning $z\sim2-9$.
However, their sample is still limited to 12+log(O/H$)<8.4$ ($<0.5$~\zsun) and thus cannot be used at higher metallicities where $\sim10^{10}$~\msun\ galaxies at $z\sim2-4$ appear to lie based on previous strong-line estimates \citep[e.g.,][]{sanders2021}.

In this work, we present new detections of [\ion{O}{3}], [\ion{O}{2}], [\ion{S}{3}], and [\ion{S}{2}] auroral emission lines for 41 star-forming galaxies at $z=1.4-7.2$ from the AURORA survey \citep{shapley2025a,sanders2024dust}.
This AURORA sample notably has direct-method metallicities that extend up to 12+log(O/H$)=8.6$ (0.9~\zsun), reaching higher metallicities than any previous $T_e$ work at high redshift.
In combination with a sample of auroral-detected sources drawn from the literature, we assemble a direct metallicity sample of 139 galaxies at $z\sim2-10$ spanning 12+log(O/H$)=7.0-8.6$ ($0.02-0.9$~\zsun).
We use this sample to construct new strong-line metallicity calibrations appropriate for use in the high-redshift Universe.

This paper is organized as follows.
In Sec.~\ref{sec:data}, we describe the AURORA observations and emission-line measurements.
The AURORA and literature auroral-detected samples are defined and the derivation of physical properties is described in Sec.~\ref{sec:metallicity}.
In Sec.~\ref{sec:results}, we investigate the relations between $T_e$ of different ionic zones and present new strong-line calibrations.
We discuss these results in Sec.~\ref{sec:discussion}, and summarize our conclusions in Sec.~\ref{sec:conclusion}.
Throughout this paper, we adopt a cosmology with $H_0=70\mbox{ km  s}^{-1}\mbox{ Mpc}^{-1}$, $\Omega_m=0.30$, and $\Omega_{\Lambda}=0.7$;
emission-line wavelengths are given in the vacuum rest frame;
and the term metallicity refers to the gas-phase oxygen abundance unless noted otherwise.
The solar oxygen abundance is taken to be 12+log(O/H$)_\odot=8.69$ \citep{asplund2021}.
Unless explicitly stated otherwise, the significance threshold for detection is 3$\sigma$ and all plotted upper and lower limits are at the 3$\sigma$ level.

\section{Observations and Measurements}\label{sec:data}

\subsection{The AURORA survey}

We use data from the Assembly of Ultradeep Rest-optical Observations Revealing Astrophysics (AURORA) survey,
 a Cycle 1 program (PID: 1914, Co-PIs: A. Shapley and R. Sanders) that obtained deep $R\sim1000$ {\it JWST}/NIRSpec Micro Shutter Assembly (MSA) observations of 97 galaxies.
Spectroscopic observations were taken in the G140M/F100LP, G235M/F190LP, and G395M/F290LP configurations, providing continuous $1-5\ \mu$m wavelength coverage.
The respective integration times in these three gratings were 44.2~ks (12.3~hr), 28.9~ks (8.0~hr), and 15.1~ks (4.2~hr), chosen to yield an approximately flat limiting line flux across all three gratings that was found to be $\approx5\times10^{-19}$ erg s$^{-1}$ cm$^{-2}$ (5$\sigma$).
Two pointings were observed with this setup, one each in the GOODS-N and COSMOS fields.
Three-shutter slitlets and a three-point nod pattern were used.
Full details of the AURORA survey design, observing setup, data reduction, calibration, spectral energy distribution (SED) fitting, and emission-line measurements can be found in \citet{shapley2025a}, \citet{sanders2024dust}, and \citet{reddy2025}.
We briefly describe these topics below, with a particular focus on the aspects most relevant to this analysis.

\subsection{Target Selection}

The main science goal of AURORA was to detect auroral emission lines in star-forming galaxies in the Cosmic Noon epoch ($z\sim2-4$) for direct method metallicity determinations, enabling the construction of new robust strong-line calibrations.
Accordingly, primary targets were selected using a scheme that prioritized galaxies for which the [\ion{O}{3}]\W4364 and/or [\ion{O}{2}]\W\W7322,7332 auroral lines were expected to be bright enough to detect within the adopted integration times.
This priority scheme, described in full detail in \citet{shapley2025a}, involved estimating the brightness of the auroral lines based on existing measurements of strong rest-optical lines from ground-based or Hubble Space Telescope ({\it HST}) near-infrared spectroscopy, or from line fluxes estimated from photometric excesses from bright [\ion{O}{3}]\W\W4960,5008 lines for some sources.
During MSA mask design, 36 primary auroral targets at $z=1.39-4.41$ were incorporated into the two masks, with 20 in GOODS-N and 16 in COSMOS.
The remaining space on the masks was devoted to filler targets that included sources with photometric redshifts at $z>6$; quiescent galaxies at $z>2$; spectroscopic emission-line galaxies at $z>5$; and objects with $z_\mathrm{phot}\ge1.5$.
The final masks included 51 and 46 targets in GOODS-N and COSMOS, respectively.

\subsection{Data reduction}

Reduced and calibrated two-dimensional spectra were obtained using a combination of the standard STScI pipeline and custom software, as described in \citet{shapley2025a}.
The applied steps included masking of saturated and bad pixels, bias and dark current subtraction, removal of cosmic ray snowballs and showers, $1/f$ noise correction, flat fielding, initial flux calibration, wavelength calibration, and combination of individual exposures from the three nod positions.
The error spectra included contributions from Poisson noise, read noise, flat fielding, and variance between exposures.
One-dimensional science and error spectra were obtained using optimal extraction \citep{horne1986}, where the spatial profile was measured from the brightest detected emission line in each grating or the integrated continuum if no lines were present (see \citealt{sanders2024dust} for details).

\subsection{Slit loss correction and flux calibration}

The correction for slit losses was computed by creating a model of the galaxy light profile as a function of wavelength based on the wavelength-dependent PSF model and {\it JWST}/NIRCam F115W imaging (or {\it HST}/WFC3 F160W imaging for 6 targets lacking NIRcam coverage).
The fraction of total light at each wavelength passing through the microshutter slitlet (including bar effects) and 1D extraction window was calculated, and the 1D science and error spectra were divided by this fraction to correct for slit losses.
Full details of the slit loss corrections can be found in \citet{reddy2025}.

The final flux calibration was achieved in two stages, described fully in \citet{sanders2024dust}.
First, the relative flux calibration between gratings was fine-tuned by comparing measured emission line fluxes and continuum flux densities in regions of spectral overlap between neighboring gratings, and scaling the G140M and G395M spectra to match G235M.
The final absolute flux calibration was then achieved by passing the spectra through NIRCam and {\it HST} imaging filters to produce mock photometry, and scaling the spectra in all three gratings by a factor that forces the median ratio of spectroscopic-to-imaging flux densities for all available filters to be unity.
Since robust \te\ and direct-method metallicity constraints rely on emission-line ratios, some of which are widely separated in wavelength or involve lines covered by different gratings, the fidelity of the relative flux calibration is of particular importance.
We tested the relative flux calibration by comparing fluxes measured in different gratings for 156 cases where an emission line was detected at $S/N\ge5$ in two gratings, finding a median offset of 0.1\% with an intrinsic scatter of 8\%.
Accordingly, flux ratios of lines covered by different gratings are reliable.

\begin{figure*}
\centering
\includegraphics[width=\textwidth]{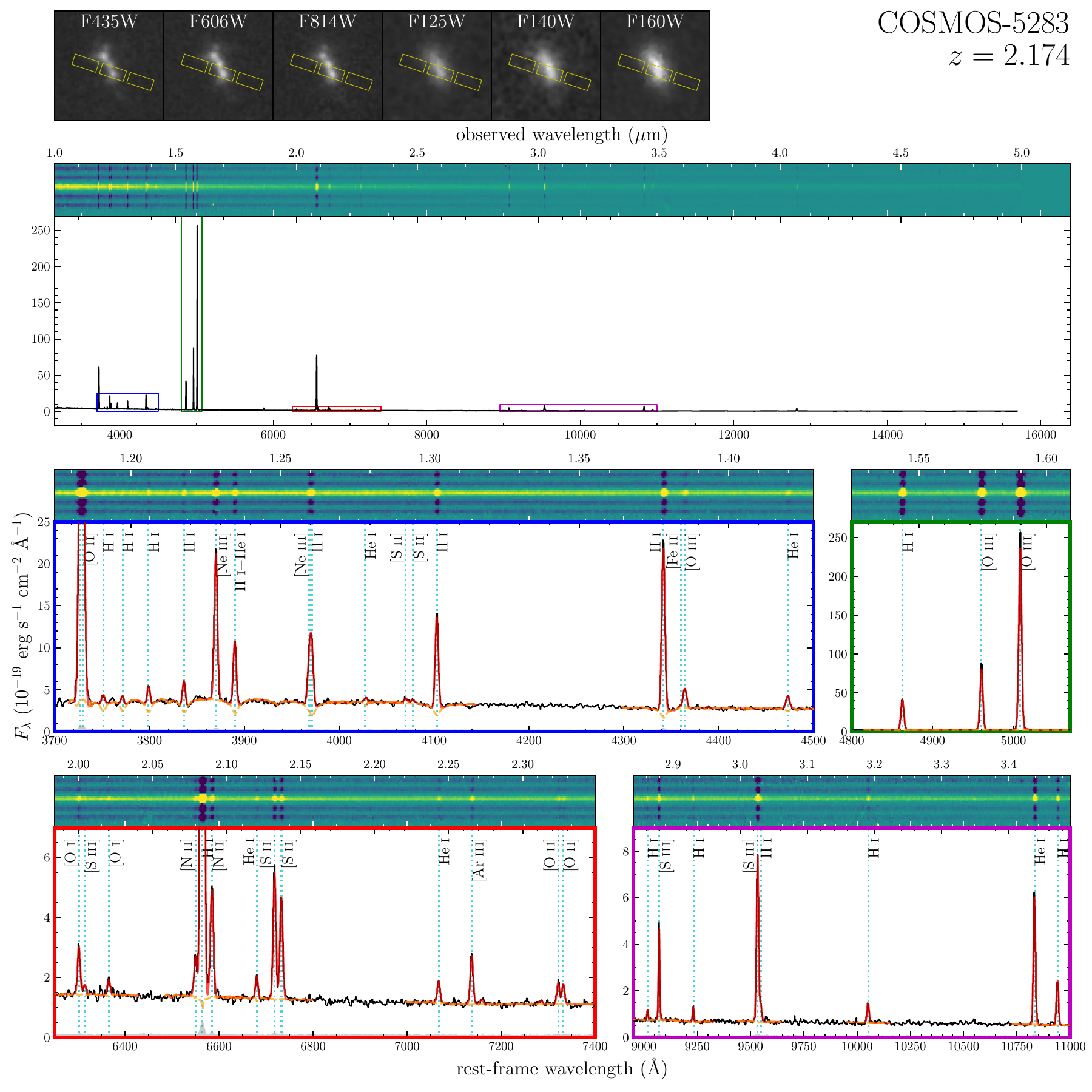}
\caption{Imaging and spectroscopic data for AURORA target COSMOS-5283, a star-forming galaxy at $z=2.174$.
The top row of panels shows 2$^{\prime\prime}$$\times$2$^{\prime\prime}$ cutout images in six {\it HST} filters, with the AURORA three-microshutter slitlet at the central nod position overlaid in yellow.
The second row displays the two- and one-dimensional AURORA {\it JWST}/NIRSpec spectra.
The 1D science spectrum is shown in black, while the 1$\sigma$ error spectrum is shown in gray shading.
The bottom four panels show zoom-ins on different wavelength regions, with axis border colors corresponding to the boxes in the full spectrum, covering specific lines of interest to this analysis including auroral emission lines.
In the zoom-in panels, the orange dashed line shows the nebular and stellar continuum model while the red line displays the full best-fit model including Gaussian emission line fits and continuum.
}\label{fig:spec5283}
\end{figure*}

\begin{figure*}
\centering
\includegraphics[width=\textwidth]{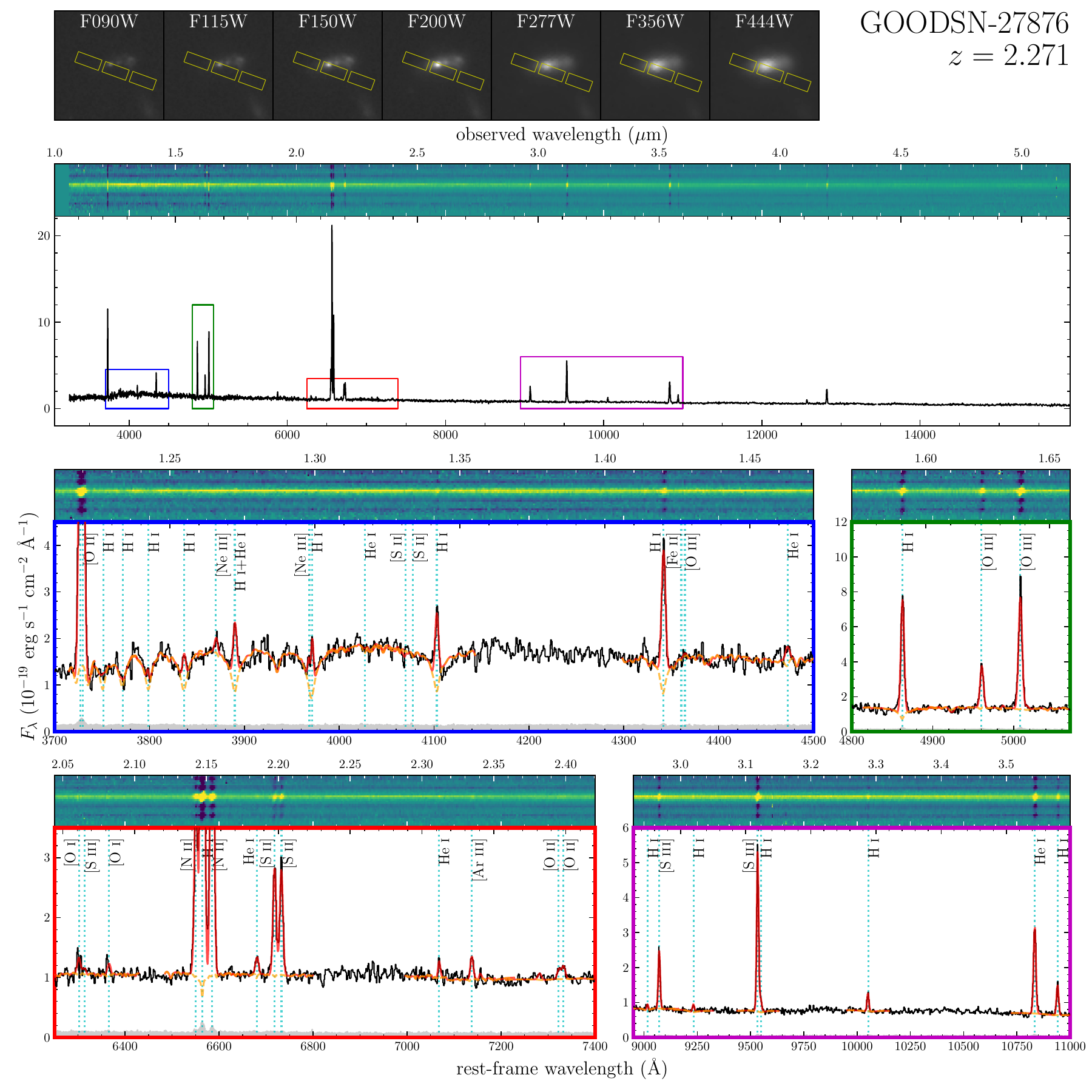}
\caption{Imaging and spectroscopic data for AURORA target GOODSN-27876, a star-forming galaxy at $z=2.271$.
The top row of panels shows 2$^{\prime\prime}$$\times$2$^{\prime\prime}$ cutout images in seven {\it JWST}/NIRCam wideband filters, with the AURORA three-microshutter slitlet at the central nod position overlaid in yellow.
The second row displays the two- and one-dimensional AURORA {\it JWST}/NIRSpec spectra.
The 1D science spectrum is shown in black, while the 1$\sigma$ error spectrum is shown in gray shading.
The bottom four panels show zoom-ins on different wavelength regions, with axis border colors corresponding to the boxes in the full spectrum, covering specific lines of interest to this analysis including auroral emission lines.
In the zoom-in panels, the orange dashed line shows the nebular and stellar continuum model while the red line displays the full best-fit model including Gaussian emission line fits and continuum.
}\label{fig:spec27876}
\end{figure*}

The final reduced and calibrated 2D and 1D spectra, with zoom-ins on certain wavelength regions covering key emission features, are shown in Figures~\ref{fig:spec5283} and \ref{fig:spec27876} for two example targets chosen to exemplify sources at moderate and high direct oxygen abundances.
COSMOS-5283 (Fig.~\ref{fig:spec5283}) is a moderate metallicity (12+log(O/H$)=8.27\pm0.03$; 0.4~\zsun) galaxy with high-equivalent width emission lines, for which multiple auroral lines are detected ([\ion{O}{3}]\W4364, [\ion{O}{2}]\W\W7322,7332, [\ion{S}{3}]\W6314, and [\ion{S}{2}]\W4070).
GOODSN-27876 (Fig.~\ref{fig:spec27876}) is a roughly solar-metallicity object (12+log(O/H$)=8.65\pm0.16$) with lower equivalent width lines, a clear Balmer break, and significant Balmer absorption, for which only the low-ionization [\ion{O}{2}]\W\W7322,7332 auroral doublet is detected.

\subsection{Stellar population properties}\label{sec:sed}

We inferred stellar population properties by fitting photometric SEDs for each AURORA target. 
All but 6 AURORA targets are covered by extensive multi-filter {\it JWST}/NIRCam and {\it HST}/ACS and WFC3 imaging spanning 4,000~\AA\ to 5~$\mu$m in the observed frame.
The {\it HST} imaging is from the CANDELS \citep{grogin2011,koekemoer2011} and 3D-HST \citep{skelton2014} surveys.
The {\it JWST} imaging was obtained as part of the PRIMER survey in COSMOS \citep{donnan2024} and the JADES, FRESCO, and JEMS programs in GOODS-N \citep{eisenstein2026,oesch2023,williams2023}.
For objects with NIRCam coverage, we make use of the imaging reductions and photometric measurements from the DAWN {\it JWST} Archive catalogs\footnote{\url{https://dawn-cph.github.io/dja/}} \citep{valentino2023}.
For the 6 sources (two in GOODS-N and four in COSMOS) that fell outside of the JADES and PRIMER NIRCam footprints, we used the combined {\it HST}, Spitzer, and ground-based photometric catalogs from 3D-HST.
Postage stamp images in selected filters with the microshutter slitlet overlaid are displayed for two example objects at the top of Figures~\ref{fig:spec5283} and \ref{fig:spec27876}.

Stellar population properties were derived using the SED fitting code FAST \citep{kriek2009} with FSPS models \citep{conroy2009}, assuming a delayed-$\tau$ star-formation history ($\mathrm{SFR}\propto t e^{-t/\tau}$) and a \citet{chabrier2003} initial mass function (IMF).
The delayed-$\tau$ parameterization was chosen for its flexibility to represent rising, falling, or approximately constant formation histories.
The SED of each target was modeled under two sets of stellar metallicity and dust curve assumptions: $Z_*=0.019$ and the \citet{calzetti2000} attenuation curve or $Z_*=0.0031$ and the \citet{gordon2003} SMC extinction curve.
The best-fit parameters and stellar continuum model were taken from the one that yielded a lower $\chi^2$ statistic.
Before SED fitting, a nebular emission model including lines and continuum was created based on the measured line fluxes in the AURORA spectra and subtracted off of the measured photometric flux densities to isolate the stellar continuum component.
The nebular continuum component was derived from a grid of Cloudy photoionization models \citep{ferland2017} with an incident ionizing spectrum drawn from BPASS v2.1 binary stellar population models \citep{eldridge2017} assuming a constant star-formation rate over 100~Myr and a $4\times[\alpha/\mbox{Fe}]_\odot$ abundance pattern in the massive stars.
The nebular continuum in each model was normalized to the H$\beta$ intensity.
Nebular metallicities were varied over 12+log(O/H$)=7.4-8.9$. The model adopted for each target is based on a metallicity estimate derived from measured strong-line ratios and the high-redshift analog calibrations of \citet{bian2018}, and was normalized to the measured H$\beta$ flux.
The nebular emission-line component is based on Gaussian line profiles measured from the NIRSpec data.
Full details of this nebular model can be found in \citet{sanders2024dust}.

\subsection{Emission-line measurements}

Full details of the emission-line fitting is described in \citet{sanders2024dust}.
Emission line centroids, fluxes, and widths were measured by fitting Gaussian profiles to the 1D science spectra.
Lines separated in wavelength by $\Delta\lambda/\lambda<0.01$ were fit simultaneously with multiple Gaussians.
Lines from closely spaced doublets of the same ion (e.g., [\ion{O}{2}]\W\W7322,7332; [\ion{S}{2}]\W\W6718,6733) were constrained to have the same width.
The [\ion{O}{2}]\W\W3727,3730 doublet components cannot be robustly separated at the resolution of the observations, but this doublet was still fit with two Gaussians with tied widths and redshifts and the sum of the doublet flux was saved, which we refer to as [\ion{O}{2}]\W3728.
The line (or set of lines) and continuum model was fit over a wavelength interval of $\pm0.01\lambda_{\mathrm{mean}}$ about the mean wavelength $\lambda_{\mathrm{mean}}$ of the emission line(s).
During fitting, line centroids were allowed to vary within 50~km~s$^{-1}$ of the systemic redshift as measured from the highest-S/N emission line (H$\alpha$ or [\ion{O}{3}]\W5008 for the majority of sources).
The observed-frame line width of each line in \AA\ was constrained to be within 20\% of the value that yields an instrument-corrected velocity width matching the intrinsic velocity width measured for the highest-S/N emission line.
We do not adopt the spectral resolution for each grating as reported in the {\it JWST}/NIRSpec documentation\footnote{\url{https://jwst-docs.stsci.edu/jwst-near-infrared-spectrograph/nirspec-instrumentation/nirspec-dispersers-and-filters}}, which assumes a uniformly illuminated slit.
Instead, we derive instrumental resolutions for each target in each grating using the \texttt{msafit} software package \citep{degraaff2024}, which forward models the effective NIRSpec resolution using a morphological light profile input.
We model the intrinsic light profile of each target in each grating by fitting multiple PSF-convolved 2D S\'{e}rsic profiles to the NIRCam imaging in a filter that overlaps the wavelength coverage of the grating.
This process yields instrumental resolutions that are on average $\sim1.4$ times higher than the uniform illumination case, though our results do not significantly change if we instead adopt the JDox spectral resolutions.

A robust continuum model is of critical importance for accurate line fluxes of weak auroral lines and to account for the effects of stellar absorption on \ion{H}{1} recombination lines.
The continuum under the emission lines was taken to be the sum of the best-fit stellar continuum model and the nebular continuum model described in Sec.~\ref{sec:sed}, convolved with a Gaussian kernel to match the instrumental resolution.
When fitting each line or set of lines, the continuum model was allowed to vary by a multiplicative factor to fine tune the local continuum fit and propagate uncertainty in the continuum level in the NIRSpec data into the line flux errors.
Across the full set of AURORA targets and lines, the median best-fit value of this multiplicative factor is 1.01 with a standard deviation of 0.32.
A median value of unity is expected because the absolute flux calibration is tied to the photometry employed in SED fitting, while the standard deviation reflects the typical continuum S/N.
We confirm that this factor is properly capturing the spectroscopic continuum uncertainty by restricting to the subset of targets and lines where the per-pixel continuum $S/N\ge10$, for which we find a median factor of 1.02 and a smaller standard deviation of 0.12.
Since the emission-line fitting depends on the SED modeling output for the underlying stellar continuum model, and the SED modeling depends on nebular corrections to the photometry based on the line fits, we performed SED modeling and line fitting iteratively to achieve convergence.

Uncertainties on all emission-line properties were estimated by perturbing the 1D science spectra based on the error spectra 1000 times, repeating the fitting process for each realization, and calculating the $1\sigma$ uncertainty as half of the 18th-64th percentile width of the resulting distributions of each parameter.
For lines covered by two gratings in overlapping wavelength regions, the final line parameters are taken to be the inverse-variance weighted mean of the values measured from each grating.
The lower panels of Figures~\ref{fig:spec5283} and \ref{fig:spec27876} display the best-fit continuum and emission-line models overlaid on the 1D spectra.
It can be seen that the spectroscopic continuum, including stellar absorption features, is accurately captured by the continuum model.

\section{Auroral-Line Sample and Metallicity Determination}\label{sec:metallicity}

\subsection{AURORA auroral-line detected sample}\label{sec:aurorasample}

We define a sample of AURORA star-forming galaxies with at least one detected auroral emission line for which direct-method oxygen abundances can be determined.
We remove objects identified as quiescent (4) or with a significant contribution from an active galactic nucleus (AGN; 5) as indicated by the presence of broad H$\alpha$ emission or [\ion{N}{2}]\W6585/H$\alpha>0.5$.
Among the remaining 89 AURORA targets, which we take to have emission lines dominated by star formation, we find 41 galaxies for which at least one auroral line is detected at $S/N\ge3$.
Of these 41 sources, 30 were primary Cosmic Noon auroral targets, indicating a high success rate of 83\% (30/36) for this target class.
The other 11 auroral-detected galaxies were filler targets.

\begin{figure*}
\centering
\includegraphics[width=\textwidth]{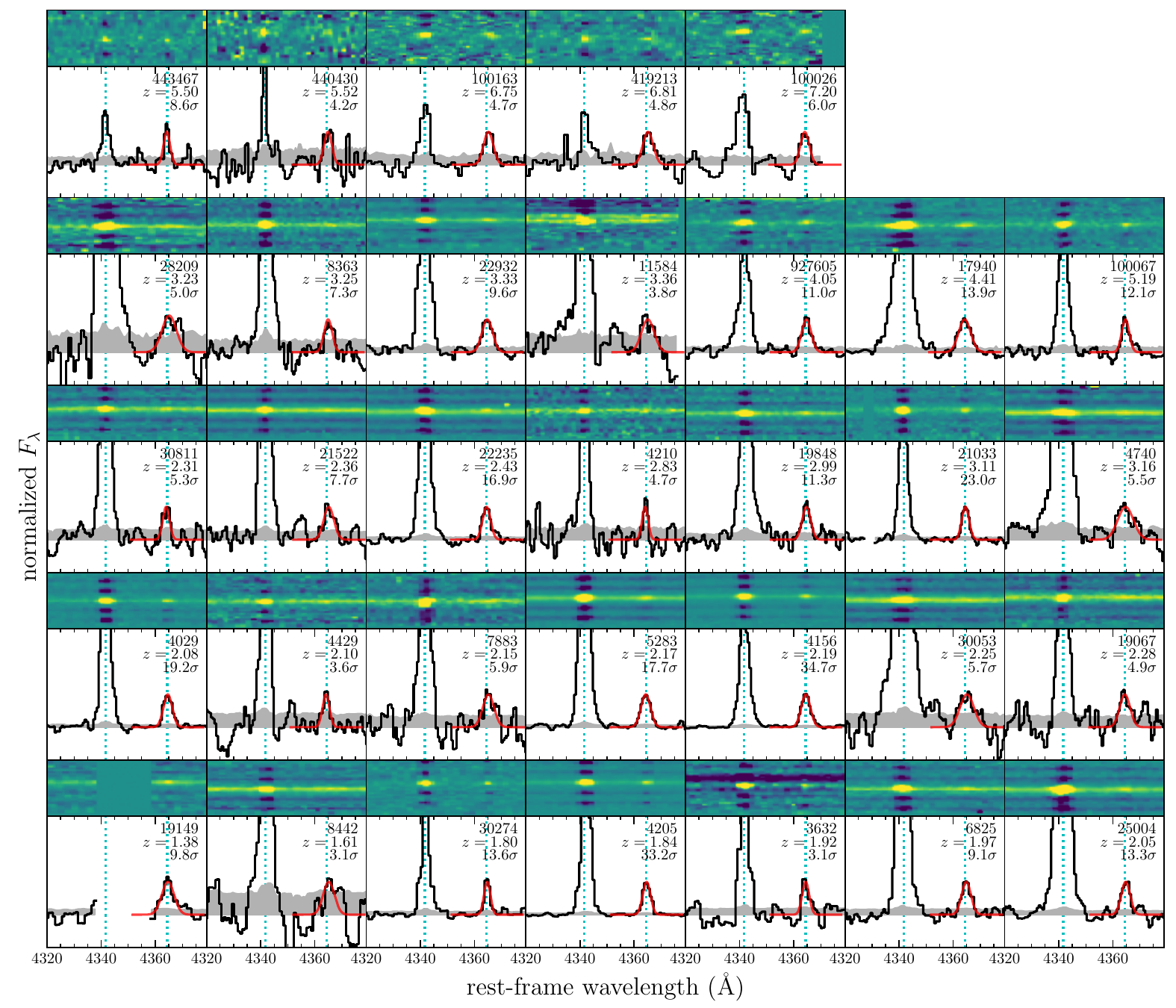}
\caption{Detections of auroral [\ion{O}{3}]\W4364 for 33 AURORA star-forming galaxies.
For each object, the 2D spectrum (top) and continuum-subtracted 1D spectrum (bottom) are displayed, with the science spectrum in black and error spectrum in gray.
The AURORA ID number, redshift, and significance of the detection is reported for each target.
These panels also cover the H$\gamma$ emission line.
The flux density has been normalized so that the peak height of the best-fit Gaussian to the auroral line is the same for each target.
}\label{fig:oiii4364}
\end{figure*}

We identify 33 AURORA star-forming galaxies spanning $z=1.38-7.20$ with detections of the [\ion{O}{3}]\W4364 line, with continuum-subtracted spectra and line fits shown in Figure~\ref{fig:oiii4364}.
The significance of these detections ranges from 3.1$\sigma$ to 34.7$\sigma$, with a median significance of 7.7$\sigma$.
These [\ion{O}{3}]\W4364 lines enable the determination of \te\ in the high-ionization nebular zone, with an ionization energy for O$^{2+}$ of 35.1~eV.
During line fitting, the [\ion{Fe}{2}]\W4361 line that lies just blueward of [\ion{O}{3}]\W4364 was included.
In $z\sim0$ galaxies, this Fe line has been found to become roughly as strong as [\ion{O}{3}]\W4364 at higher metallicities \citep[12+log(O/H$)\gtrsim8.5$;][]{curti2017}.
However, [\ion{Fe}{2}]\W4361 was not detected in any of the AURORA [\ion{O}{3}]\W4364 emitters and thus does not introduce significant contamination, and may be weaker in high-redshift galaxies due to their expected deficit in Fe abundance at fixed O/H as a result of super-solar O/Fe ratios \citep[e.g.,][]{steidel2016,topping2020a,sanders2020,cullen2021}.
Detections of [\ion{Fe}{2}]\W4361 have been reported in a few high-redshift galaxies up to $z\sim6.7$ \citep{shapley2025b,cataldi2025}.

\begin{figure*}
\centering
\includegraphics[width=\textwidth]{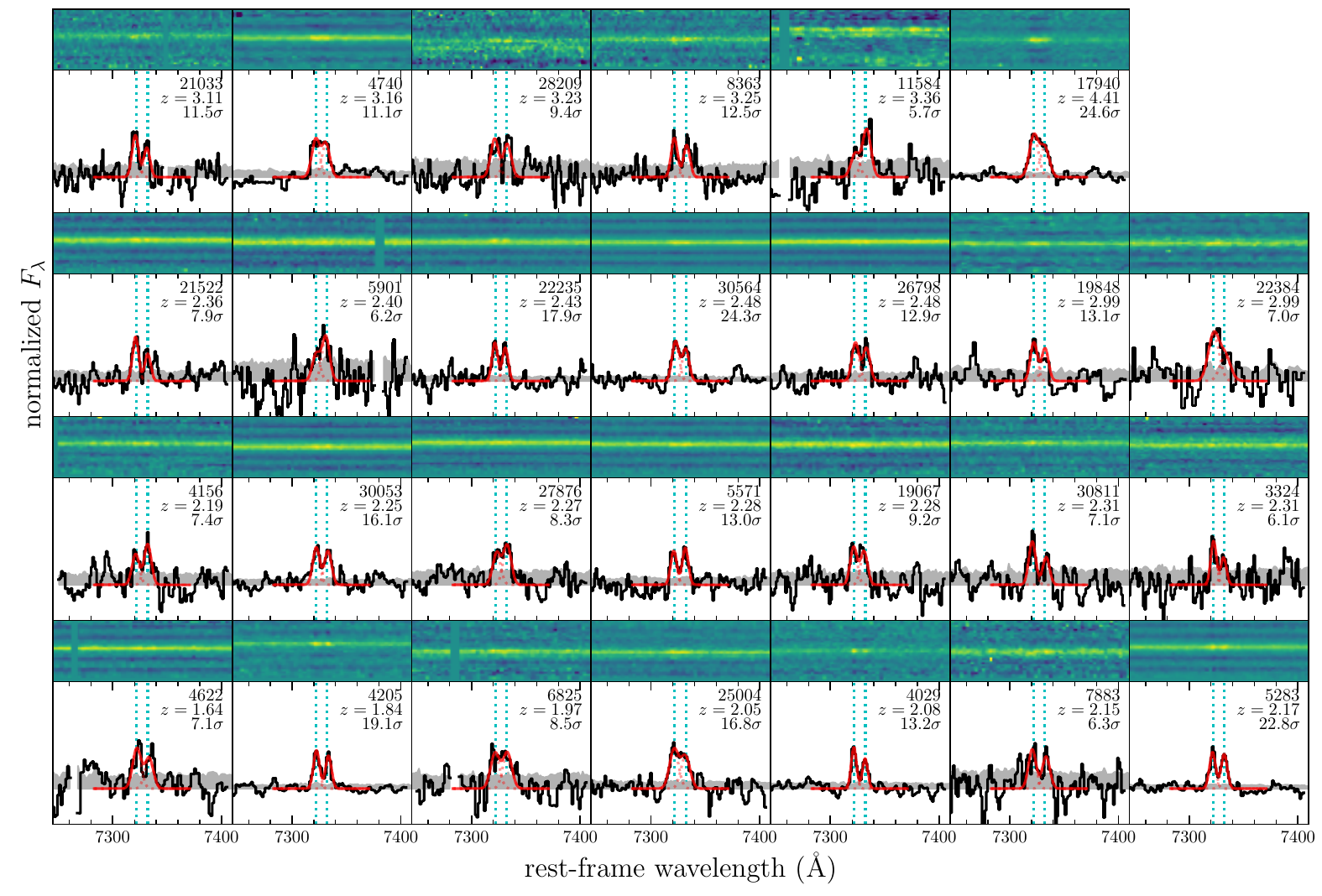}
\caption{Detections of auroral [\ion{O}{2}]\W7322,7332 for 27 AURORA star-forming galaxies, with lines and information as described in Figure~\ref{fig:oiii4364}.
The significance reported is on the total doublet flux.
}\label{fig:oii7327}
\end{figure*}

In Figure~\ref{fig:oii7327}, we present detections of the [\ion{O}{2}]\W\W7322,7332 auroral doublet for 27 star-forming galaxies at $z=1.38-7.20$, providing constraints on \te\ in the low-ionization nebular zone (the O$^+$ ionization energy is 13.6~eV).
The significance of the summed doublet flux ranges from 5.7$\sigma$ to 24.6$\sigma$, with a median value of 11.1$\sigma$.
This apparent doublet, which is actually a quadruplet at rest-frame 7320.9~\AA, 7322.0~\AA, 7331.7~\AA, and 7332.7~\AA, has a theoretical ratio of [\ion{O}{2}]\W\W7322/$7332\approx1.2$ as calculated from \texttt{pyneb} \citep{luridiana2015}.
It can be seen that, for objects with the highest S/N, the blue component is slightly larger than the red component in good agreement with the expected intrinsic ratio.
In lower-S/N objects, there are some cases where the red component is equal to or stronger than the blue one.
Collectively, 67\% lie within 1$\sigma$ of the theoretical ratio, 85\% lie within 2$\sigma$, and all but 1 are within 3$\sigma$, with the largest outlier at 3.2$\sigma$, approximately as expected from Gaussian noise.
We find 19 AURORA sources have both [\ion{O}{3}]\W4364 and [\ion{O}{2}]\W\W7322,7332 detections, for which the high-ionization (\temotp) and low-ionization (\temop) temperatures are simultaneously constrained from observations.

\begin{figure}
\centering
\includegraphics[width=\columnwidth]{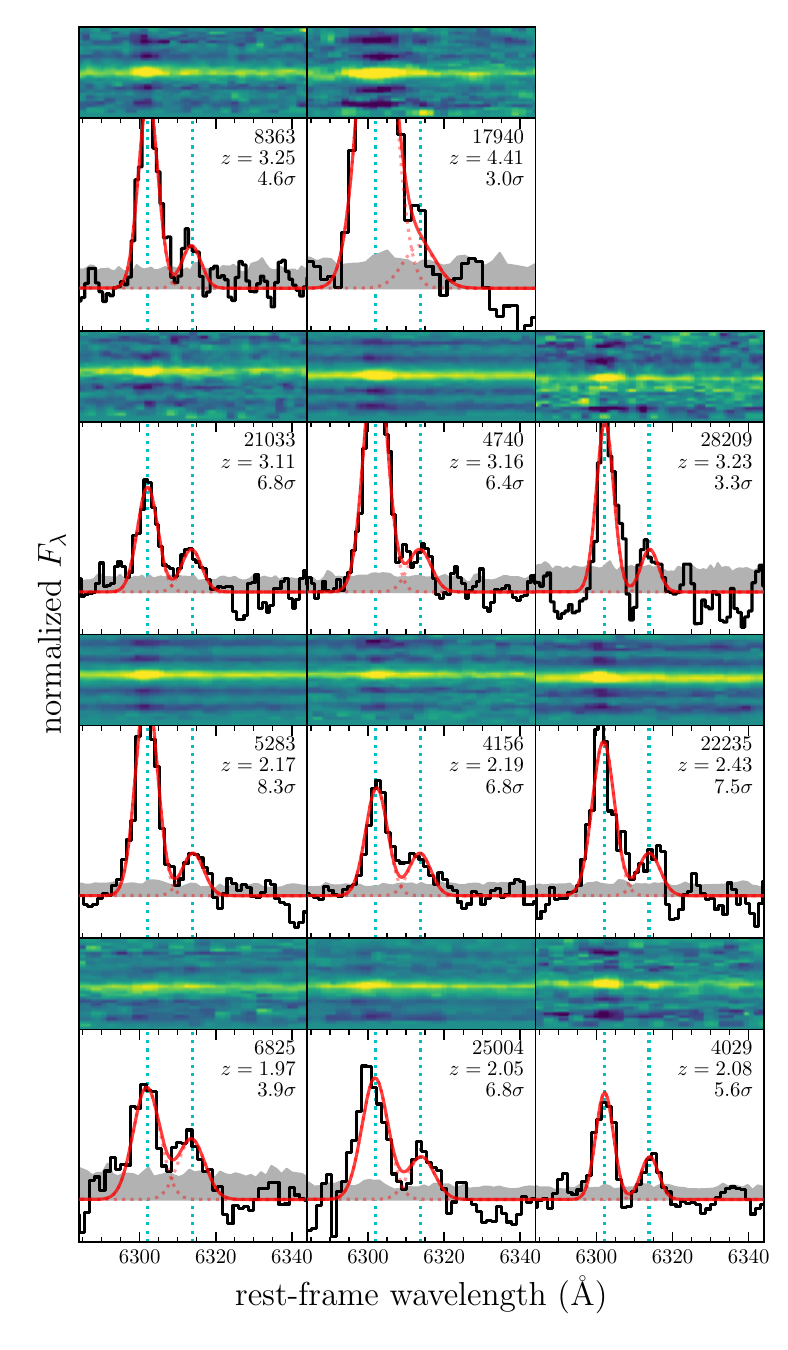}
\caption{Detections of auroral [\ion{S}{3}]\W6314 for 11 AURORA star-forming galaxies, with lines and information as described in Figure~\ref{fig:oiii4364}.
We also show the fit to [\ion{O}{1}]\W6302 given its close proximity to the [\ion{S}{3}] auroral line.
}\label{fig:siii6314}
\end{figure}

\begin{figure}
\centering
\includegraphics[width=\columnwidth]{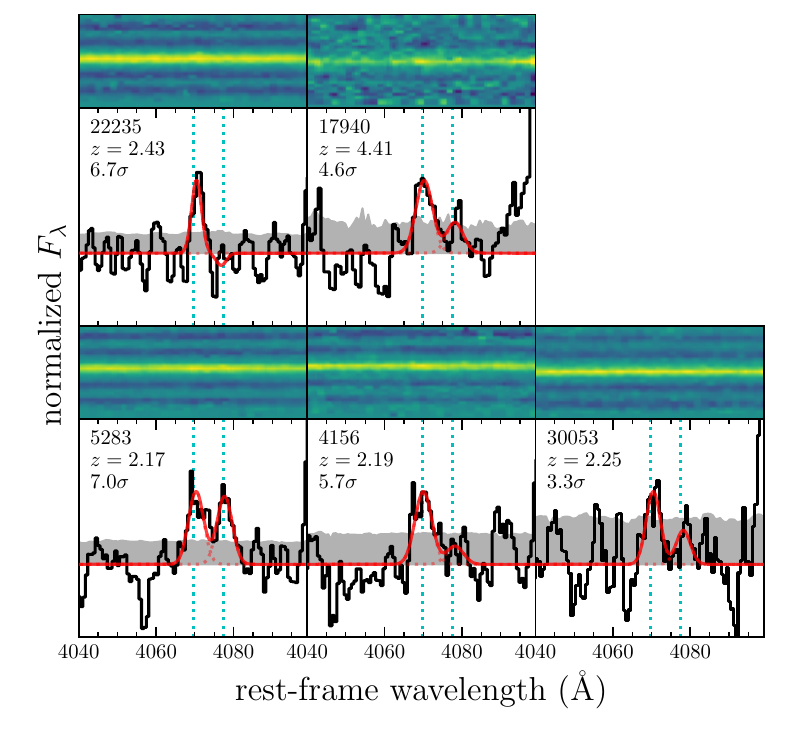}
\caption{Detections of auroral [\ion{S}{2}]\W4070 for 5 AURORA star-forming galaxies, with lines and information as described in Figure~\ref{fig:oiii4364}.
}\label{fig:sii4070}
\end{figure}

We also find detections of sulfur auroral emission lines, including 11 detections of [\ion{S}{3}]\W6314 and 5 detections of [\ion{S}{2}]\W4070, shown in Figures~\ref{fig:siii6314} and \ref{fig:sii4070}.
With an ionization energy of 23.3~eV, \temstp\ derived from [\ion{S}{3}]\W6314 probes the temperature in the intermediate-ionization zone, while [\ion{S}{2}]\W4070 (10.4~eV) traces the low-ionization zone similar to O$^+$.
All of the objects with [\ion{S}{3}] and/or [\ion{S}{2}] auroral detections also have detections of both [\ion{O}{3}]\W4364 and [\ion{O}{2}]\W\W7322,7332, and four out of five of those with [\ion{S}{2}]\W4070 detections are also detected in [\ion{S}{3}]\W6314.
Given the range of ionization energies these lines probe, the AURORA auroral-line detected sample thus offers valuable insight into the nebular temperature structure at high redshift.

\subsection{Literature auroral-line detected sample}\label{sec:litsample}

We supplement the sample drawn from AURORA with a sample of high-redshift auroral-line detected star-forming galaxies selected from the literature.
We searched for sources at $z>1$ with published detections of at least one auroral emission line at S/N$\ge$3, either [\ion{O}{2}]\W\W7322,7332, [\ion{O}{3}]\W4364, or the rest-ultraviolet \ion{O}{3}]\W1666.
We rejected sources that had a significant probability of a non-stellar ionizing source based on broad \ion{H}{1} line emission, very high-ionization lines, or line ratio modeling \citep[e.g.,][]{ubler2023,ji2024,cullen2025,curti2025}.
We searched both ground-based and {\it JWST} spectroscopic studies.
For objects with {\it JWST}/NIRSpec spectroscopy, we required that the spectral resolution was sufficient to cleanly separate H$\gamma$ and [\ion{O}{3}]\W4364, which effectively limited our search to sources with medium- or high-resolution NIRSpec grating observations (i.e., excluding the prism mode).
The final requirement was the availability of cataloged emission-line fluxes so that the physical properties of the literature sample could be recomputed using the same analysis pipeline as for the AURORA sample.
The final literature auroral-line detected sample includes 98 galaxies at $z=1.33-10.60$, with a median redshift of 4.24.
In this sample, 73 galaxies have detections of [\ion{O}{3}]\W4364 only, 2 have detections of [\ion{O}{2}]\W\W7322,7332 only, 9 have detections of both [\ion{O}{3}]\W4364 and [\ion{O}{2}]\W\W7322,7332, and the remaining 14 have detections of \ion{O}{3}]\W1666 without any detected rest-optical auroral lines.
Out of the total of 98, 17 are based on observations from the ground and the other 81 have {\it JWST} spectroscopy.
When available, stellar masses and SED-based SFRs were also drawn from literature sources for this sample, and placed onto a \citet{chabrier2003} IMF scale if a different IMF was assumed in the original reference.
Further details about the literature sample can be found in Appendix~\ref{app:lit} and Table~\ref{tab:lit}.

\subsection{Combined high-redshift auroral sample}\label{sec:sample}

\begin{figure*}
\centering
\includegraphics[width=\textwidth]{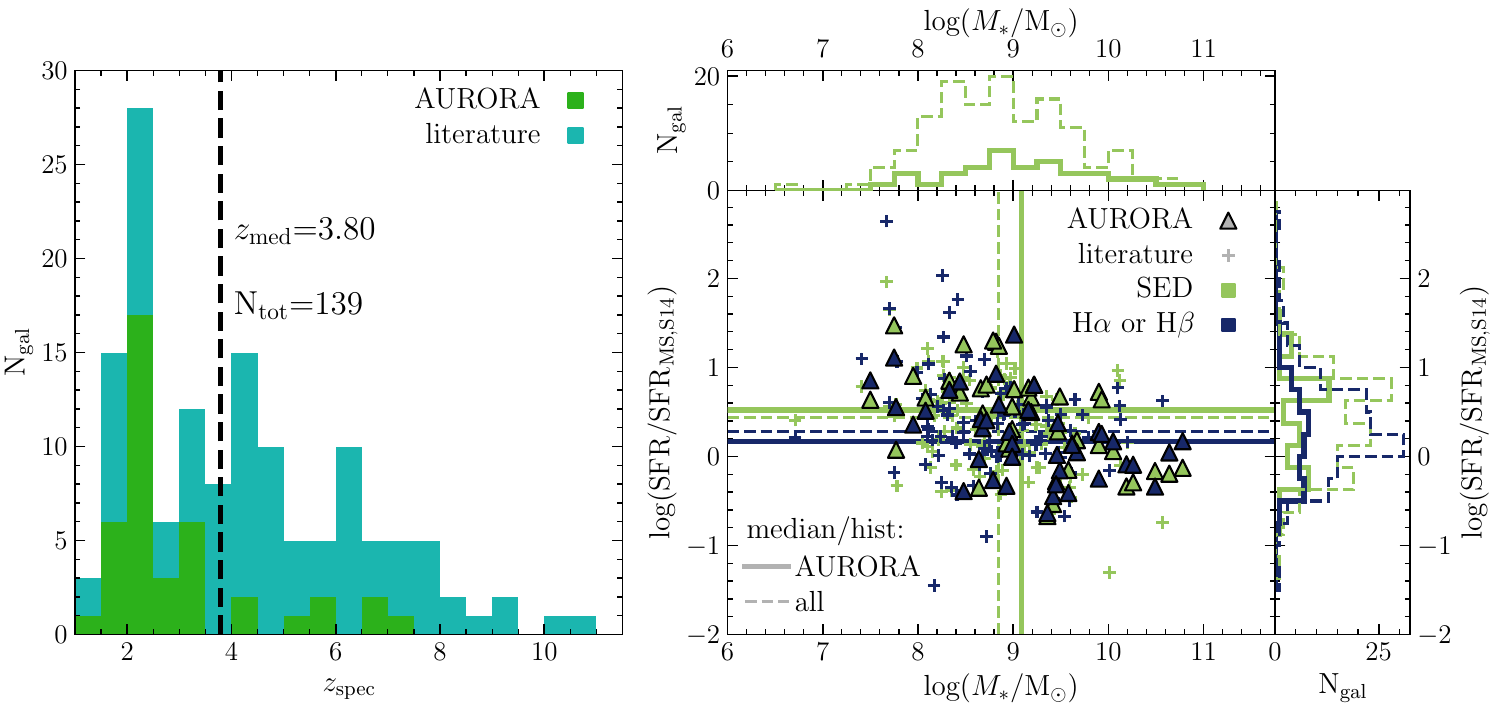}
\caption{Properties of the auroral-line detected high-redshift samples from AURORA and drawn from the literature.
The left panel shows the redshift distributions.
The right panel displays offset from the parameterized star-forming main sequence of \citet{speagle2014} vs.\ stellar mass, using SFRs derived from SED fitting and from dust-corrected H$\alpha$ or H$\beta$ luminosities.
The top and right panels display histograms in stellar mass and main-sequence offset, respectively, with solid lines denoting AURORA and dashed lines denoting the combined sample.
Median lines in the main panel follow the same presentation.
}\label{fig:sample}
\end{figure*}

The properties of the 139 galaxies in the combined AURORA and literature high-redshift auroral-line sample are shown in Figure~\ref{fig:sample}.
The sample spans a redshift range of $z=1.3-10.6$ with a median redshift of $z_\mathrm{med}=3.80$.
The majority of the sample lies within and slightly earlier than the Cosmic Noon epoch ($z\sim1.5-4.5$), but there is a non-negligible tail toward higher redshifts with $\sim30$ galaxies at $z=5-8$.
Stellar masses span $\log(M_*/\mathrm{M}_\odot)\sim7.5-10.5$ with a median mass of $10^{8.8}$~\msun.
The AURORA sample has a median stellar mass of $10^{9.2}$~\msun\ and includes more high-mass galaxies than the literature sample, extending to higher metallicities.

The vertical axis of the right panel of Fig.~\ref{fig:sample} displays the offset from the star-forming main sequence based on the parameterization of \citet{speagle2014}, for SFRs based on both SED fitting and dust-corrected H$\alpha$ luminosity (Sec.~\ref{sec:sfr}).
The median main-sequence offset of the combined sample is 0.25~dex based on SFR(H$\alpha$) and 0.4~dex when using SFR(SED).
The auroral-line sample is thus biased toward higher specific SFRs on average than the typical star-forming population at these redshifts, a consequence of selection based on the detection of faint auroral lines.
However, the median offset is approximately equal to the intrinsic $1\sigma$ scatter in the main sequence \citep[][]{whitaker2012,clarke2024} such that roughly half of the sample falls within $\pm1\sigma$ of the main sequence.
This is a significant improvement over the selection bias affecting ground-based and early {\it JWST} auroral samples that lay nearly an order of magnitude above the main-sequence on average \citep[e.g.,][]{sanders2020,sanders2024,laseter2024}.

\begin{figure}
\centering
\includegraphics[width=0.49\textwidth]{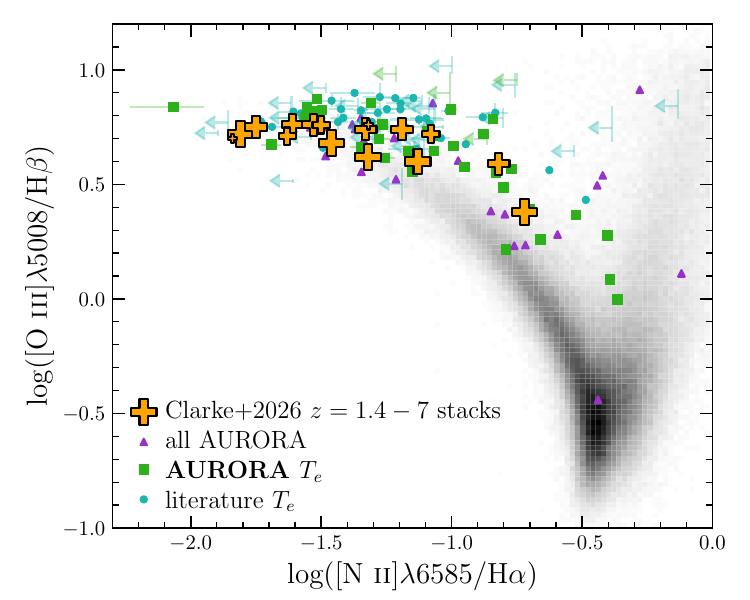}
\caption{[\ion{O}{3}]$\lambda$5008/H$\beta$ vs.\ [\ion{N}{2}]$\lambda$6585/H$\alpha$ BPT diagram.  The AURORA and literature auroral-line detected samples are shown as green squares and turquoise circles, respectively.
Sources from the full AURORA survey without auroral-line detections are displayed as purple triangles.
The distribution of $z\sim0$ galaxies and AGN from SDSS is presented in the grayscale histogram.
Orange plus signs denote stacked JADES spectra of 693 galaxies at $z=1.4-7$ from \citet{clarke2026}, with smaller point size corresponding to higher redshift.
The second-lowest redshift bin at $z=2.7-4.0$ is most closely matched to the median redshift of the combined auroral sample ($z_\mathrm{med}=3.80$).
}\label{fig:bpt}
\end{figure}

We present the combined high-redshift auroral sample in the [\ion{O}{3}]$\lambda$5008/H$\beta$ vs.\ [\ion{N}{2}]$\lambda$6585/H$\alpha$ ``BPT'' diagram in Figure~\ref{fig:bpt}.
We compare these sources to the AURORA galaxies without auroral-line detections (purple triangles), and $z\sim0$ galaxies and AGN based on spectroscopic measurements from the MPA-JHU Sloan Digital Sky Survey catalog \citep[SDSS;][]{brinchmann2004,tremonti2004}\footnote{\url{https://wwwmpa.mpa-garching.mpg.de/SDSS/DR7/}}.
We also show stacked JADES spectra of 693 star-forming galaxies at $z=1.4-7.0$ from \citet{clarke2026}, divided into subsets at $z=1.4-2.7$, $2.7-4.0$, $4.0-5.0$, and $5.0-7.0$ (orange plus signs; decreasing point size with increasing redshift).
The high-redshift auroral sample displays the well-established offset from the $z\sim0$ sequence of star-forming galaxies.
The auroral sample displays a larger offset on average from the $z\sim0$ sequence than typical $z=1.4-2.7$ galaxies, but follows a similar locus to that of the $z=2.7-4.0$ stacks that are better matched the median redshift of the auroral sample ($z_\mathrm{med}=3.80$).
This agreement suggests that the auroral sample has ionization conditions that are similar to typical star-forming galaxies at $z\sim3-4$, but distinct from local galaxies.

\subsection{Physical property derivation}\label{sec:properties}

Physical properties derived from the AURORA and literature emission line measurements include the dust reddening, electron temperatures, electron density, SFR(H$\alpha$), ionic abundances, and total oxygen abundance.
Our procedure for calculating these properties is described below.
The same procedure is applied to both the AURORA and literature sources, starting with the measured line fluxes to yield a self-consistent set of physical properties.
The uncertainty on each parameter is estimated using a Monte Carlo technique in which the measured line fluxes are perturbed according to the flux errors assuming Gaussian noise, and all properties are recomputed as described below.
This process is repeated for 1000 realizations to sample the posterior distribution of each parameter, and the 1$\sigma$ uncertainty is taken to be half of the 16th-84th percentile width of the distribution for each parameter.
This approach properly accounts for covariances among derived properties, under the assumption that the line fluxes are independent.
The derived physical properties for the AURORA auroral-detected sample are presented in Appendix~\ref{app:aurora} and Table~\ref{tab:aurora}.

\subsubsection{Reddening, temperature, and density}\label{sec:temden}

Dust reddening, parameterized as the color excess \ebvgas, was determined from flux ratios of the detected \ion{H}{1} recombination lines relative to H$\beta$.
We calculated \ebvgas\ by finding the value that minimizes the expression
\begin{equation}\label{eq:ebvgas}
\chi^2=\sum_i{\frac{\left(R_{\mathrm{obs},i} - R_{\mathrm{int},i}10^{-0.4E(B-V)_\mathrm{gas}[k(\lambda_i)-k(\lambda_{\mathrm{H}\beta})]}\right)^2}{\sigma_{\mathrm{obs},i}^2}} 
\end{equation}
where $R_{\mathrm{obs},i}$ is the measured flux ratio of \ion{H}{1}$(\lambda_i)$/H$\beta$,
 $\sigma_{\mathrm{obs},i}$ is the uncertainty on the measured flux ratio,
 $R_{\mathrm{int},i}$ is the theoretical intrinsic intensity ratio of these two \ion{H}{1} lines,
and $k(\lambda)$ is the adopted dust curve, for which we adopt the \citet{cardelli1989} Milky Way extinction law.
This expression was summed over all Balmer and Paschen series \ion{H}{1} lines detected at $S/N\ge5$.
The intrinsic intensity ratios were calculated using \texttt{pyneb} using Case B recombination rates from \citet{storey1995} at the measured \te\ and \den\ values described below.
All emission-line flux ratios are dust-corrected using the best-fit \ebvgas, assuming the \citet{cardelli1989} extinction law.

\citet{sanders2024dust} and \citet{reddy2025} have recently shown that the measured \ion{H}{1} line ratios suggest that the nebular dust attenuation curves for some AURORA sources differ from the \citet{cardelli1989} Milky Way extinction law, and that this result may be true on average for high-redshift galaxies.
We have assumed the \citet{cardelli1989} curve for all sources in our analysis for uniformity as well as consistency with past studies.
If we instead adopt the average high-redshift attenuation curve from \citet{reddy2025}, the median and mean direct oxygen abundances derived for the AURORA sample decrease by $0.01\pm0.03$~dex and $0.06\pm0.04$~dex, respectively.
There is thus not a large shift in the derived metallicities on average if the \citet{reddy2025} dust curve is used.
This result is largely due to the fact the [\ion{O}{3}]\W4364/\W5008 and [\ion{O}{3}]\W5008/H$\beta$ ratios used to derive \temotp\ and O$^{2+}$/H are not strongly sensitive to the reddening correction, coupled to the fact that the majority of O is in O$^{2+}$ for moderate- and low-metallicity galaxies.
In contrast, [\ion{O}{2}]\W\W7322,7332/\W3728 and [\ion{O}{2}]\W3728/H$\beta$ are much more sensitive to the reddening correction, such that \temop\ and O$^{+}$/H are more strongly influenced by the choice of dust curve.
There is thus larger systematic uncertainty due to the adopted dust law for high-metallicity, low-excitation galaxies, especially those for which the only detected auroral line is [\ion{O}{2}]\W\W7322,7332.
Indeed, all 8 galaxies in AURORA for which [\ion{O}{2}]\W\W7322,7332 is the only detected auroral line have relatively high degrees of dust reddening under our fiducial set of assumptions (\ebvgas=$0.24-0.77$).
Future work using larger samples of high-metallicity galaxies with auroral line detections and simultaneous constraints on the nebular attenuation curve is required to robustly asses the impact of dust curve systematics on the derived strong-line calibrations.

The density-sensitive [\ion{O}{2}]\W\W3727,3730 doublet is not resolved in the medium-resolution NIRSpec gratings.
We instead derived electron densities from the [\ion{S}{2}]\W\W6718,6733 doublet.
We used \texttt{pyneb} to compute \densp\ from the measured [\ion{S}{2}]\W\W6718/\W6733 flux ratio, adopting the S$^+$ collision strengths from \citet{tayal2010}.
If [\ion{S}{2}]\W\W6718,6733 measurements are not available or both doublet components are not detected at $S/N\ge3$, then we adopt the density from the best-fit scaling of [\ion{S}{2}] and [\ion{O}{2}] densities with redshift over $z=0-10$ from \citet{topping2025}:
\begin{equation}\label{eq:zden}
n_\mathrm{e}=(1+z)^{1.5}\ 40\ \mathrm{cm}^{-3}
\end{equation}
We add 0.1~dex random scatter about this relation in Monte Carlo realizations.
Densities directly constrained from [\ion{S}{2}] were used for 45 objects, while the scaling relation was employed to estimate \den\ for the other 94 sources.
See \citet{topping2025} for a detailed analysis of electron densities in the full AURORA sample.

Electron temperatures were computed from each auroral emission line detected at $S/N\ge3$ using \texttt{pyneb} assuming \den\ as described above.
The high-ionization temperature \temotp\ was calculated from the [\ion{O}{3}]\W4364/\W5008 ratio or \ion{O}{3}]\W1666/\W5008 ratio using the O$^{2+}$ collision strengths of \citet{aggarwal1999} that go up to energy level $n=6$, necessary to compute \te\ from the 6$\rightarrow$3 transition \ion{O}{3}]\W1666.
Derived metallicities change by $\lesssim10\%$ (0.04~dex) if we instead assume the collision strengths from \citet{tayal2017} or \citet{storey2014} (for [\ion{O}{3}]$\lambda$4364, only up to $n=5$).
The low-ionization temperature \temop\ was derived from the [\ion{O}{2}]\W\W7322,7332/\W3728 ratio using O$^+$ collision strengths from \citet{kisielius2009}.
We adopted these O$^+$ collision strengths over those of \citet{tayal2007} because the latter cannot be used above $T_\mathrm{e}=20,000$~K, into which some very metal-poor galaxies in our sample extend.
The alternative low-ionization tracer \temsp\ was computed based on the [\ion{S}{2}]\W4070/\W\W6718,6733 ratio, assuming the collision strengths from \citet{tayal2010}.
The intermediate-ionization temperature \temstp\ was derived from the [\ion{S}{3}]\W6314/\W9533 ratio, adopting the collision strengths of \citet{hudson2012}.
The derived \temsp\ and \temstp\ values are reported in Table~\ref{tab:tems} for the small number of source with a detection of at least one of the S auroral lines.
Energy levels and transition probabilities were from the default \texttt{pyneb} atomic data sources for all ions.

Of the 139 galaxies in the sample, 28 have both \temotp\ and \temop\ constrained from detected auroral lines, while the other 111 have either \temotp\ or \temop.
For those objects with constraints on only one of these temperatures, we infer the other using the relation from \citet{campbell1986}:
\begin{equation}\label{eq:t2t3}
T_{\mathrm{e}}(\mathrm{O}^{+}) = 0.7\times T_{\mathrm{e}}(\mathrm{O}^{2+}) + 3000~\mathrm{K}
\end{equation}

Since fitting \ebvgas\ requires \te\ and \den\ as input, and \te\ calculations require a dust correction and \den, we solve for these three parameters iteratively.
We start with an initial assumption of $T_\mathrm{e}=10,000$~K and $n_\mathrm{e}(z)$ from the \citet{topping2025} relation.
We then iteratively compute \ebvgas, density, and temperatures until convergence is reached, which requires less than 5 iterations for all objects.
The \te\ used in the \ebvgas\ and \den\ calculations is \temotp\ when available, otherwise \temop.

\subsubsection{Ionic and total oxygen abundances}\label{sec:oh}

Ionic and total oxygen abundances were computed with \texttt{pyneb} using the collision strengths adopted in the \te\ calculations and \den\ as described above.
The O$^2+$/H$^+$ abundance ratio was derived from [\ion{O}{3}]\W5008/H$\beta$ using \temotp.
The O$^+$/H$^+$ abundance ratio was derived from [\ion{O}{2}]\W3728/H$\beta$ assuming \temop.
Neutral O and higher ionization states were assumed to be negligible, such that the total oxygen abundance is the sum of these two ionic abundance ratios.

\subsubsection{Star-formation rate}\label{sec:sfr}

SFR is inferred from the dust-corrected luminosity of H$\alpha$ when available.
If H$\alpha$ is not covered, then the H$\beta$ luminosity is scaled to that of H$\alpha$ using the intrinsic ratio calculated with \texttt{pyneb} using \temotp\ and \den\ for each object.
The conversion factor, $C$, from $\mathrm{SFR}/\mathrm{M}_\odot\ \mathrm{yr}^{-1}=10^C\times L(\mathrm{H}\alpha)/\mathrm{erg\ s}^{-1}$ is metallicity dependent since the ionizing photon production rate depends on the abundance of metals in the stellar photospheres.
To obtain a metallicity-dependent conversion factor, we fit a function to the conversion factors $C=[-41.67, -41.59, -41.37]$ at $Z_*=[0.007, 0.004, 0.02]$ derived from BPASS v2.2.1 binary models \citep{stanway2018} with a 100~\msun\ IMF cutoff \citep{reddy2022,shapley2023,clarke2024}:
\begin{equation}\label{eq:sfr}
C(Z_*) = -40.26 + 0.89\log(Z_*) + 0.14\log(Z_*)^2
\end{equation}
For each source, we convert the direct gas-phase oxygen abundance to a bulk metallicity based on the solar value for these properties (12+log(O/H$)_\odot=8.69$; Z$_\odot=0.014$; \citealt{asplund2021}).
We thus infer the bulk metallicity of the young and massive stars using $Z_*=0.014\times 10^{12+\log(\mathrm{O/H}) - 8.69}$ and obtain the conversion factor to compute SFR(H$\alpha$) from equation~\ref{eq:sfr}.

\section{Results}\label{sec:results}

Using the derived \te\ and direct-method oxygen abundances for the sample of 139 high-redshift galaxies, we explore the relations among \te\ in different nebular ionic zones in Sec.~\ref{sec:ttrelations} and present new strong-line metallicity calibrations in Sec.~\ref{sec:cal}.

\subsection{Ionic temperature relations}\label{sec:ttrelations}

\begin{figure*}
\centering
\includegraphics[width=0.49\textwidth]{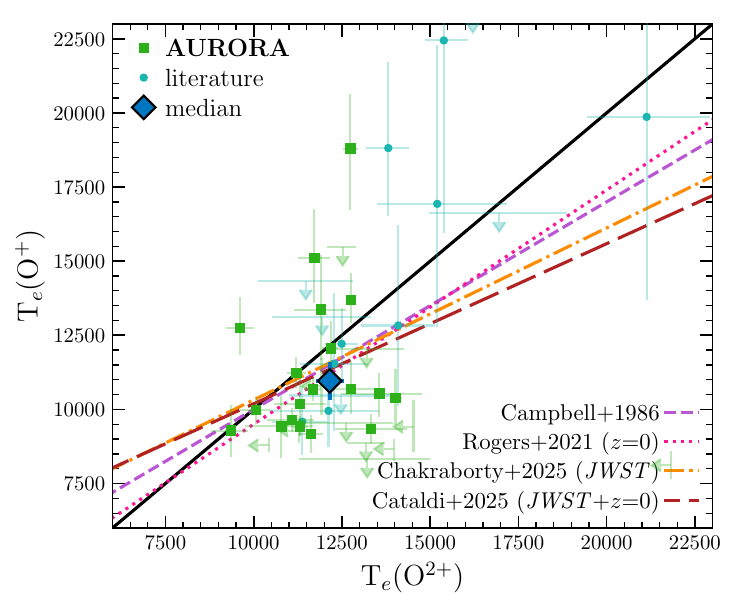}
\includegraphics[width=0.49\textwidth]{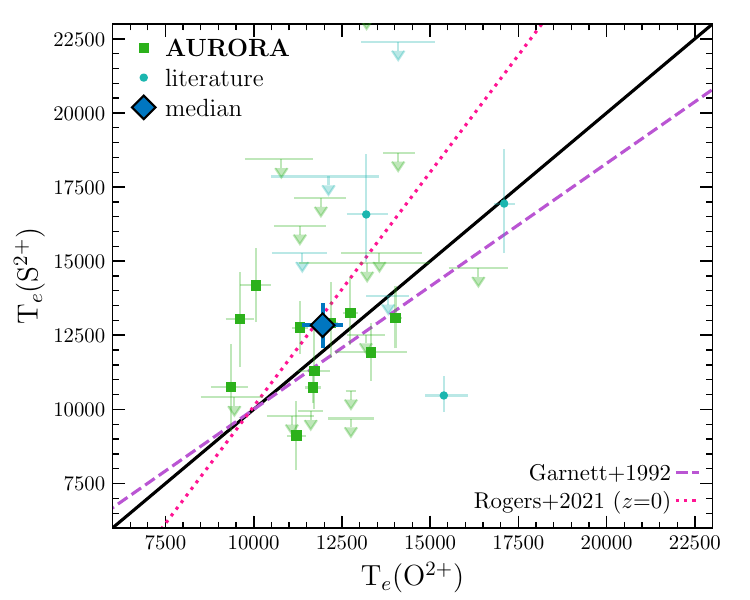}
\includegraphics[width=0.49\textwidth]{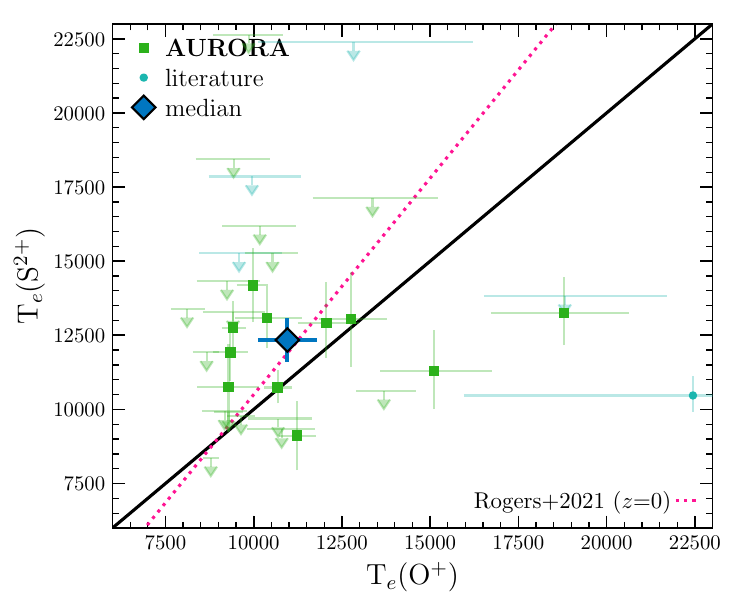}
\includegraphics[width=0.49\textwidth]{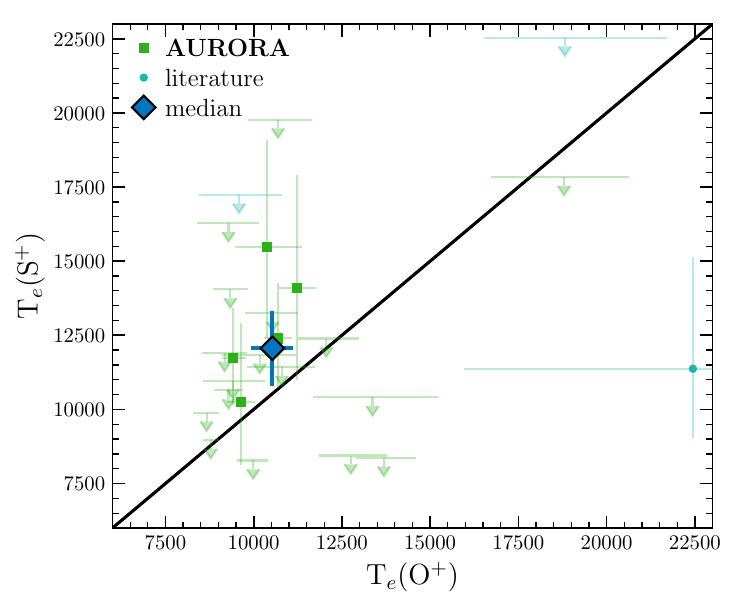}
\caption{Relations among electron temperatures measured from different ions, tracing nebular zones with differing degrees of ionization.
In each panel, the solid black line shows a one-to-one relation.
The dashed and dotted colored lines display different \te-\te\ relations from the literature.
Blue diamonds show median values for the high-redshift sample.
Median ionic electron temperature relations from the high-redshift sample are largely consistent with those of $z=0$ \ion{H}{2} regions.
}\label{fig:tt}
\end{figure*}

We present the relations among \te\ derived from different ions in Figure~\ref{fig:tt}.
In the top left panel, we show \temop\ vs.\ \temotp\ for 28 galaxies in our sample, tracing the low- and high-ionization zones, respectively.
We find significant scatter among the 28 galaxies and that the dynamic \te\ range is not large enough to resolve a correlation, with most sources lying at $T_{\mathrm{e}}(\mathrm{O}^{2+})=11,000-14,000$~K.
The large scatter is similar to what is seen at $z=0$, where intrinsic scatters in $T_\mathrm{e}-T_\mathrm{e}$ relations are $\approx$1,000~k \citep{rogers2021}.
The median value of these 28 sources closely matches relations calibrated to $z=0$ \ion{H}{2} regions \citep{campbell1986,rogers2021}.
Recently calibrated relations including {\it JWST} high-redshift observations also agree well with both the $z=0$ relations and our sample on average \citep{chakraborty2025,cataldi2025}.
The median of the 28 galaxies in our sample is consistent at the 1$\sigma$ level with all four of these relations.

The top right panel of Fig.~\ref{fig:tt} compares the intermediate-ionization \temstp\ to the high-ionization \temotp.
We find that the 14 high-redshift galaxies in our sample with constraints on both temperatures scatter approximately about a one-to-one relation, which may be expected given that the difference in ionization potential and thus the radial range within the \ion{H}{2} regions that these two ions inhabit is smaller than for O$^+$ vs.\ O$^{2+}$.
We again find good agreement with $z=0$ \ion{H}{2} region relations.
The median of the 14 galaxies lies closest to the \citet{rogers2021} line, but remains $\approx$1$\sigma$ consistent with the \citet{garnett1992} relation based on photoionization models designed for use at $z=0$.
A similar result was found based on 6 high-redshift sources by \citet{cataldi2025}.

The lower left panel displays \temstp\ vs.\ \temop\ for 12 high-redshift galaxies, comparing the intermediate- and low-ionization zone temperatures.
The sample median lies $\approx$2$\sigma$ above the one-to-one line, in excellent agreement with the $z=0$ \ion{H}{2} region relation from \citet{rogers2021}.
The two low-ionization tracers, \temsp\ and \temop, are compared in the bottom right panel for 6 galaxies.
All 5 AURORA sources lie above the one-to-one line, though 4 of these are 1$\sigma$ consistent with this line.
The one detected literature source is the Sunburst Arc \citep{welch2025}, which has a large uncertainty due to its high, but significantly uncertain, \den.
The median of these 6 sources lies within 1.1$\sigma$ of the one-to-one relation, as expected if O$^+$ and S$^+$ occupy similar radii near the edges of \ion{H}{2} regions and as seen at $z=0$ \citep[e.g.,][]{mendez2023}.
Using the largest sample of high-redshift galaxies with multi-ion \te\ constraints to date, we thus find no evidence that relations among temperatures in the high, intermediate, and low ionization zones differ at $z=0$ and $z>1$.

\subsection{Metallicity calibrations}\label{sec:cal}

Using the combined high-redshift auroral-line sample of 139 galaxies, we calibrate relations between emission-line ratios and direct-method oxygen abundance.
In total, we provide metallicity calibration coefficients for 19 line ratios.
We first present calibrations of strong-line ratios involving only lines of H and O (Sec.~\ref{sec:ocal}), then continue with line ratios involving Ne (Sec.~\ref{sec:necal}), N (Sec.~\ref{sec:ncal}), S (Sec.~\ref{sec:scal}), and Ar (Sec.~\ref{sec:arcal}).
The relations between line ratios and direct-method metallicity are represented as polynomials with the functional form
\begin{equation}\label{eq:cal}
\log(R) = \sum_i c_i x^i
\end{equation}
where $x=12+\log(\mathrm{O/H})-8.0$, $R$ is the emission-line ratio, and $c_i$ are the best-fit coefficients.
For each line ratio, we use the lowest-order polynomial that can match the shape implied by binned medians of the data to minimize the number of free parameters.
Table~\ref{tab:cal} presents the line ratio abbreviations and definitions, best-fit calibration coefficients, metallicity range over which the calibration is valid, and intrinsic scatters about the best-fit calibrations.
The measured line ratios and direct oxygen abundances of the calibrating samples and the best-fit calibrations are shown in Figures~\ref{fig:ocal} to~\ref{fig:arcal}.

\begin{table*}
 \centering
 \setlength{\tabcolsep}{2pt}
 \caption{Line ratio definitions, best-fit calibration coefficients, and intrinsic scatters about the best--fit calibrations.}\label{tab:cal}
 \begin{tabular}{ l l | r | r r r r r | l l | l l l }
\hline
$R$ & {\it Definition} & N$_\mathrm{gal}$ & $c_0$ & $c_1$ & $c_2$ & $c_3$ & $c_4$ & $Z_\mathrm{min}$ & $Z_\mathrm{max}$ & $\sigma_{R\mathrm{,fit}}$ & $\sigma_{R\mathrm{,int}}$ & $\sigma_{\mathrm{O/H,int}}$ \\
\hline\hline
O3  &  [O \textsc{iii}]$\lambda$5008/H$\beta$  &  139  &  $0.852$  &  $-0.162$  &  $-1.149$  &  $-0.553$  &    ---  &  7.3  &  8.6  &  0.04  &  0.13  &  0.14  \\
O2  &  [O \textsc{ii}]$\lambda\lambda$3727,3730/H$\beta$  &  123  &  $0.172$  &  $0.954$  &  $-0.832$  &    ---  &    ---  &  7.3  &  8.6  &  0.03  &  0.25  &  0.22  \\
R23  &  ([O \textsc{iii}]$\lambda$4960,5008 + [O \textsc{ii}]$\lambda\lambda$3727,3730)/H$\beta$  &  123  &  $0.998$  &  $0.053$  &  $-0.141$  &  $-0.493$  &  $-0.774$  &  7.3  &  8.6  &  0.03  &  0.07  &  0.13  \\
O32  &  [O \textsc{iii}]$\lambda$5008/[O \textsc{ii}]$\lambda\lambda$3727,3730  &  123  &  $0.697$  &  $-1.245$  &  $-0.869$  &    ---  &    ---  &  7.3  &  8.6  &  0.09  &  0.29  &  0.25  \\
O32\tablenotemark{a}  &  [O \textsc{iii}]$\lambda$5008/[O \textsc{ii}]$\lambda\lambda$3727,3730  &  123  &  $0.640$  &  $-1.478$  &    ---  &    ---  &    ---  &  7.3  &  8.6  &  0.04  &  0.34  &  0.23  \\
$\hat{R}$\tablenotemark{b}  &  0.47$\times$log(O2) + 0.88$\times$log(O3)  &  123  &  $0.779$  &  $0.263$  &  $-0.849$  &  $-0.493$  &    ---  &  7.3  &  8.6  &  0.03  &  0.12  &  0.18  \\
Ne3  &  [Ne \textsc{iii}]$\lambda$3870/H$\delta$  &  111  &  $0.306$  &  $-0.281$  &  $-0.844$  &    ---  &    ---  &  7.4  &  8.6  &  0.03  &  0.12  &  0.18  \\
Ne3O2  &  [Ne \textsc{iii}]$\lambda$3870/[O \textsc{ii}]$\lambda\lambda$3727,3730  &  105  &  $-0.333$  &  $-1.459$  &  $-1.127$  &    ---  &    ---  &  7.4  &  8.6  &  0.14  &  0.28  &  0.16  \\
Ne3O2\tablenotemark{a}  &  [Ne \textsc{iii}]$\lambda$3870/[O \textsc{ii}]$\lambda\lambda$3727,3730  &  105  &  $-0.408$  &  $-1.454$  &    ---  &    ---  &    ---  &  7.4  &  8.6  &  0.04  &  0.28  &  0.19  \\
RO2Ne3  &  ([Ne \textsc{iii}]$\lambda$3870 + [O \textsc{ii}]$\lambda\lambda$3727,3730)/H$\delta$  &  105  &  $0.842$  &  $0.486$  &    ---  &    ---  &    ---  &  7.4  &  8.6  &  0.02  &  0.14  &  0.26  \\
N2  &  [N \textsc{ii}]$\lambda$6585/H$\alpha$  &  68  &  $-1.356$  &  $1.532$  &    ---  &    ---  &    ---  &  7.8  &  8.6  &  0.08  &  0.30  &  0.20  \\
O3N2  &  ([O \textsc{iii}]$\lambda$5008/H$\beta$)/([N \textsc{ii}]$\lambda$6585/H$\alpha$)  &  68  &  $2.294$  &  $-1.411$  &  $-3.077$  &    ---  &    ---  &  7.8  &  8.6  &  0.17  &  0.32  &  0.13  \\
N2S2  &  [N \textsc{ii}]$\lambda$6585/[S \textsc{ii}]$\lambda\lambda$6718,6733  &  55  &  $-0.332$  &  $1.259$  &    ---  &    ---  &    ---  &  7.8  &  8.6  &  0.13  &  0.29  &  0.23  \\
S3  &  [S \textsc{iii}]$\lambda\lambda$9071,9533/H$\alpha$  &  51  &  $-0.798$  &  $0.381$  &  $-0.447$  &  $0.128$  &    ---  &  7.9  &  8.6  &  0.07  &  0.08  &  0.20  \\
S2  &  [S \textsc{ii}]$\lambda\lambda$6718,6733/H$\alpha$  &  59  &  $-1.139$  &  $0.723$  &    ---  &    ---  &    ---  &  7.9  &  8.6  &  0.06  &  0.24  &  0.33  \\
S23  &  ([S \textsc{iii}]$\lambda\lambda$9071,9533 + [S \textsc{ii}]$\lambda\lambda$6718,6733)/H$\alpha$  &  47  &  $-0.193$  &  $0.389$  &    ---  &    ---  &    ---  &  7.9  &  8.6  &  0.02  &  0.09  &  0.23  \\
S32  &  [S \textsc{iii}]$\lambda\lambda$9071,9533/[S \textsc{ii}]$\lambda\lambda$6718,6733  &  47  &  $0.374$  &  $-0.556$  &    ---  &    ---  &    ---  &  7.9  &  8.6  &  0.05  &  0.16  &  0.28  \\
S3O3  &  [S \textsc{iii}]$\lambda\lambda$9071,9533/[O \textsc{iii}]$\lambda$5008  &  53  &  $-1.215$  &  $0.889$  &  $0.688$  &    ---  &    ---  &  7.9  &  8.6  &  0.05  &  0.17  &  0.18  \\
O3S2  &  ([O \textsc{iii}]$\lambda$5008/H$\beta$)/([S \textsc{ii}]$\lambda$6718,6733/H$\alpha$)  &  59  &  $1.997$  &  $-1.981$  &    ---  &    ---  &    ---  &  7.9  &  8.6  &  0.13  &  0.45  &  0.23  \\
Ar3  &  [Ar \textsc{iii}]$\lambda$7137/H$\alpha$  &  48  &  $-1.682$  &  $0.502$  &  $-0.947$  &    ---  &    ---  &  8.0  &  8.6  &  0.07  &  0.09  &  0.04  \\
Ar3O3  &  [Ar \textsc{iii}]$\lambda$7137/[O \textsc{iii}]$\lambda$5008  &  49  &  $-2.076$  &  $0.914$  &    ---  &    ---  &    ---  &  8.0  &  8.6  &  0.06  &  0.18  &  0.19  \\
\hline
\end{tabular}
\tablecomments{$R$: Line-ratio abbreviation. {\it Definition}: The explicit definition of the line ratio including constituent lines and rest-frame wavelengths. N$_\mathrm{gal}$: The number of galaxies with detections of the line ratio used in fitting the calibration. $c_i$: Best-fit polynomial coefficients for use in equation~\ref{eq:cal}. $Z_\mathrm{min}$ and $Z_\mathrm{max}$: Minimum and maximum values of 12+log(O/H) over which the calibration is valid. $\sigma_{R\mathrm{,fit}}$: The median uncertainty in log($R$) of the best-fit polynomial over the valid metallicity range (c.f. gray shading in Figs.~\ref{fig:ocal}--\ref{fig:arcal}). $\sigma_{R\mathrm{,int}}$: Intrinsic scatter in log($R$) at fixed O/H estimated with equation~\ref{eq:chi2}. $\sigma_{\mathrm{O/H,int}}$: Intrinsic scatter in log(O/H) at fixed $R$ estimated with equation~\ref{eq:chi2}.}
\tablenotetext{a}{Alternative linear fit to O32 and Ne3O2, though we recommend using the quadratic fits for these line ratios.}
\tablenotetext{b}{Since $\hat{R}$ is defined in the logarithm, the left side of equation~\ref{eq:cal} should instead read $R$ for this ratio only.}
\end{table*}

For all line ratios, we follow the fitting methodology developed by \citet{sanders2024}.
This approach is designed to account for the fact that the intrinsic scatter in strong-line calibrations (typically $0.1-0.2$~dex in O/H at fixed $R$) is now larger than the statistical uncertainty on metallicity for many sources thanks to the sensitivity of NIRSpec, and the size of the statistical uncertainty varies widely across the sample.
A standard inverse-variance weighted fit would not account for this dominant intrinsic scatter and would skew the fit toward the objects with the smallest uncertainties (often gravitationally lensed sources) rather than a true sample-averaged relation.

The best-fit polynomial coefficients are derived as follows.
We fit polynomial functions to the line ratios and direct oxygen abundances of the individual galaxies using an orthogonal distance regression (ODR) without including any weighting according to the uncertainties on either parameter.
Only galaxies with detections of all lines required for the particular line ratio are included in fitting, while limits are not included.
We perform ODR fitting on 1000 realizations of the data by perturbing log($R$) and 12+log(O/H) according to their statistical errors.
We then compute the median $R$ at fixed O/H among the distribution of 1000 polynomials, and derive the final best-fit coefficients by performing ODR fitting on the resulting median curve.
This approach accounts for the relative uncertainties between different objects, as galaxies with larger errors will have larger shifts between realizations, and also allows us to quantify the typical uncertainty of the best-fit relation in log($R$) at fixed O/H ($\sigma_{R,\mathrm{fit}}$) using the spread among the 1000 polynomials.
The resulting relations should agree well with binned medians of the data, though this match may not be exact since ODR minimizes the distance in both variables while medians are affected by the choice of which variable to bin in.
Accordingly, we also display median line ratios in bins of O/H containing equal numbers of galaxies to guide the eye and to aid in determining the lowest-degree polynomial required to match the shape implied by the data.

The metallicity range over which each calibration is valid differs for each line ratio since the subsample of galaxies with detections of all necessary lines differs, especially for ratios involving the fainter lines ([\ion{N}{2}], [\ion{S}{2}], [\ion{Ar}{3}]).
Using the subsample of detections for each line ratio, we define the valid metallicity range, $Z_\mathrm{min}$ to $Z_\mathrm{max}$, over which the calibrations are reliable and can be confidently applied.
To determine the valid range, we compute the uncertainty on the running median O/H in 0.2~dex wide bins in O/H across the full range of the sample.
We then require this uncertainty to be less than 0.1~dex, and at least 5 galaxies to fall within each bin to avoid regions strongly affected by sample variance.
The range in 12+log(O/H) over which these two criteria are satisfied defines the valid metallicity range of each calibration.
We recommend that these calibrations only be used within the valid ranges, and urge caution if extrapolating beyond these limits.
To show the results of potential extrapolation, we plot the best-fit polynomials beyond the valid range in dotted black lines.

There is significant intrinsic scatter present in the calibrating samples about strong-line calibrations, reflecting variations in gas and ionization conditions at fixed O/H.
This scatter can be stated as either the intrinsic scatter in log($R$) at fixed O/H ($\sigma_{R,\mathrm{int}}$) or the intrinsic scatter in log(O/H) at fixed $R$ ($\sigma_{\mathrm{O/H,int}}$).
We estimate both quantities using the following method, and report them in Tab.~\ref{tab:cal}.
We compute the reduced $\chi^2$ statistic, including measurement uncertainties in both parameters and an intrinsic scatter term, as:
\begin{equation}\label{eq:chi2}
\chi^2_\nu = \frac{1}{\nu} \sum_i{\frac{[O_i - f(x_i)]^2}{\sigma_{O,i}^2 + \left(\frac{df}{dx}(x_i)\sigma_{x,i}\right)^2 + \sigma_{\mathrm{int}}^2}}
\end{equation}
where the sum is over all objects used to fit the calibration and $\nu$ is the degrees of freedom equal to the difference between the number of data points and number of parameters in the polynomial minus one.
We then find the value of $\sigma_{\mathrm{int}}$ that results in $\chi^2_\nu=1$.
When computing $\sigma_{R,\mathrm{int}}$, $O_i$ is the measured log($R$) with uncertainty $\sigma_{O,i}$, $f$ is the best-fit calibration polynomial (eq.~\ref{eq:cal}), and $x_i$ is 12+log(O/H) with uncertainty $\sigma_{x,i}$.
When deriving $\sigma_{\mathrm{O/H,int}}$, $O_i$ is 12+log(O/H) with uncertainty $\sigma_{O,i}$, $f$ is the best-fit calibration polynomial inverted to solve for metallicity as a function of log($R$), and $x_i$ is log($R$) with uncertainty $\sigma_{x,i}$.
During $\sigma_{\mathrm{O/H,int}}$ estimation for ratios that are double-valued as a function of O/H (e.g., O3, R23), if a galaxy has measured log($R$) higher than the peak value reached by the calibration, then $f(x_i)$ is undefined.
In these cases, we take $f(x_i)$ to be the metallicity at the highest value of $R$ reached by the calibration.
For galaxies with log($R$) below the peak value, $f(x_i)$ has two solutions, and we adopt the one that lies closer to the measured metallicity.

\subsubsection{Oxygen line-ratio calibrations}\label{sec:ocal}

\begin{figure*}
\centering
\includegraphics[width=0.8\textwidth]{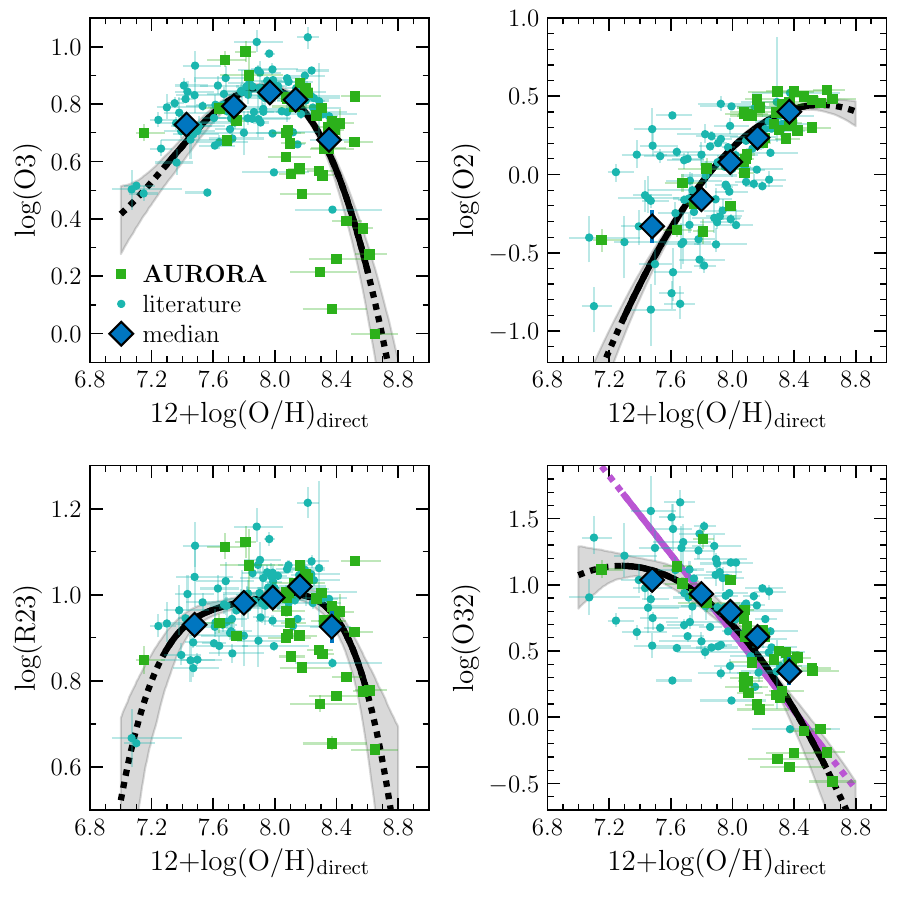}
\caption{Relations between direct-method metallicity and line ratios involving lines of O, including O3, O2, R23, and O32.
Green squares show AURORA galaxies, while turquoise circles denote objects drawn from the literature.
Blue diamonds display median values of the combined sample in bins of O/H, with an equal number of galaxies per bin.
The thick black line shows the best-fit polynomial, which is solid over the valid metallicity range and dotted at extrapolations outside of that range.
Gray shading shows the 1$\sigma$ uncertainty bound on the best-fit polynomial.
The purple line shows an alternative linear fit for O32.
Best-fit coefficients and intrinsic scatters about the best-fit polynomial are reported in Tab.~\ref{tab:cal}.
}\label{fig:ocal}
\end{figure*}

We present metallicity calibrations for line ratios involving only \ion{H}{1} and ionized oxygen lines in Figure~\ref{fig:ocal}, featuring 139 galaxies for O3 and 123 galaxies for ratios that require [\ion{O}{2}].
These O-based calibrations are valid over 12+log(O/H$)=7.3-8.6$.
With the large size and dynamic range in O/H of our sample, it is clear that the qualitative shapes of these relations agree with what is observed at $z\sim0$.
The ratios O3 and R23 peaks at roughly 12+log(O/H$)\sim8.0$ and decreasing both above and below that value.
R23 displays a notably flatter and wider peak as a function of O/H than O3, potentially limiting the use of R23 in constraining high-redshift metallicities except at very low and high O/H.
O2 is largely monotonic over this metallicity range, increasing until a peak is reached at relatively high metallicities (12+log(O/H$)\sim8.5$).
O32 monotonically decreases with increasing metallicity.
The O32 binned medians display a flattening in the lowest-metallicity bin.
Accordingly, we have fit O32 with a second-degree polynomial to capture this flattening.
It is also common to fit O32 with a linear function \citep[e.g.,][]{bian2018,sanders2024,chakraborty2025}.
We thus provide an alternative linear fit (purple line) and report the linear coefficients in Tab.~\ref{tab:cal}, though we recommend using the quadratic fit that more closely matches the binned medians.
We also see hints of increased scatter in O32 and O2 below 12+log(O/H$)\sim7.8$ that may suggest a larger dispersion in ionization parameter at fixed metallicity in metal-poor galaxies.

\begin{figure}
\centering
\includegraphics[width=\columnwidth]{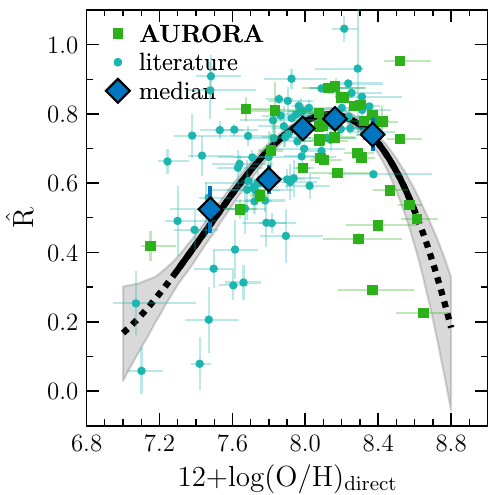}
\caption{Relation between direct-method metallicity and $\hat{R}$.
Points and lines are as in Fig.~\ref{fig:ocal}.
Best-fit coefficients and intrinsic scatters about the best-fit polynomial are reported in Tab.~\ref{tab:cal}.
}\label{fig:rhatcal}
\end{figure}

The $\hat{R}$ indicator, introduced by \citet{laseter2024}, is a linear combination of the logarithm of O3 and O2 defined as $\hat{R}=0.47\log(\mathrm{O}2)+0.88\log(\mathrm{O}3)$.
These authors derived the coefficients by finding the projection in O3-O2-O/H space that minimized the scatter in $\hat{R}$ at fixed O/H for a sample of $z\sim0$ galaxies and \ion{H}{2} regions, thought to be a result of minimizing the secondary dependence on ionization parameter at fixed O/H.
\citet{scholte2025} also found that $\hat{R}$ minimized scatter in a $z\sim0$ sample drawn from the DESI Early Data Release.
We present a calibration of $\hat{R}$ in Figure~\ref{fig:rhatcal} for the high-redshift auroral-line sample.
As with O3 and R23, $\hat{R}$ is double-valued as a function of O/H.
We find no significant reduction in intrinsic scatter of the high-redshift sample about our best-fit calibration for $\hat{R}$ relative to O3 or R23 ($\sigma_{R,\mathrm{int}}=[0.13, 0.12, 0.07]$ and $\sigma_{\mathrm{O/H},\mathrm{int}}=[0.14, 0.13, 0.18]$ for O3, R23, and $\hat{R}$, respectively; Tab.~\ref{tab:cal}).
Indeed, at low metallicities (12+log(O/H$)\lesssim8.0$), the scatter in line ratio at fixed O/H is clearly increased for $\hat{R}$ relative to O3.
This likely reflects the increase in O2 scatter at low metallicities noted above.
Furthermore, similar to O3 and R23, $\hat{R}$ provides minimal leverage on the metallicity determination near the peak of the double-valued curve at 12+log(O/H$)\sim7.8-8.4$.
The $\hat{R}$ indicator thus appears to provide no quantitative advantage over O3 or R23 for deriving metallicities at high redshifts, and all three would require the use of an additional line ratio to break the degeneracy between the upper and lower branches.

\citet{chakraborty2025} proposed an alternate projection based on their analysis of a high-redshift sample, defined as $\hat{R}_{Ch25}=0.18\log(\mathrm{O}2)+0.98\log(\mathrm{O}3)$.
However, since the coefficient for O2 is much smaller than that for O3 and O3$>$O2 for most high-redshift galaxies, the $\hat{R}_{Ch25}$ projection is only marginally different from employing O3 or R23 alone.
This fact is reflected in the similar scatters \citet{chakraborty2025} report about their calibrations of 0.07, 0.067, and 0.06~dex for O3, R23, and $\hat{R}_{Ch25}$, respectively, demonstrating that using $\hat{R}_{Ch25}$ does not provide a significant advantage over O3 or R23.
We do not investigate $\hat{R}_{Ch25}$ further for this reason.

\subsubsection{Neon line-ratio calibrations}\label{sec:necal}

\begin{figure*}
\centering
\includegraphics[width=\textwidth]{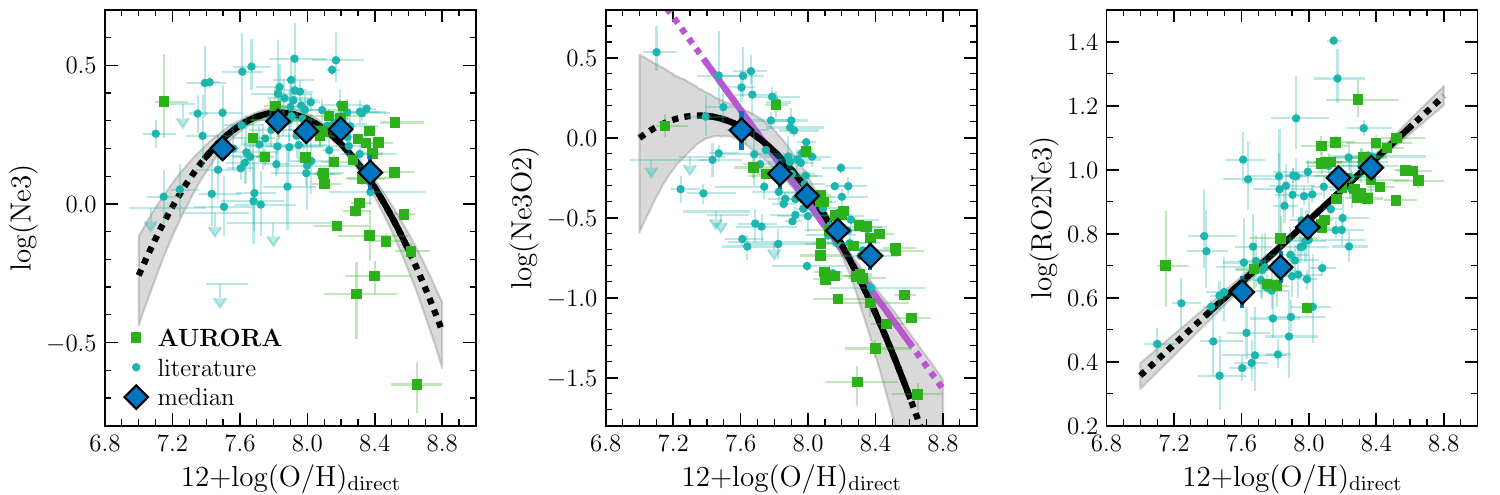}
\caption{Relations between direct-method metallicity and line ratios involving lines of Ne, including Ne3, Ne3O2, and RO2Ne3.
Points and lines are as in Fig.~\ref{fig:ocal}.
The purple line shows an alternative linear fit for Ne3O2.
Best-fit coefficients and intrinsic scatters about the best-fit polynomial are reported in Tab.~\ref{tab:cal}.
}\label{fig:necal}
\end{figure*}

We show calibrations of Ne-based line ratios in Figure~\ref{fig:necal} based on more than 100 high-redshift galaxies, valid over 12+log(O/H$)=7.4-8.6$.
These line ratios are combinations of [\ion{Ne}{3}]\W3870, [\ion{O}{2}]\W3728, and H$\delta$, defined as Ne3=[\ion{Ne}{3}]/H$\delta$, Ne3O2=[\ion{Ne}{3}]/[\ion{O}{2}], and RO2Ne3=([\ion{Ne}{3}]+[\ion{O}{2}])/H$\delta$.
A great advantage of these particular ratios is that the required lines are all at $\lambda_\mathrm{rest}\lesssim4100$~\AA, such that they can be covered by {\it JWST}/NIRSpec at redshifts up to $z\approx11.7$.
These Ne-based indicators will be of particular use at $z>9.7$ where [\ion{O}{3}]\W\W4960,5008 and H$\beta$ shift redward of NIRSpec's wavelength coverage, and can thus probe chemical abundances in the very early Universe.
Due to their relatively close wavelength spacing ($\Delta\lambda_\mathrm{rest}=374$~\AA\ from [\ion{O}{2}] to H$\delta$), these ratios are also less sensitive to dust reddening than other commonly employed ratios like O32 and R23.
These Ne-based ratios can be thought of as analogs of more familiar O-based ratios and used in similar ways, where the high-ionization [\ion{O}{3}] lines (ionization energy 35.1~eV) has been replaced by similarly high-ionization [\ion{Ne}{3}] (41.0~eV) and H$\beta$ by H$\delta$: Ne3 is analogous O3, Ne3O2 is analogous to O32, and RO2Ne3 is analogous to R23.
This point is reinforced by the fact that the Ne/O abundance ratio is not expected to vary as a function of redshift, metallicity, or across a variety of star-formation histories as production of both of these elements is dominated by core-collapse supernovae, supported by measurements at high-redshift with {\it JWST} \citep[e.g.,][]{arellano2022,arellano2025,welch2025} and at $z\sim0$ \citep[e.g.,][]{izotov2006,guseva2011}.

We find that Ne3 is double-valued with O/H and Ne3O2 monotonically decreasing with increasing O/H, similar to O3 and O32 respectively.
RO2Ne3 shows a monotonic increase with increasing O/H rather than a plateau and turnover like R23, indicating that it provides more distinguishing power for high-redshift metallicity determinations than R23 over a wider range of metallicities.
Similar to O32, Ne3O2 displays a tentative flattening and increase in scatter at very low metallicities, though better statistics at very low metallicities are needed.
We provide both quadratic polynomial and linear fit coefficients for Ne3O2 in Tab.~\ref{tab:cal}, again recommending the use of the quadratic as it provides a better match across the full metallicity range of the sample.

The relatively weak H$\delta$ line used in Ne3 and RO2Ne3 is not always detected even in cases where [\ion{Ne}{3}] and [\ion{O}{2}] may be.
However, if line fluxes have been corrected for dust reddening, the brighter H$\gamma$ or H$\beta$ lines can be used to reliably infer the H$\delta$ strength based on their theoretical intrinsic intensity ratios.
Indeed, we have adopted this approach for 11 galaxies in the sample for which H$\delta$ fell in the chip gap to increase the number of galaxies used in creating the Ne calibrations.
Thus, the Ne3 and RO2Ne3 indicators can still be used even in the case that H$\delta$ is too faint to detect or falls in a gap in wavelength coverage.

\subsubsection{Nitrogen line-ratio calibrations}\label{sec:ncal}

\begin{figure*}
\centering
\includegraphics[width=0.8\textwidth]{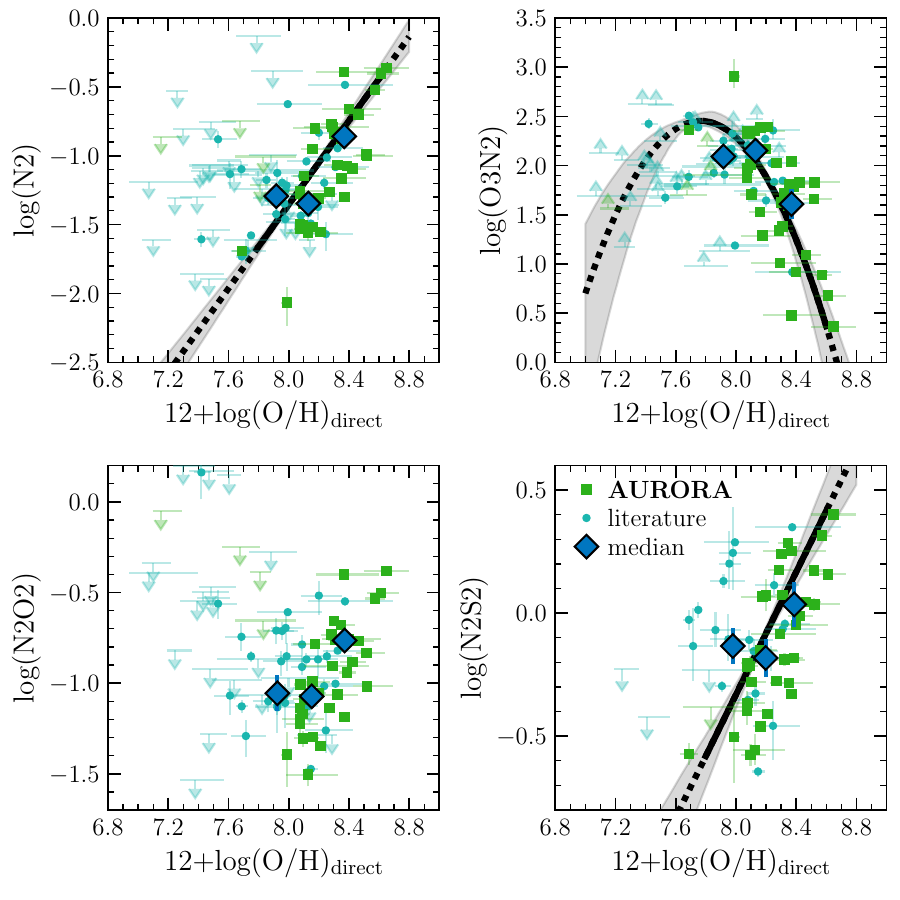}
\caption{Relations between direct-method metallicity and line ratios involving lines of N, including N2, O3N2, N2O2, and N2S2.
Points and lines are as in Fig.~\ref{fig:ocal}.
Best-fit coefficients and intrinsic scatters about the best-fit polynomial are reported in Tab.~\ref{tab:cal}.
A calibration is not fit to N2O2 because the correlation between N2O2 and O/H is not statistically significant.
}\label{fig:ncal}
\end{figure*}

We present N-based line ratios as a function of direct oxygen abundance in Figure~\ref{fig:ncal}, including 68 galaxies for N2 and O3N2, 64 for N2O2, and 55 for N2S2.
We note that AURORA makes up over half of the [\ion{N}{2}]-detected sample.
Among all four line ratios, it is clear that the [\ion{N}{2}]\W6585 detection rate falls precipitously at low metallicities, with only a handful of detections at 12+log(O/H$)\lesssim7.7$ (0.1~\zsun) despite the significant depth of AURORA and other literature observations.
The high [\ion{N}{2}] non-detection rate among high-redshift, low-mass sources was noted in early {\it JWST} spectroscopic studies \citep[e.g.,][]{shapley2023mass,sanders2023}.
Therefore, N-based strong-line ratios for metallicity determinations can only potentially be useful at relatively high metallicities, otherwise [\ion{N}{2}] becomes fainter than [\ion{O}{3}]\W4364 and a robust direct metallicity could be derived instead.
Additionally, galaxies at low metallicity (12+log(O/H$)\lesssim8.0$) with [\ion{N}{2}] detections may be biased toward higher N/H on average (or, equivalently, N/O at fixed O/H) which boosts the [\ion{N}{2}]\W6585 flux.

We find significant trends of N2 and O3N2 with metallicity.
N2 increases with increasing O/H, while O3N2 decreases, as seen in the local Universe \citep[e.g.,][]{pettini2004,marino2013}.
O3N2 appears to flatten at 12+log(O/H$)\lesssim8.0$, as observed in $z\sim0$ samples \citep[e.g.,][]{marino2013,curti2017,sanders2021} and some high-redshift studies that report [\ion{N}{2}] detections at such low metallicities \citep{chakraborty2025,cataldi2025}.
Accordingly, we fit O3N2 with a quadratic polynomial while using a linear form for N2.

The bottom row of Fig.~\ref{fig:ncal} displays the N2O2 and N2S2 ratios.
These ratios are often used as proxies for the N/O and N/S abundance ratios, respectively.
Based on the assumption of little variation in S/O with O/H \citep[e.g.,][]{berg2020}, both ratios would be sensitive to N/O.
We find that N2O2, the most direct proxy of N/O, does not display a significant correlation with O/H and we thus do not fit a calibration to these data.
In a spearman correlation test for N2O2, the correlation coefficient is $\rho=0.242$ and the $p$-value is 0.054 indicating $<2\sigma$ significance.
In contrast, the same test for N2S2 yields $\rho=0.356$ and $p=0.0076$, indicating a $\approx3\sigma$ correlation.
We thus report coefficients for a linear fit to the N2S2 data in Tab.~\ref{tab:cal}.
For comparison, the $p$-values for N2 and O3N2 are both $<1\times10^{-6}$, indicating highly significant correlations.

A significant correlation with O/H can be present in N2S2 but absent in N2O2 because, despite their common dependence on N/O, N2S2 also depends on the ionization parameter due to differing ionization potentials between the ions in the numerator and denominator while N2O2 has little sensitivity to the ionization parameter (ionization energies: 10.4~eV for S$^+$, 13.6~eV for O$^+$, 14.5~eV for N$^+$).
This dependence (or lack thereof) on ionization parameter is confirmed in theoretical photoionization models \citep[e.g.,][]{kewley2019}.
Thus, without any correlation between N/O and O/H, an anticorrelation between ionization parameter and metallicity produces a positive correlation between N2S2 and O/H without producing one for N2O2.
This same argument holds for the O3N2 ratio, which has an even larger ionization potential difference.
N2 also depends on ionization parameter (which modulates the fraction of N in N$^+$), such that decreasing ionization parameter with increasing O/H also contributes to a positive correlation between N2 and metallicity.
We thus find that N2, O3N2, and N2S2 maintain utility as empirical tracers of metallicity at high redshift largely due to their sensitivity to ionization parameter, while N2O2 does not appear to be a reliable indicator.
However, relative to the O- and Ne-based calibrations discussed above, all N-based calibrations will have an additional systematic dependence on variations of N/O at fixed O/H.
In Sec.~\ref{sec:properties}, we discuss the physical implications of the relation between N-based ratios and O/H, and their limited utility in deriving accurate high-redshift metallicities.

\subsubsection{Sulfur line-ratio calibrations}\label{sec:scal}

\begin{figure*}
\centering
\includegraphics[width=\textwidth]{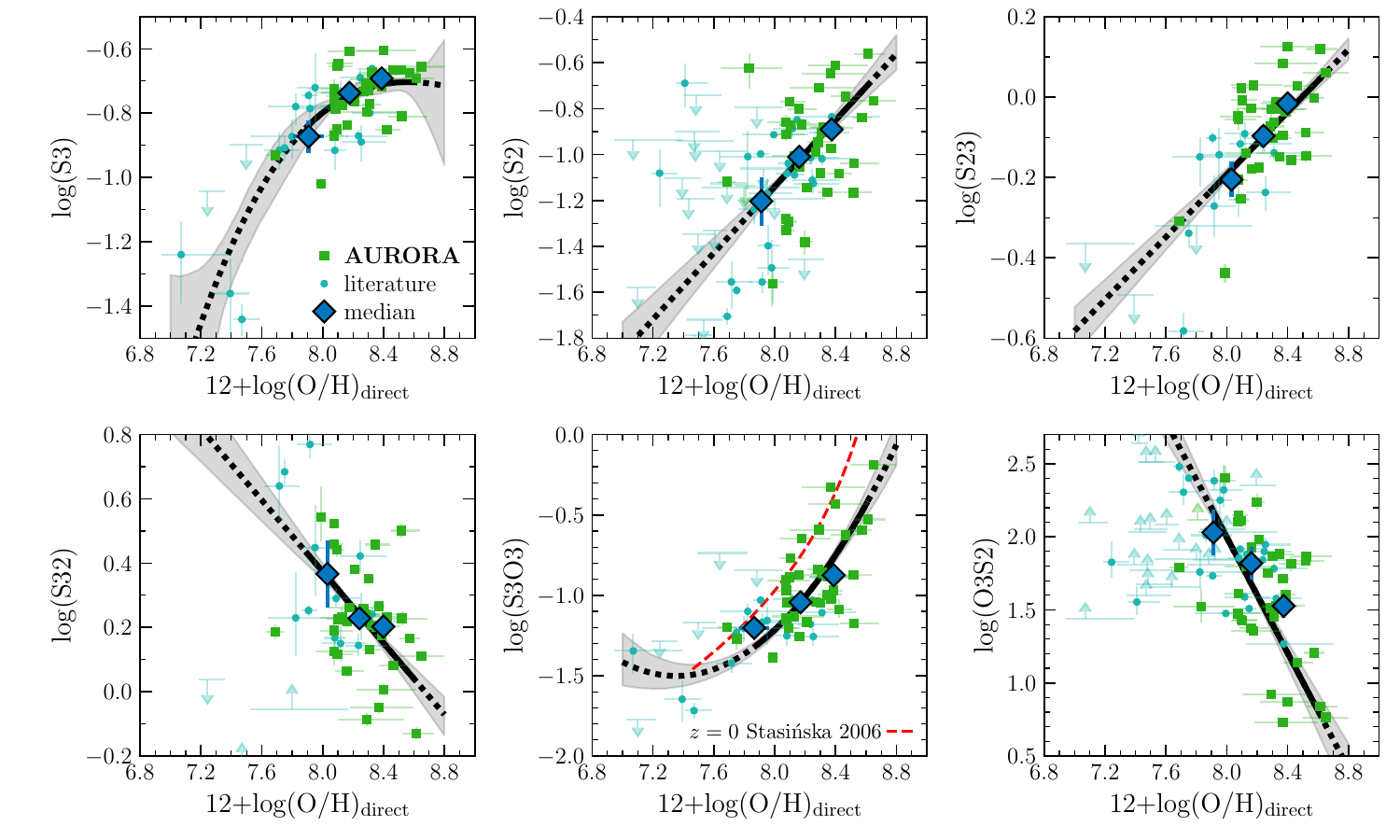}
\caption{Relations between direct-method metallicity and line ratios involving lines of S, including S3, S2, S23, S32, S3O3, and O3S2.
Points and lines are as in Fig.~\ref{fig:ocal}.
Best-fit coefficients and intrinsic scatters about the best-fit polynomial are reported in Tab.~\ref{tab:cal}.
}\label{fig:scal}
\end{figure*}

In Figure~\ref{fig:scal}, we present ratios including the sulfur lines [\ion{S}{2}] and [\ion{S}{3}] for $\sim50$ high-redshift galaxies.
The first four of these ratios (S3, S2, S23, S32) represent analogs of the commonly used O-based ratios (O3, O2, R23, O32) substituting [\ion{S}{2}] or [\ion{S}{3}] for the O line of corresponding ionization-state as well as H$\alpha$ for H$\beta$ to reduce the sensitivity to the dust correction.
However, due to the different ionization and excitation energies of these S and O strong lines, their exact dependence on O/H will differ.
We find that S3, S2, and S23 increase with increasing metallicity, while S32 decreases.
The behavior of S32 is analogous to what is observed in O32 and Ne3O2, likely driven by an anticorrelation between ionization parameter and O/H.
S3, S2, and S23 are expected to be double-valued and turn over at some metallicity \citep[see, e.g., the photoionization models of][]{kewley2019}, but our sample does not extend to high enough metallicities to see this effect except for some flattening in S3.
As with [\ion{N}{2}], the S lines are predominantly detected at 12+log(O/H$)\gtrsim7.8$, and AURORA comprises the majority of the S-detected sample.
[\ion{S}{3}]\W9533 in particular should be detectable at lower metallicities, and indeed is for a small number of metal-poor literature sources, but requires sufficiently deep spectroscopy at longer wavelengths than the rest-optical lines ($\lambda_\mathrm{rest}>9000$~\AA) that often fall in a different NIRSpec grating.

S3O3 and O3S2 have also been proposed as metallicity indicators \citep[e.g.,][]{stasinska2006,curti2020}, and we find a significant dependence of these line ratios on metallicity.
S3O3 displays the smallest scatter in O/H at fixed line ratio among the S-based indicators considered here.
It is thus a promising tool for high-redshift metallicities (though only out to $z\sim4.5$ with NIRSpec), with hints that this ratio could be more readily detected at very low metallicities with spectroscopy covering a suitable wavelength range.
The increased number of [\ion{S}{2}] detections at higher metallicities offered by AURORA allows us to resolve a significant anticorrelation between O3S2 and O/H and to fit a high-redshift calibration, which was not possible for the \citet{cataldi2025} sample that is $\sim2$ times smaller.
However, we find that O3S2 has very large scatter at moderately low metallicities, similar to what is seen for S2 and O2 and possibly driven by increased scatter in ionization parameter in this regime.
Among the S-based indicators, S23 and S3O3 both have a monotonic metallicity dependence and smaller scatter and thus may be of the most utility.

\subsubsection{Argon line-ratio calibrations}\label{sec:arcal}

\begin{figure}
\centering
\includegraphics[width=\columnwidth]{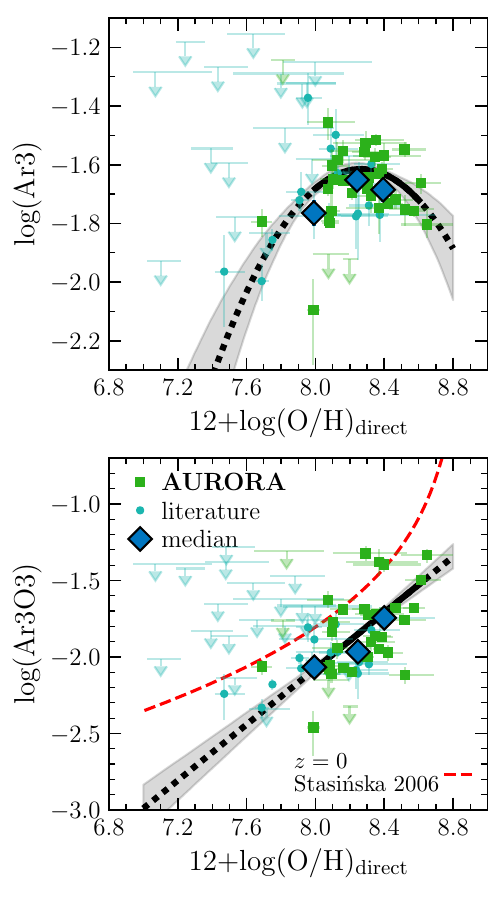}
\caption{Relations between direct-method metallicity and line ratios involving lines of Ar, including Ar3 and Ar3O3.
Points and lines are as in Fig.~\ref{fig:ocal}.
Best-fit coefficients and intrinsic scatters about the best-fit polynomial are reported in Tab.~\ref{tab:cal}.
}\label{fig:arcal}
\end{figure}

We present calibrations for two lines ratios based on [\ion{Ar}{3}]\W7137 in Figure~\ref{fig:arcal}.
Ar3 appears to be double-valued with O/H, as expected analogous to O3, based on a slight decrease at high metallicities and the few low-metallicity detections all having low Ar3.
However, more detections of [\ion{Ar}{3}] are needed to robustly determine the shape of this curve, especially at lower metallicities.
The majority of the current Ar sample, dominated by AURORA galaxies, lies near the apparent peak of the best-fit relation in a regime where the line ratio does not vary much with O/H.
Ar3O3 was proposed as a metallicity indicator by \citet{stasinska2006}, and performs well over the full metallicity range of the Ar-detected sample.
However, [\ion{Ar}{3}]\W7137 is the weakest of the lines considered in our calibrations and thus requires a deep spectrum to detect, such that an auroral line may also be detected in the same spectrum.
The practical utility of Ar-based metallicity calibrations is thus unclear, though they may be useful in cases where [\ion{O}{3}]\W4364 is not in the wavelength coverage but [\ion{Ar}{3}] is.

\section{Discussion}\label{sec:discussion}

\subsection{Comparison with other calibrations}

\begin{figure*}
\centering
\includegraphics[width=\textwidth]{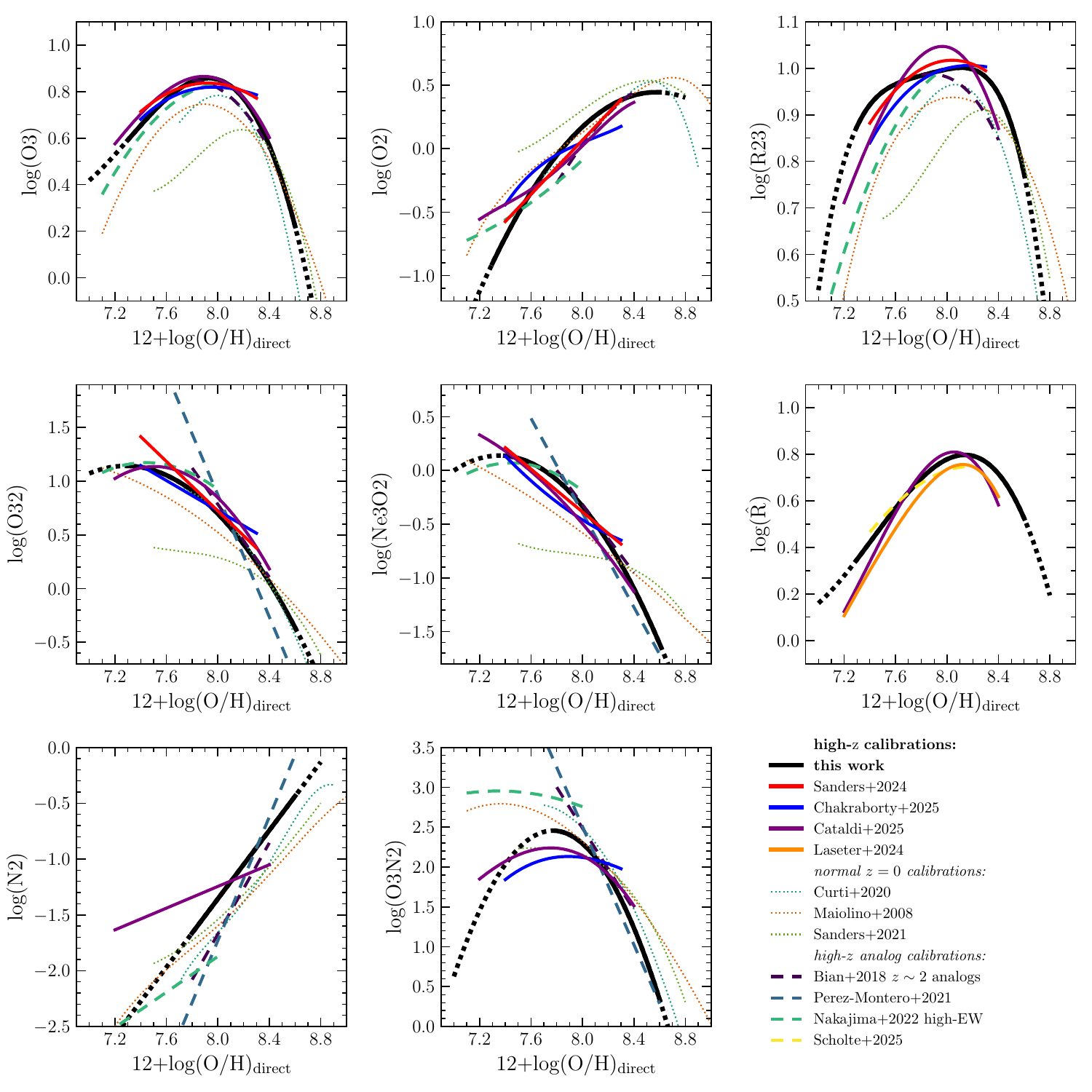}
\caption{Comparison of the new high-redshift metallicity calibrations derived in this work (black lines) to calibrations drawn from the literature.
Solid colored lines show calibrations based on high-redshift galaxy samples \citep{sanders2024,laseter2024,chakraborty2025,cataldi2025}.
Dotted lines display relations based on calibration sets representative of typical $z\sim0$ star-forming galaxies at \ion{H}{2} regions \citep{maiolino2008,curti2020,sanders2021}.
Long-dashed lines denote calibrations based on samples of extreme local objects that have properties analogous to those of high-redshift galaxies \citep{bian2018,perezmontero2021,nakajima2022,scholte2025}.
The new high-redshift calibrations display distinct evolution relative to normal $z=0$ calibrations, but generally agree with calibrations based on past {\it JWST} high-redshift studies and those derived from local analogs of high-redshift galaxies.
}\label{fig:calcomp}
\end{figure*}

Figure~\ref{fig:calcomp} compares our new high-redshift calibrations (black lines) with relations drawn from the literature for a selection of O-, Ne-, and N-based line ratios.
For line ratios involving O and Ne, our new calibrations generally agree well with past studies using {\it JWST} high-redshift measurements \citep[solid lines;][]{sanders2024,laseter2024,chakraborty2025,cataldi2025}, typically within 0.05~dex in O/H at fixed line ratio.
The agreement across O-based ratios (O3, O2, R23, O32, $\hat{R}$) and Ne3O2 is particularly good.

However, we find differences between our calibrations and literature high-redshift relations using N-based indicators.
For O3N2, both the \citet{cataldi2025} and \citet{chakraborty2025} relations are relatively flat as a function of O/H, with a peak value of log(O3N2$)\sim2.0$.
In contrast, we find that O3N2 is steep as a function of O/H at higher metallicities and shows signs of flattening toward log(O3N2$)\sim2.5$, closer to what is observed in local samples \citep[e.g.,][]{pettini2004,marino2013,curti2020}.
Our N2 calibration increases much more steeply with O/H than that of \citet{cataldi2025}, who find a shallow slope that leads to much higher N2 values at low metallicity (12+log(O/H$)\lesssim7.8$) than are seen in $z\sim0$ representative or high-redshift analog samples.
We note that many of the [\ion{N}{2}]-detected low-metallicity galaxies in the \citet{chakraborty2025} and \citet{cataldi2025} samples are not included in our sample.
A selection bias may be at play where metal-poor galaxies with [\ion{N}{2}] detections are preferentially nitrogen-enhanced, having higher N/O at fixed O/H.
Deeper spectroscopy of low metallicity sources is required to understand the behavior of N-based ratios and N/O variations below $\sim0.1$~\zsun.

The sample used in this work has key advantages over these past high-redshift calibration studies, namely a larger sample size and wider metallicity range.
Our sample is $2-3$ times larger than those presented in \citet{sanders2024}, \citet{laseter2024}, and \citet{chakraborty2025}.
Compared to the recent sample used by \citet{cataldi2025}, ours is 20\%\ larger for O-based and N-based ratios, and has double the number of detections for Ne- and S-based ratios.
All of the aforementioned {\it JWST} $T_e$ studies contain virtually no galaxies at 12+log(O/H$)>8.4$ (0.5~\zsun), while our sample extends up to 12+log(O/H$)=8.65$ (0.9~\zsun), a significant improvement in metallicity range.
We also note that our sample of 139 high-redshift aurora-detected sources is approximately equal in size to $z=0$ \ion{H}{2} region samples with which foundational strong-line calibrations were constructed, such as that of \citet{pettini2004} ($N=137$).
We have thus reached an era of robust statistical precision in high-redshift strong-line metallicity calibrations.

We find clear systematic offsets between our high-redshift calibrations and those based on representative $z\sim0$ galaxies or \ion{H}{2} regions \citep[dotted lines;][]{maiolino2008,curti2020,sanders2021}.
At fixed O/H, the high-redshift calibrations have higher O3, R23, O32, Ne3O2, and N2; and lower O2.
These trends hold across the majority of the covered metallicity range, though for some ratios the offset from $z\sim0$ appears to lessen at high metallicities (e.g., O3, R23, O32, Ne3O2).
These offsets in line ratio at fixed O/H imply an evolution in ISM ionization conditions between $z\sim0$ and $z>2$, which we further discuss in Sec.~\ref{sec:properties}.
The local calibrations show less of an offset in O3N2 at fixed O/H, perhaps indicating that the evolution of underlying properties leading to higher O3 and N2 at fixed O/H reduces the magnitude of an offset in this space.
It is clear that calibrations based on typical $z\sim0$ HII regions or star-forming galaxies would return systematically biased metallicities when applied to high-redshift samples, with offsets most often in excess of 0.1~dex in O/H at fixed line ratio.

We also compare to calibrations constructed from local-Universe analogs of high-redshift galaxies \citep[dashed lines;][]{bian2018,perezmontero2021,nakajima2022,scholte2025}, often applied to high-redshift spectroscopic samples in the pre-{\it JWST} era.
While \citet{scholte2025} studied direct metallicities of high-redshift galaxies from the EXCELS survey, we note that their $\hat{R}$ calibration was fit to a sample of $z\sim0$ galaxies drawn from DESI with relatively high H$\beta$ equivalent widths and we thus classify this work as an analog calibration.
We find that the high-redshift analog calibrations perform better than the representative $z\sim0$ calibrations across virtually all the line ratios shown here.
The analogs reach higher R23 and O3 peaks than the normal $z\sim0$ samples, evidence that their ISM ionization conditions more closely match what is now seen in-situ in the early Universe.
This result suggests that high-redshift metallicity evolution work based on these analog calibrations was not strongly biased \citep[e.g.,][]{sanders2021,topping2021}.
However, due to the increased number of high-redshift galaxies with robust auroral line detections, calibrations constructed with actual high-redshift samples have now reached a level where the use of local analog calibrations no longer provides an advantage.

\subsection{Implications for ionized gas properties}\label{sec:properties}

The redshift evolution of strong-line metallicity calibrations can provide valuable insight into the evolution of the underlying physical conditions of the ionized ISM.
While evolving ISM conditions can be resolved from line-ratio excitation diagrams \citep[e.g., the BPT diagram;][]{shapley2015,shapley2019,steidel2014,steidel2016,strom2017}, differences in line ratio at fixed O/H can more clearly disentangle how gas properties change at fixed metallicity.
The higher O3 and R23 peak values reached in the high-redshift sample can be explained by the presence of a harder ionizing spectrum (i.e., hotter stellar effective temperatures) at fixed O/H compared to $z\sim0$.
Such a shift increases the strength of high-ionization collisionally excited lines relative to \ion{H}{1} recombination lines.
While, at face value, the increased O32 and Ne3O2 could be read as an increase in ionization parameter at fixed O/H between $z\sim0$ and $z>2$, the O32 ratio at fixed ionization parameter and fixed O/H increases with increasing stellar effective temperature \citep[e.g.,][]{kewley2002,sanders2016den}.
We thus find that an ionizing spectrum that increases in hardness (i.e., hotter effective temperatures) at fixed O/H with increasing redshift can qualitatively describe the shifts in the O- and Ne-based calibrations.
Such evolution is in accord with studies that have found evidence for an increase in the $\alpha$/Fe abundance ratio with increasing redshift among star-forming galaxies, leading to more Fe-poor stars at fixed O/H at high redshifts \citep[e.g.,][]{steidel2016,shapley2019,sanders2020,topping2020a,topping2020b,cullen2021,stanton2024}.
The resulting decrease in stellar opacity from Fe-peak metal line blanketing leads to hotter effective temperatures and harder ionizing spectra.

The anticorrelations between metallicity and both O32 and Ne3O2 provide strong evidence that ionization parameter and metallicity are anticorrelated at high redshift, qualitatively consistent with trends found in local \ion{H}{2} regions \citep[e.g.,][]{perez2014}.
Other line ratios made up of only metal lines in both the numerator and denominator and with a difference between the ionization energies of the two species (e.g., O3N2, N2S2, S32, S3O3, O3S2, Ar3O3) follow a behavior wherein the higher-ionization line gets weaker relative to the lower-ionization line as metallicity increases.
These trends provide further evidence that, on average, ionization parameter decreases with increasing gas-phase metallicity.

Nitrogen-based metallicity indicators have an additional level of complication given that their relationship with O/H depends on the N/O abundance ratio.
With all other parameters kept fixed, a harder ionizing spectrum leads to relatively little change in N2 at fixed O/H, with a decrease in the fraction of N in N$^+$ being balanced out by an increase in the collisional excitation rate due to a higher equilibrium \te.
O3N2, on the other hand, would increase at fixed O/H due to a larger fraction of O in O$^{2+}$ and higher \te.
We in fact observe an increase in N2 and little evolution in O3N2 at fixed O/H from $z\sim0$ to $z>2$.
A possible solution is that there is an accompanying increase in N/O at fixed O/H with increasing redshift on average, which would further increase N2 and decrease O3N2 leading to the trends that we observe.

An additional clue is the fact that N2O2 and O/H do not show a statistically significant correlation, unlike at $z\sim0$.
The large scatter in N2O2 at fixed O/H observed in the high-redshift sample is likely driven by large scatter in N/O at fixed O/H even at low metallicities, in departure from the ordered relation between N/O and O/H seen at $z\sim0$ wherein N/O follows a low plateau at low metallicities and begins rising with increasing O/H at 12+log(O/H$)\gtrsim8.0$ \citep[e.g.,][]{perez2009,berg2020}.
High N/O ratios approaching the solar value or larger have been reported for several high-redshift sources spanning $z\sim2-10$ based on emission-line spectroscopy \citep[e.g.,][]{bunker2023,cameron2023,isobe2023b,navarro2024,castellano2024,marquez2024,schaerer2024,arellano2025,topping2025highz,stiavelli2025,zhang2025}, suggesting that enhanced N/O at fixed O/H relative to the $z\sim0$ relation may become common at $z\gtrsim2$.
These studies have proposed a number of physical drivers of N enhancement, including globular cluster formation, enrichment from supermassive stars, Wolf-Rayet wind enrichment, and massive pristine gas accretion events.
In contrast, absorption-line based abundance measures of damped Lyman-$\alpha$ systems at $z\sim2-4$ do not show elevated N/O at low O/H \citep[e.g.][]{pettini2008,cooke2011}, suggesting the N enhancement may be present in ionized gas but absent in neutral \ion{H}{1} gas.
Regardless of its physical origin, the high dispersion in N2O2 (and plausibly N/O) at fixed O/H suggests that the use of N-based strong-line ratios to derive oxygen abundances at high redshift should be avoided if possible, and that results based on lines of $\alpha$ elements (e.g., O, Ne) will generally be more reliable as they avoid any systematic dependence on N/O.

In Figures~\ref{fig:scal} and~\ref{fig:arcal}, we show the S3O3 and Ar3O3 calibrations of \citet{stasinska2006} based on $z=0$ \ion{H}{2} regions.
We find that the high-redshift calibrations are shifted $\sim0.2-0.3$~dex lower in S3O3 and Ar3O3 at fixed O/H than the $z=0$ calibrations.
One possible explanation of this shift is that high-redshift galaxies have systematically lower S/O and Ar/O abundance ratios than typical $z\sim0$ galaxies.
Sub-solar S/O and Ar/O has been seen in a small number of high-redshift sources at $z\sim2-7$ \citep{rogers2024,stanton2025,stiavelli2025,bhattacharya2025}.
While typically thought to behave similarly to O and Ne with prompt enrichment from core-collapse supernovae, S and Ar have an additional significant contribution to their enrichment from Type~Ia supernovae on $\gtrsim100$~Myr timescales \citep{kobayashi2020}.
The rapidity of galaxy formation at high redshifts thus may result in sub-solar S/O and Ar/O.
The offset observed in our high-redshift S3O3 and Ar3O3 calibrations suggests that this deficit in S/O and Ar/O is widespread among the high-redshift star-forming population.

\subsection{Can the same strong-line calibrations be used at $z\sim2$ and $z>6$?}

\begin{figure*}
\centering
\includegraphics[width=0.49\textwidth]{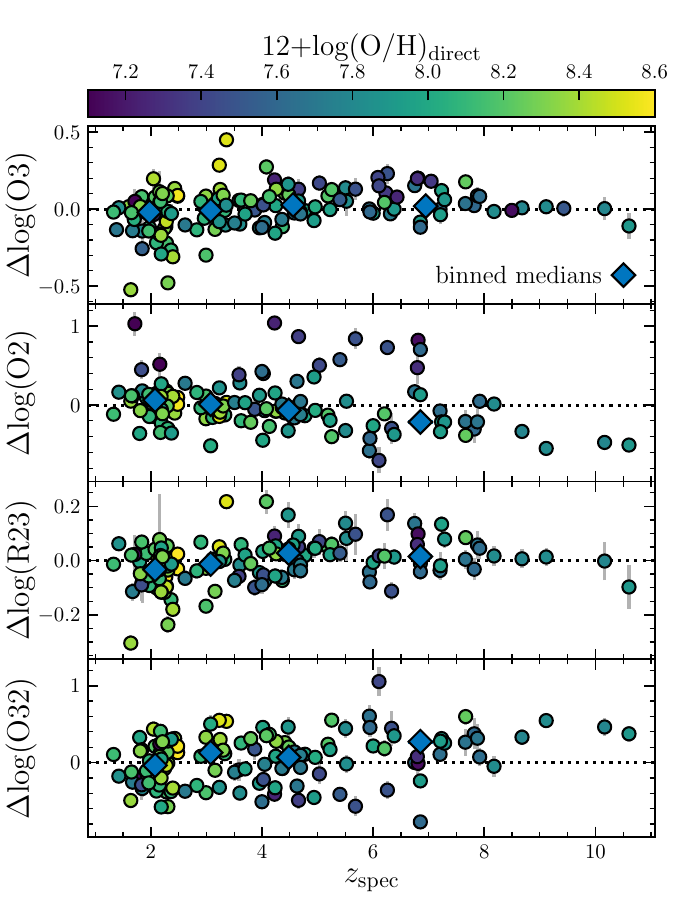}
\includegraphics[width=0.49\textwidth]{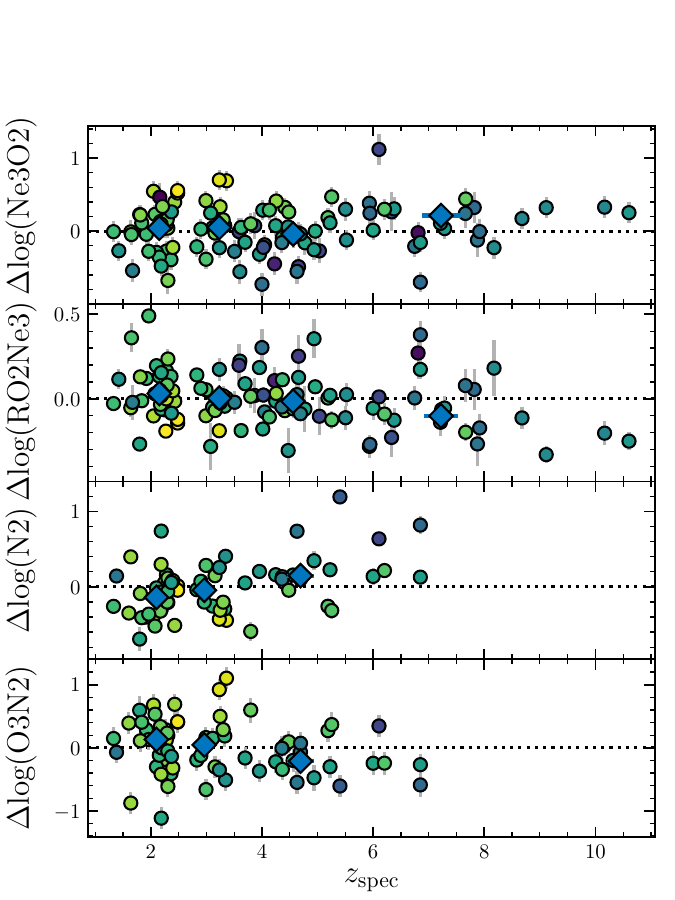}
\caption{Residuals in line ratio at fixed O/H relative to the best-fit metallicity calibration polynomials (Tab.~\ref{tab:cal}) as a function of redshift, color-coded by O/H.
Medians in bins of redshift are displayed as blue diamonds.
}\label{fig:zcal}
\end{figure*}

Given the wide redshift range ($z=1.4-10.6$) of our auroral-detected sample, it is natural to ask whether the calibrations presented here can robustly be applied to strong-line samples over such a wide redshift range.
In Figure~\ref{fig:zcal}, we show the offset in line ratio at fixed O/H relative to the best-fit calibration as a function of redshift for a selection of line ratios, color-coded by metallicity.
This plot allows us to address whether there appears to be systematic redshift evolution relative to the best-fit calibrations.
We find that none of the line ratios show a significant ($>3\sigma$ based on the error on the medians) offset from zero in the binned medians out to $z\approx6$, suggesting that all of these ratios can be applied in this redshift range with reasonable confidence within current constraints.
N2 and O3N2 display tentative evidence of evolution between $z\sim3$ and $z>4$, potentially associated with evolution of the [\ion{N}{2}]-BPT diagram star-forming sequence observed across this redshift range \citep{shapley2025a}.
A larger [\ion{N}{2}]-detected sample is needed to confirm this trend.

At $z>6$, we find tentative evidence of evolution with $\approx3\sigma$ deviations based on the binned median toward higher O32 and Ne3O2 and lower O2 at fixed O/H, consistent with trends seen by \citet{cataldi2025}.
However, O3, R23, and RO2Ne3 do not show significant offsets from zero even at such high redshifts.
We thus do not find evidence for strong evolution of metallicity calibrations between $z\sim2$ and $z\sim10$.
If the evolution of strong-line calibrations between $z\sim0$ and $z\sim2$ is primarily driven by an increasing degree of $\alpha$-enhancement in massive stars, then this evolution may plausibly slow down or halt at $z\gtrsim2$ as galaxies approach the maximum value of $\alpha/\mathrm{Fe}\approx5\times\alpha/\mathrm{Fe}_\odot$ from pure core-collapse supernovae enrichment, assuming a standard IMF.
Typical star-forming populations at $z\sim2$ already lie at $3-4\times\alpha/\mathrm{Fe}_\odot$, such that galaxies at higher redshifts cannot achieve significantly higher $\alpha$-enhancement without altering the IMF or supernova yields \citep[e.g.,][]{steidel2016,strom2018,sanders2020,topping2020a,cullen2021,stanton2024,sanders2024}.
However, our evaluation is ultimately limited by the significant decrease in the number of auroral-detected galaxies at $z>6$, comprising less than 25\% of our sample.
Additional deep {\it JWST}/NIRSpec observations of $z>6$ galaxies is needed to improve statistics and robustly assess the case for calibration evolution.
Even if calibrations do evolve between $z\sim2$ and $z>6$, the new calibrations in this work still likely provide more accurate results than local-Universe calibrations for $z>6$ sources.

\subsection{Recommendations for applying the calibrations}

Lastly, we provide some recommendations for applying these calibrations to high-redshift strong-line samples.
It is strongly recommended that these calibrations only be used within the valid metallicity range for any particular line ratio, reported in Tab.~\ref{tab:cal}.
Extrapolating beyond this range may yield unreliable results, especially for higher-order polynomial forms, and great caution should be exercised if interpreting results from an extrapolation.

These calibrations should only be used with galaxies for which the line emission is predominantly powered by star formation, and with ionized ISM properties similar to the calibrating sample used here.
They will likely not yield reliable results if applied to objects powered by other ionizing sources (e.g., AGN, shocks).
They also may prove unreliable for sources with particularly extreme or anomalous properties, such as very high Lyman continuum escape fractions or electron densities ($n_\mathrm{e}\gtrsim10^4$~cm$^{-3}$).
At such high densities, collisional de-excitation will significantly impact the strength of low-ionization lines ([\ion{O}{2}], [\ion{S}{2}], [\ion{N}{2}]).
For the subset with available constraints, the vast majority of our sample has $n_\mathrm{e}\sim10^2-10^3$~cm$^{-3}$ based on [\ion{S}{2}].

As discussed in Sec.~\ref{sec:properties}, the apparent high dispersion in N/O at O/H among high-redshift sources makes N-based line ratios less reliable tracers of the gas-phase oxygen abundance, such that deviations from the mean N/O--O/H relation of the calibrating sample would directly bias the inferred O/H.
Even though the calibration scatters ($\sigma_{\mathrm{O/H,int}}$; Tab.~\ref{tab:cal}) for N2, O3N2, and N2S2 are quantitatively comparable to those of the O- and Ne-based ratios, their systematic sensitivity to N/O argues against their use if other line ratios that do not include [\ion{N}{2}] are available.
The resulting directional bias on O/H can be seen by comparing the panels of Fig.~\ref{fig:ncal}, where at fixed O/H the literature-sample sources that display the highest N2O2 lie above the N2 and N2S2 calibration lines and below the best-fit O3N2 relation.
We thus caution against the use of N2 and O3N2 (or other ratios involving [\ion{N}{2}]), and instead suggest that O- and Ne-based line ratios will have the most utility and reliability for high-redshift metallicity studies.
On a practical level, N-based indicators will have limited utility for metal-poor high-redshift galaxies due to the faintness of the [\ion{N}{2}]\W6585 line, for which there are high non-detection rates among such sources even with deep spectroscopy.

As with all strong-line calibrations, the intrinsic scatter of individual sources about these best-fit mean relations is large such that this source of systematic uncertainty often dominates the measurement uncertainty when inferring the metallicity of any single galaxy using its strong-line ratios.
The practical result of this scatter is that a strong-line metallicity will always carry a large total uncertainty for any individual object, but precision can be obtained for sample-averaged metallicities.
The intrinsic scatter estimates included in Tab.~\ref{tab:cal} can be used to account for this source of systematic uncertainty.
Simultaneously fitting multiple line ratios to derive the metallicity is desirable if sufficient measurements are available and is required if using double-valued ratios line O3, R23, or $\hat{R}$, and reduces the magnitude of systematic uncertainty on the metallicity estimate.
For an example of the multi-line ratio approach, see \citet{sanders2021}.

\section{Summary and Conclusions}\label{sec:conclusion}

In this paper, we reported the detection of auroral emission lines for 41 star-forming galaxies at $z=1.4-7.2$ in deep {\it JWST}/NIRSpec observations from the AURORA survey.
The detected auroral lines include 33 detections of [\ion{O}{3}]\W4364, 27 detections of [\ion{O}{2}]\W7322,7332, 11 detections of [\ion{S}{3}]\W6314, and 5 detections of [\ion{S}{2}]\W4070.
We combined the AURORA sources with 98 high-redshift galaxies with auroral-line detections drawn from the literature to form a combined sample of 139 galaxies at $z=1.3-10.6$ ($z_\mathrm{med}=3.80$) with electron temperature constraints and direct-method oxygen abundance determinations.
This sample spans 12+log(O/H$)=7.0-8.6$, extending to higher metallicities than previous {\it JWST} samples thanks to the depth of the AURORA observations.
We used this combined high-redshift \te\ sample to construct new strong-line metallicity calibrations that can be reliably applied to high-redshift star-forming galaxy samples.
Our main results and conclusions are as follows.

\begin{itemize}
    \item We find no evidence for redshift evolution of the relations between electron temperatures in the low-, intermediate-, and high-ionization nebular zones as traced by auroral lines of [\ion{O}{2}] and [\ion{S}{2}], [\ion{S}{3}], and [\ion{O}{3}], respectively.
    \item We provide metallicity calibration coefficients for 19 emission-line ratios involving \ion{H}{1} lines and metal lines of O, Ne, N, S, and Ar (Table~\ref{tab:cal}; Figs.~\ref{fig:ocal} to \ref{fig:arcal}), along with valid metallicity ranges and estimates of intrinsic scatter. The calibrating sample of 139 high-redshift galaxies provides robust statistics across a wide metallicity range for accurate chemical evolution studies in the early Universe.
    \item We find that the proposed $\hat{R}$ indicator \citep{laseter2024} provides no quantitative advantage for metallicity determination over [\ion{O}{3}]\W5008/H$\beta$ or R23 alone. The intrinsic scatter about the $\hat{R}$ calibration (Fig.~\ref{fig:rhatcal}) is not smaller than that of O3 or R23 for the high-redshift sample, and the inclusion of [\ion{O}{2}]/H$\beta$ in $\hat{R}$ leads to an increase in scatter at low metallicities.
    \item The Ne-based calibrations presented here (Ne3, Ne3O2, RO2Ne3; Fig.~\ref{fig:necal}) can be of particular utility in the very early Universe. {\it JWST}/NIRSpec can access the required [\ion{O}{2}]\W3728, [\ion{Ne}{3}]\W3870, and H$\delta$ lines out to $z\sim11.7$.
    \item Calibrations based on line ratios involving [\ion{N}{2}]\W6585 are significantly affected by variations in N/O at fixed O/H (Fig.~\ref{fig:ncal}). We find a large dispersion in N2O2 at fixed O/H and no significant correlation between these two parameters, implying that N/O displays more variation at fixed O/H among high-redshift sources than is seen locally. Consequently, N-based line ratios are less reliable tracers of the oxygen abundance at high redshifts ($z\gtrsim2$). We instead recommend the use of metallicity indicators based on $\alpha$ elements (O, Ne, S, or Ar) that do not have a systematic dependence on N/O.
    \item Our new strong-line calibrations show good agreement with those of previous studies based on high-redshift {\it JWST}/NIRSpec observations of smaller samples covering a more limited metallicity range, and reasonable agreement with calibrations based on local-Universe analogs of high-redshift galaxies (Fig.~\ref{fig:calcomp}). There is clear evolution of strong-line calibrations relative to those based on typical $z\sim0$ samples, with the direction of the shifts in qualitative agreement with the scenario where increasing $\alpha$/Fe at fixed O/H leads to hotter stellar effective temperatures and harder ionizing spectra with increasing redshift. Evolution of the N-based indicators additionally suggests an increase in N/O at fixed O/H.
\end{itemize}

The transformational leap in spectroscopic capabilities provided by {\it JWST}/NIRSpec has produced a sample of well over 100 galaxies at high redshifts ($z\gtrsim2$) with detections of auroral emission lines within its first few years of operation, a several-fold increase relative to ground-based efforts.
The possibility of \te\ constraints for significant numbers of high-redshift sources has opened the door to accurate studies of galaxy metallicity evolution across cosmic history, either through direct-method chemical abundances or (for much larger samples) strong-line techniques recalibrated for use at high redshifts.
The new metallicity calibrations presented in this work will enable accurate oxygen abundance determinations for samples of thousands of galaxies with suitable strong-line measurements in the {\it JWST} spectroscopic archive, including sources at $z>10$ probing the earliest generations of galaxies.
The resulting improved metallicity constraints will lead to refined characterizations of metallicity scaling relations and an enhanced understanding of baryon cycling and galaxy growth and formation at early times.

\begin{acknowledgments}
This work is based on observations made with the NASA/ESA/CSA James Webb Space Telescope. The data were
obtained from the Mikulski Archive for Space Telescopes at
the Space Telescope Science Institute, which is operated by the
Association of Universities for Research in Astronomy, Inc.,
under NASA contract NAS5-03127 for JWST.
Some of the data presented in this article were obtained from the Mikulski Archive for Space Telescopes (MAST) at the Space Telescope Science Institute. The specific observations analyzed can be accessed via \dataset[doi:10.17909/hvne-7139]{https://doi.org/10.17909/hvne-7139}.
Some of the data products presented herein were retrieved from the Dawn JWST Archive (DJA). DJA is an initiative of the Cosmic Dawn Center (DAWN), which is funded by the Danish National Research Foundation under grant DNRF140.
We acknowledge support from NASA grant JWST-GO-01914.
F.C. acknowledges support from a UKRI Frontier Research Guarantee Grant (PI Cullen; grant reference EP/X021025/1).
A.C.C. acknowledges support from a UKRI Frontier Research Guarantee Grant (PI Carnall; grant reference EP/Y037065/1).
J.S.D. acknowledges the support of the Royal Society via the award of a Royal Society Research Professorship.
This work has received funding from the Swiss State Secretariat for Education, Research and Innovation (SERI) under contract number MB22.00072, as well as from the Swiss National Science Foundation (SNSF) through project grant 200020\_207349.
\end{acknowledgments}

%

\facilities{JWST(NIRSpec and NIRCam), HST(ACS and WFC3)}


\appendix

\section{Properties of galaxies in the AURORA survey with auroral-line detections}\label{app:aurora}

The derived physical properties of the star-forming galaxies in the AURORA survey with auroral-line detections are presented in Table~\ref{tab:aurora}.
The emission-lines [\ion{O}{2}]\W3728 and [\ion{O}{3}]\W\W5008 are necessary for deriving O/H.
One of these lines was not covered for four of the AURORA sources with detections of at least one auroral line.
To derive metallicities for these four sources, we inferred the strength of these lines indirectly as described below.
The indirectly inferred line fluxes are only used for metallicity calculation.
None of our results significantly change if these sources are removed.

GOODSN-19149 ($z=1.38$) and COSMOS-8442 ($z=1.60$) do not have coverage of [\ion{O}{2}]\W3728 in the F100LP/G140M configuration due to their low redshifts.
Both of these sources display relatively high excitation line ratios (log(O3$)=0.64$ and 0.67 for 19149 and 8442, respectively), suggesting that O$^{2+}$ is the dominant O ion.
Based on the median value and range of O32 ratios measured for sources with similar O3 ratios that have [\ion{O}{2}] coverage, we infer [\ion{O}{2}]\W3728 from [\ion{O}{3}]\W5008 assuming $\mathrm{O32}=6$ and compute O$^+$/H using this value, adopting a uniform distribution of $\mathrm{O32}=3-13$ to estimate the uncertainty.
Objects with [\ion{O}{2}] coverage with log(O3$)=0.6-0.75$ have O$^+/\mathrm{O}=0.07-0.49$ with a median value of 0.24 and a standard deviation of 0.14, providing further evidence that O$^{+}$ is subdominant for 19149 and 8442. Consequently, the lack of a direct [\ion{O}{2}]\W3728 constraint is unlikely to significantly bias the inferred metallicities for these targets.

\begin{figure}
\centering
\includegraphics[width=0.49\textwidth]{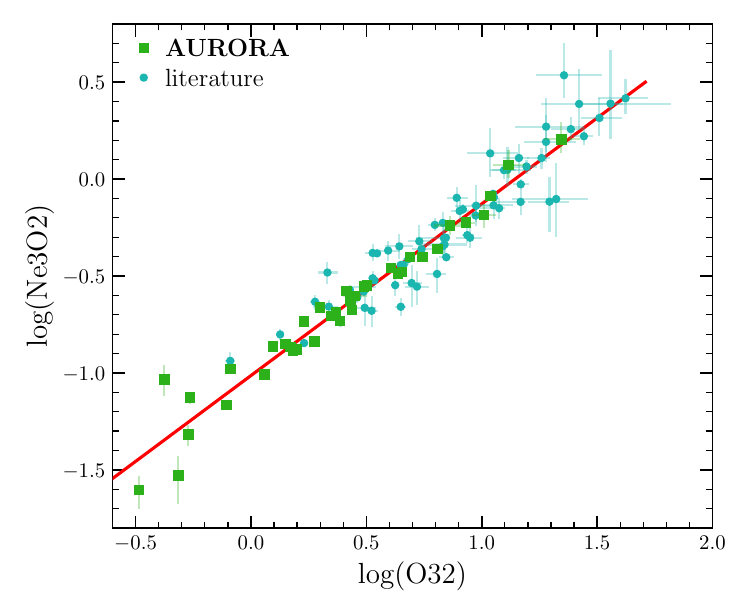}
\caption{Ne3O2 vs.\ O32 for the combined high-redshift auroral-detected sample.
The red line shows the best-fit relation.
After accounting for measurement uncertainties, the intrinsic scatter about this best-fit relation was found to be 0.05~dex.
}\label{fig:ne3o2_o32}
\end{figure}

GOODSN-25004 and COSMOS-440430 lack [\ion{O}{3}]\W\W4960,5008 coverage because these lines fell in the chip gap between the two NIRSpec detectors.
The line ratios Ne3O2 and O32 are known to be tightly related \citep[e.g.,][]{levesque2014,jeong2020}, such that the measured Ne3O2 ratio can be used to estimate the dust-corrected O32 ratio and the [\ion{O}{3}]\W5008 flux can then be inferred from the measured [\ion{O}{2}]\W3728 flux.
Figure~\ref{fig:ne3o2_o32} displays Ne3O2 vs.\ O32 for objects in the auroral-line detected sample with [\ion{O}{2}]\W3728 coverage.
We fit a linear relationship to these data, obtaining $\log(\mathrm{O32})=1.14+1.13\log(\mathrm{Ne3O2})$ (red line).
We determine the intrinsic scatter about this best-fit line to be 0.05~dex (12\%) after accounting for measurement uncertainties.
We use this relationship and the measured log(Ne3O2) ratios ($-0.60\pm0.01$ for 25004 and $-0.24\pm0.05$ for 440430) and [\ion{O}{2}]\W3728 fluxes to estimate to infer the [\ion{O}{3}]\W5008 strength and compute O$^{2+}$/H for 25004 and 440430, including the 0.05~dex intrinsic scatter when calculating uncertainties.

\startlongtable
\begin{deluxetable*}{ l l l l l l l l l l }
 \setlength{\tabcolsep}{3pt}
 \tablecaption{Derived properties of the AURORA star-forming galaxies with auroral-line detections.\label{tab:aurora}}
 \tablehead{ \colhead{ID} & \colhead{$z_{\mathrm{spec}}$} & \colhead{\ebvgas} & \colhead{$\log(M_*/\mathrm{M}_\odot)$} & \colhead{$\log\left(\frac{\mathrm{SFR(SED)}}{\mathrm{M}_\odot\ \mathrm{yr}^{-1}}\right)$} & \colhead{$\log\left(\frac{\mathrm{SFR(H}\alpha)}{\mathrm{M}_\odot\ \mathrm{yr}^{-1}}\right)$} & \colhead{\densp} & \colhead{\temotp} & \colhead{\temop} & \colhead{12+log$\left(\frac{\mathrm{O}}{\mathrm{H}}\right)$} \\ & & \colhead{\scriptsize mag} & & & & \colhead{\scriptsize{cm$^{-3}$}} & \colhead{\scriptsize{K}} & \colhead{\scriptsize{K}} & }
\startdata
  GOODSN-11584  &  3.3616  &  $0.015^{+0.014}_{-0.014}$  &  $10.78^{+0.01}_{-0.05}$  &  $2.01^{+0.01}_{-0.01}$  &  $2.31^{+0.01}_{-0.01}$  &  $170^{+140}_{-120}$  &  $10790^{+910}_{-910}$  &  $9430^{+1010}_{-1060}$  &  $8.52^{+0.17}_{-0.09}$  \\
  GOODSN-17940  &  4.4115  &  $0.388^{+0.006}_{-0.007}$  &  $9.01^{+0.03}_{-0.03}$  &  $1.58^{+0.05}_{-0.01}$  &  $2.19^{+0.01}_{-0.01}$  &  $840^{+120}_{-110}$  &  $11210^{+270}_{-250}$  &  $11230^{+510}_{-540}$  &  $8.30^{+0.03}_{-0.03}$  \\
  GOODSN-19067  &  2.2813  &  $0.193^{+0.011}_{-0.011}$  &  $9.58^{+0.06}_{-0.10}$  &  $0.92^{+0.11}_{-0.13}$  &  $0.66^{+0.01}_{-0.01}$  &  $<210$  &  $13560^{+1220}_{-1080}$  &  $10530^{+730}_{-800}$  &  $8.16^{+0.11}_{-0.08}$  \\
  GOODSN-19149  &  1.3833  &  $0.000^{+0.000}_{-0.000}$  &  $7.95^{+0.08}_{-0.04}$  &  $0.59^{+0.11}_{-0.16}$  &  $0.04^{+0.01}_{-0.01}$  &  $320^{+180}_{-170}$  &  $16370^{+880}_{-800}$  &  ---  &  $7.69^{+0.05}_{-0.06}$  \\
  GOODSN-19848  &  2.9918  &  $0.240^{+0.008}_{-0.008}$  &  $9.19^{+0.05}_{-0.04}$  &  $1.53^{+0.12}_{-0.01}$  &  $1.37^{+0.01}_{-0.01}$  &  $1050^{+270}_{-250}$  &  $11620^{+370}_{-370}$  &  $9180^{+740}_{-650}$  &  $8.35^{+0.09}_{-0.08}$  \\
  GOODSN-21033  &  3.1120  &  $0.093^{+0.007}_{-0.007}$  &  $8.96^{+0.05}_{-0.08}$  &  $0.79^{+0.11}_{-0.06}$  &  $0.98^{+0.01}_{-0.01}$  &  $50^{+90}_{-50}$  &  $12740^{+240}_{-210}$  &  $18790^{+1810}_{-2140}$  &  $8.07^{+0.02}_{-0.02}$  \\
  GOODSN-21522  &  2.3634  &  $0.035^{+0.008}_{-0.008}$  &  $9.49^{+0.11}_{-0.06}$  &  $1.69^{+0.03}_{-0.13}$  &  $0.86^{+0.01}_{-0.01}$  &  $140^{+80}_{-60}$  &  $12760^{+630}_{-680}$  &  $10690^{+850}_{-880}$  &  $8.10^{+0.10}_{-0.07}$  \\
  GOODSN-22235  &  2.4298  &  $0.224^{+0.006}_{-0.006}$  &  $9.16^{+0.12}_{-0.05}$  &  $1.55^{+0.03}_{-0.14}$  &  $1.30^{+0.01}_{-0.01}$  &  $370^{+60}_{-60}$  &  $11320^{+240}_{-260}$  &  $9420^{+390}_{-360}$  &  $8.37^{+0.04}_{-0.04}$  \\
  GOODSN-22384  &  2.9935  &  $0.246^{+0.010}_{-0.010}$  &  $9.67^{+0.13}_{-0.05}$  &  $1.42^{+0.02}_{-0.17}$  &  $1.29^{+0.01}_{-0.01}$  &  $190^{+60}_{-50}$  &  $<11920$  &  $10790^{+1050}_{-960}$  &  $8.18^{+0.18}_{-0.16}$  \\
  GOODSN-22932  &  3.3305  &  $0.047^{+0.010}_{-0.009}$  &  $8.99^{+0.10}_{-0.09}$  &  $1.05^{+0.13}_{-0.14}$  &  $0.89^{+0.01}_{-0.01}$  &  $310^{+200}_{-170}$  &  $13190^{+540}_{-550}$  &  $<8850$  &  $8.09^{+0.05}_{-0.05}$  \\
  GOODSN-25004  &  2.0487  &  $0.219^{+0.004}_{-0.004}$  &  $8.82^{+0.01}_{-0.01}$  &  $1.75^{+0.01}_{-0.13}$  &  $1.39^{+0.01}_{-0.01}$  &  $350^{+80}_{-50}$  &  $10060^{+440}_{-420}$  &  $9980^{+400}_{-460}$  &  $8.42^{+0.09}_{-0.07}$  \\
  GOODSN-26798  &  2.4831  &  $0.387^{+0.009}_{-0.010}$  &  $10.49^{+0.01}_{-0.04}$  &  $1.63^{+0.22}_{-0.01}$  &  $1.46^{+0.01}_{-0.01}$  &  $450^{+60}_{-50}$  &  $<13990$  &  $8650^{+380}_{-340}$  &  $8.62^{+0.10}_{-0.10}$  \\
  GOODSN-27876  &  2.2709  &  $0.531^{+0.010}_{-0.009}$  &  $10.19^{+0.02}_{-0.40}$  &  $1.19^{+0.66}_{-0.02}$  &  $1.44^{+0.01}_{-0.01}$  &  $550^{+90}_{-90}$  &  $<21840$  &  $8110^{+520}_{-520}$  &  $8.65^{+0.17}_{-0.15}$  \\
  GOODSN-28209  &  3.2325  &  $0.297^{+0.011}_{-0.010}$  &  $9.22^{+0.06}_{-0.12}$  &  $1.72^{+0.18}_{-0.02}$  &  $1.72^{+0.01}_{-0.01}$  &  $1050^{+380}_{-290}$  &  $9350^{+520}_{-540}$  &  $9290^{+950}_{-910}$  &  $8.52^{+0.13}_{-0.08}$  \\
  GOODSN-30053  &  2.2454  &  $0.468^{+0.006}_{-0.005}$  &  $9.90^{+0.05}_{-0.26}$  &  $2.03^{+0.55}_{-0.04}$  &  $1.58^{+0.01}_{-0.01}$  &  $300^{+60}_{-50}$  &  $11080^{+600}_{-680}$  &  $9640^{+410}_{-420}$  &  $8.31^{+0.07}_{-0.05}$  \\
  GOODSN-30274  &  1.7997  &  $0.037^{+0.011}_{-0.012}$  &  $7.50^{+0.01}_{-0.01}$  &  $0.08^{+0.01}_{-0.10}$  &  $0.30^{+0.01}_{-0.01}$  &  $310^{+1300}_{-310}$  &  $14090^{+420}_{-520}$  &  $<25570$  &  $7.99^{+0.05}_{-0.03}$  \\
  GOODSN-30564  &  2.4828  &  $0.523^{+0.006}_{-0.005}$  &  $9.93^{+0.12}_{-0.11}$  &  $2.01^{+0.17}_{-0.11}$  &  $1.62^{+0.01}_{-0.01}$  &  $550^{+50}_{-50}$  &  $<10430$  &  $8790^{+260}_{-240}$  &  $8.57^{+0.07}_{-0.07}$  \\
  GOODSN-30811  &  2.3067  &  $0.103^{+0.012}_{-0.012}$  &  $9.42^{+0.02}_{-0.04}$  &  $0.42^{+0.09}_{-0.08}$  &  $0.51^{+0.01}_{-0.01}$  &  $590^{+190}_{-170}$  &  $11310^{+700}_{-740}$  &  $10170^{+1210}_{-1020}$  &  $8.32^{+0.13}_{-0.10}$  \\
  GOODSN-100026  &  7.2043  &  $0.000^{+0.114}_{-0.000}$  &  $8.64^{+0.17}_{-0.27}$  &  $0.25^{+0.05}_{-0.08}$  &  $0.57^{+0.17}_{-0.02}$  &  ---  &  $18570^{+2270}_{-1500}$  &  ---  &  $7.64^{+0.09}_{-0.11}$  \\
  GOODSN-100067  &  5.1875  &  $0.223^{+0.018}_{-0.017}$  &  $8.85^{+0.07}_{-0.02}$  &  $1.97^{+0.13}_{-0.22}$  &  $1.31^{+0.02}_{-0.02}$  &  $20^{+470}_{-20}$  &  $12520^{+370}_{-420}$  &  $<15450$  &  $8.20^{+0.05}_{-0.04}$  \\
  GOODSN-100163  &  6.7479  &  $0.295^{+0.078}_{-0.068}$  &  $7.75^{+0.16}_{-0.04}$  &  $1.35^{+0.27}_{-0.24}$  &  $0.99^{+0.08}_{-0.07}$  &  ---  &  $21800^{+3430}_{-3130}$  &  ---  &  $7.68^{+0.14}_{-0.11}$  \\
  GOODSN-927605  &  4.0480  &  $0.064^{+0.078}_{-0.064}$  &  $8.68^{+0.02}_{-0.08}$  &  $1.03^{+0.04}_{-0.11}$  &  $0.83^{+0.11}_{-0.09}$  &  ---  &  $16090^{+880}_{-750}$  &  ---  &  $7.76^{+0.05}_{-0.06}$  \\
  COSMOS-3324  &  2.3077  &  $0.438^{+0.019}_{-0.018}$  &  $10.64^{+0.09}_{-0.01}$  &  $1.68^{+0.13}_{-0.30}$  &  $1.93^{+0.02}_{-0.02}$  &  $120^{+100}_{-90}$  &  $<30690$  &  $9810^{+1030}_{-950}$  &  $8.30^{+0.22}_{-0.20}$  \\
  COSMOS-3632  &  1.9201  &  $0.165^{+0.013}_{-0.015}$  &  $8.08^{+0.01}_{-0.06}$  &  $0.55^{+0.05}_{-0.01}$  &  $0.40^{+0.01}_{-0.02}$  &  $<590$  &  $13220^{+1580}_{-1810}$  &  $<8320$  &  $8.13^{+0.18}_{-0.13}$  \\
  COSMOS-4029  &  2.0765  &  $0.147^{+0.006}_{-0.006}$  &  $8.44^{+0.01}_{-0.07}$  &  $0.90^{+0.10}_{-0.01}$  &  $1.02^{+0.01}_{-0.01}$  &  $160^{+90}_{-80}$  &  $12190^{+230}_{-210}$  &  $12060^{+900}_{-870}$  &  $8.21^{+0.03}_{-0.03}$  \\
  COSMOS-4156  &  2.1897  &  $0.000^{+0.117}_{-0.000}$  &  $8.66^{+0.09}_{-0.03}$  &  $1.13^{+0.02}_{-0.12}$  &  $0.78^{+0.01}_{-0.01}$  &  $420^{+110}_{-110}$  &  $14020^{+170}_{-170}$  &  $10380^{+1010}_{-950}$  &  $8.08^{+0.06}_{-0.04}$  \\
  COSMOS-4205  &  1.8368  &  $0.149^{+0.004}_{-0.004}$  &  $8.33^{+0.01}_{-0.01}$  &  $0.91^{+0.01}_{-0.10}$  &  $0.81^{+0.01}_{-0.01}$  &  $170^{+60}_{-60}$  &  $12760^{+140}_{-150}$  &  $13690^{+910}_{-770}$  &  $8.16^{+0.02}_{-0.02}$  \\
  COSMOS-4210  &  2.8280  &  $0.185^{+0.026}_{-0.027}$  &  $9.36^{+0.01}_{-0.13}$  &  $0.31^{+0.27}_{-0.02}$  &  $0.36^{+0.03}_{-0.03}$  &  $<690$  &  $13240^{+1120}_{-1080}$  &  $<11820$  &  $8.07^{+0.11}_{-0.09}$  \\
  COSMOS-4429  &  2.1023  &  $0.150^{+0.018}_{-0.018}$  &  $8.93^{+0.29}_{-0.15}$  &  $0.70^{+0.13}_{-0.48}$  &  $0.23^{+0.02}_{-0.02}$  &  $310^{+310}_{-240}$  &  $12640^{+1330}_{-1420}$  &  $<9470$  &  $8.07^{+0.16}_{-0.12}$  \\
  COSMOS-4622  &  1.6400  &  $0.775^{+0.013}_{-0.011}$  &  $9.46^{+0.58}_{-0.21}$  &  $2.76^{+0.05}_{-3.16}$  &  $0.86^{+0.01}_{-0.01}$  &  $760^{+190}_{-170}$  &  $<29290$  &  $9240^{+870}_{-910}$  &  $8.37^{+0.24}_{-0.19}$  \\
  COSMOS-4740  &  3.1556  &  $0.495^{+0.009}_{-0.009}$  &  $10.05^{+0.09}_{-0.47}$  &  $1.61^{+1.10}_{-0.10}$  &  $1.72^{+0.01}_{-0.01}$  &  $720^{+90}_{-100}$  &  $13320^{+1000}_{-1010}$  &  $9340^{+540}_{-540}$  &  $8.29^{+0.10}_{-0.08}$  \\
  COSMOS-5283  &  2.1742  &  $0.103^{+0.004}_{-0.004}$  &  $9.47^{+0.03}_{-0.11}$  &  $1.25^{+0.13}_{-0.04}$  &  $1.34^{+0.01}_{-0.01}$  &  $260^{+40}_{-40}$  &  $11680^{+220}_{-240}$  &  $10680^{+350}_{-370}$  &  $8.27^{+0.03}_{-0.02}$  \\
  COSMOS-5571  &  2.2784  &  $0.296^{+0.008}_{-0.009}$  &  $10.26^{+0.02}_{-0.01}$  &  $1.29^{+0.01}_{-0.01}$  &  $1.49^{+0.01}_{-0.01}$  &  $160^{+40}_{-40}$  &  $<11280$  &  $9310^{+420}_{-410}$  &  $8.46^{+0.10}_{-0.10}$  \\
  COSMOS-5901  &  2.3963  &  $0.263^{+0.014}_{-0.014}$  &  $9.90^{+0.13}_{-0.08}$  &  $1.46^{+0.17}_{-0.09}$  &  $1.08^{+0.01}_{-0.02}$  &  $100^{+60}_{-50}$  &  $<14520$  &  $9440^{+900}_{-760}$  &  $8.39^{+0.19}_{-0.18}$  \\
  COSMOS-6825  &  1.9748  &  $0.133^{+0.013}_{-0.013}$  &  $8.72^{+0.03}_{-0.07}$  &  $1.18^{+0.05}_{-0.09}$  &  $0.78^{+0.02}_{-0.02}$  &  ---  &  $11720^{+460}_{-480}$  &  $15110^{+1590}_{-1580}$  &  $8.08^{+0.05}_{-0.04}$  \\
  COSMOS-7883  &  2.1532  &  $0.011^{+0.011}_{-0.010}$  &  $9.45^{+0.02}_{-0.31}$  &  $0.67^{+0.41}_{-0.01}$  &  $0.64^{+0.01}_{-0.01}$  &  $220^{+110}_{-100}$  &  $11900^{+790}_{-750}$  &  $13370^{+2010}_{-1730}$  &  $8.10^{+0.10}_{-0.06}$  \\
  COSMOS-8363  &  3.2475  &  $0.140^{+0.008}_{-0.008}$  &  $9.62^{+0.16}_{-0.14}$  &  $1.37^{+0.13}_{-0.20}$  &  $1.36^{+0.01}_{-0.01}$  &  $300^{+80}_{-90}$  &  $9610^{+370}_{-380}$  &  $12750^{+980}_{-910}$  &  $8.39^{+0.06}_{-0.05}$  \\
  COSMOS-8442  &  1.6053  &  $0.125^{+0.008}_{-0.008}$  &  $8.99^{+0.13}_{-0.15}$  &  $1.05^{+0.15}_{-0.14}$  &  $0.49^{+0.01}_{-0.01}$  &  $10^{+50}_{-10}$  &  $9460^{+810}_{-980}$  &  ---  &  $8.35^{+0.16}_{-0.13}$  \\
  COSMOS-419213  &  6.8090  &  $0.118^{+0.070}_{-0.064}$  &  $7.77^{+0.36}_{-0.48}$  &  $-0.03^{+0.59}_{-0.18}$  &  $0.43^{+0.07}_{-0.07}$  &  ---  &  $30220^{+6370}_{-5860}$  &  ---  &  $7.16^{+0.17}_{-0.05}$  \\
  COSMOS-440430  &  5.5207  &  $0.000^{+0.000}_{-0.000}$  &  $8.48^{+0.65}_{-0.09}$  &  $1.70^{+0.85}_{-1.71}$  &  $0.05^{+0.02}_{-0.02}$  &  $1390^{+3710}_{-1180}$  &  $17610^{+2800}_{-2610}$  &  ---  &  $7.84^{+0.25}_{-0.16}$  \\
  COSMOS-443467  &  5.5039  &  $0.041^{+0.068}_{-0.041}$  &  $8.79^{+1.17}_{-0.49}$  &  $1.99^{+0.69}_{-3.97}$  &  $0.42^{+0.07}_{-0.04}$  &  ---  &  $18840^{+1450}_{-1240}$  &  ---  &  $7.81^{+0.07}_{-0.07}$  \\
\enddata
\end{deluxetable*}

\section{Literature sample with auroral-line detections}\label{app:lit}

In Table~\ref{tab:lit}, we report the original references for objects in the literature sample alongside our derived electron temperatures and direct-method oxygen abundances.
Seven literature sources lack [\ion{O}{2}]\W3728 coverage (30055, 20028, 10010, SL2SJ02176-0513, GLASS 150008, A1703-zd5.2, and A1703-23).
Given the similarity in degree of excitation between this subsample (median log(O3$)=0.76$ with a standard deviation of 0.12) and the two AURORA objects lacking [\ion{O}{2}]\W3728 coverage, we adopt the same approach used above to estimate the contribution of O$^+$ to the total oxygen abundance.
The [\ion{O}{2}]\W3728 flux is inferred from [\ion{O}{3}]\W5008 assuming $\mathrm{O32}=6$ when computing O$^+$/H, and a uniform O32 distribution between 3 and 13 is used when calculating uncertainties.
These seven sources are not included in the subsamples used to construct calibrations involving ratios that include [\ion{O}{2}] (e.g., R23, O32, Ne3O2, etc.).
Our results do not significantly change if these seven sources are excluded.

\startlongtable
\begin{deluxetable*}{ l l l l l l }
 \setlength{\tabcolsep}{2pt}
 \tablecaption{ID, spectroscopic redshift, literature reference, and the electron temperatures and metallicities derived in this work for the literature auroral-detected sample of star-forming galaxies.\label{tab:lit}}
 \tablehead{ \colhead{ID} & \colhead{$z_{\mathrm{spec}}$} & \colhead{Reference} & \colhead{\temotp\tablenotemark{a}} & \colhead{\temop} & \colhead{12+log$\left(\frac{\mathrm{O}}{\mathrm{H}}\right)$} \\ & & & \colhead{\scriptsize{K}} & \colhead{\scriptsize{K}} & }
 \startdata
  10058975  &  9.431  &  \citet{laseter2024}\tablenotemark{b}  &  $21040^{+4370}_{-3530}$  &  ---  &  $7.51^{+0.17}_{-0.18}$  \\
  18846  &  6.332  &  \citet{laseter2024}\tablenotemark{b}  &  $19770^{+3080}_{-2960}$  &  ---  &  $7.47^{+0.16}_{-0.13}$  \\
  9422  &  5.933  &  \citet{laseter2024}\tablenotemark{b}  &  $17930^{+2010}_{-1640}$  &  ---  &  $7.66^{+0.10}_{-0.10}$  \\
  18090  &  4.773  &  \citet{laseter2024}\tablenotemark{b}  &  $14380^{+3010}_{-3120}$  &  ---  &  $8.00^{+0.33}_{-0.19}$  \\
  7892  &  4.227  &  \citet{laseter2024}\tablenotemark{b}  &  $30290^{+5120}_{-4120}$  &  ---  &  $7.24^{+0.12}_{-0.05}$  \\
  19519  &  3.603  &  \citet{laseter2024}\tablenotemark{b}  &  $16440^{+3400}_{-3520}$  &  ---  &  $7.82^{+0.28}_{-0.18}$  \\
  10035295  &  3.587  &  \citet{laseter2024}\tablenotemark{b}  &  $22590^{+3590}_{-2930}$  &  ---  &  $7.39^{+0.13}_{-0.11}$  \\
  19607  &  1.846  &  \citet{laseter2024}\tablenotemark{b}  &  $15700^{+5080}_{-3600}$  &  ---  &  $7.80^{+0.37}_{-0.28}$  \\
  21598  &  1.714  &  \citet{laseter2024}\tablenotemark{b}  &  $26610^{+10940}_{-7720}$  &  ---  &  $7.07^{+0.33}_{-0.13}$  \\
  30055  &  3.214  &  \citet{morishita2024}  &  $15640^{+2130}_{-2240}$  &  ---  &  $7.98^{+0.17}_{-0.14}$  \\
  20028  &  3.345  &  \citet{morishita2024}  &  $15070^{+860}_{-800}$  &  ---  &  $7.85^{+0.07}_{-0.07}$  \\
  150880  &  4.247  &  \citet{morishita2024}  &  $15010^{+1140}_{-1250}$  &  ---  &  $7.99^{+0.10}_{-0.08}$  \\
  320108  &  4.257  &  \citet{morishita2024}  &  $10820^{+370}_{-440}$  &  ---  &  $8.35^{+0.06}_{-0.05}$  \\
  320002  &  4.658  &  \citet{morishita2024}  &  $26610^{+6590}_{-5390}$  &  ---  &  $7.38^{+0.19}_{-0.10}$  \\
  150903  &  4.659  &  \citet{morishita2024}  &  $16650^{+1770}_{-1860}$  &  ---  &  $7.90^{+0.14}_{-0.11}$  \\
  10010  &  6.311  &  \citet{morishita2024}  &  $17730^{+3180}_{-2870}$  &  ---  &  $7.73^{+0.20}_{-0.16}$  \\
  2  &  7.232  &  \citet{morishita2024}  &  $16370^{+480}_{-520}$  &  ---  &  $7.96^{+0.04}_{-0.04}$  \\
  3  &  9.114  &  \citet{morishita2024}  &  $16490^{+1120}_{-1200}$  &  ---  &  $7.82^{+0.08}_{-0.07}$  \\
  1019  &  8.679  &  \citet{sanders2024}  &  $16990^{+2040}_{-1660}$  &  ---  &  $7.78^{+0.11}_{-0.11}$  \\
  1149  &  8.175  &  \citet{sanders2024}  &  $16720^{+2930}_{-2680}$  &  ---  &  $7.83^{+0.20}_{-0.16}$  \\
  1027  &  7.819  &  \citet{sanders2024}  &  $19000^{+2550}_{-2010}$  &  ---  &  $7.61^{+0.12}_{-0.12}$  \\
  792  &  6.257  &  \citet{sanders2024}  &  $28280^{+8440}_{-5600}$  &  ---  &  $7.48^{+0.17}_{-0.08}$  \\
  397  &  6.000  &  \citet{sanders2024}  &  $14230^{+1290}_{-1350}$  &  ---  &  $7.98^{+0.12}_{-0.09}$  \\
  1536  &  5.033  &  \citet{sanders2024}  &  $25700^{+5250}_{-4830}$  &  ---  &  $7.43^{+0.17}_{-0.08}$  \\
  1477  &  4.631  &  \citet{sanders2024}  &  $18750^{+2280}_{-2260}$  &  ---  &  $7.69^{+0.13}_{-0.11}$  \\
  1746  &  4.560  &  \citet{sanders2024}  &  $15570^{+1900}_{-1950}$  &  ---  &  $7.95^{+0.17}_{-0.12}$  \\
  1665  &  4.482  &  \citet{sanders2024}  &  $11940^{+1400}_{-1420}$  &  $<13120$  &  $8.25^{+0.17}_{-0.13}$  \\
  1559  &  4.471  &  \citet{sanders2024}  &  $18130^{+3160}_{-2670}$  &  ---  &  $7.88^{+0.17}_{-0.14}$  \\
  11728  &  3.869  &  \citet{sanders2024}  &  $19680^{+2600}_{-2240}$  &  ---  &  $7.50^{+0.11}_{-0.11}$  \\
  11088  &  3.302  &  \citet{sanders2024}  &  $12120^{+1440}_{-1620}$  &  $9950^{+1400}_{-1230}$  &  $8.31^{+0.22}_{-0.12}$  \\
  3788  &  2.295  &  \citet{sanders2024}  &  $12280^{+970}_{-980}$  &  $11540^{+2390}_{-2250}$  &  $8.24^{+0.19}_{-0.09}$  \\
  3537  &  2.162  &  \citet{sanders2024}  &  $24970^{+3510}_{-3310}$  &  ---  &  $7.10^{+0.12}_{-0.08}$  \\
  AEGIS-11452  &  1.671  &  \citet{sanders2020}  &  $16090^{+2670}_{-2790}$  &  ---  &  $7.72^{+0.21}_{-0.15}$  \\
  COSMOS-1908  &  3.077  &  \citet{sanders2020}  &  $13630^{+1740}_{-1150}$  &  ---  &  $8.02^{+0.11}_{-0.14}$  \\
  CSWA 141  &  1.425  &  \citet{stark2013}  &  $16990^{+1270}_{-1230}$  &  ---  &  $7.86^{+0.08}_{-0.07}$  \\
  A1689 ID 31.1  &  1.834  &  \citet{christensen2012}  &  $20870^{+6130}_{-3590}$  &  ---  &  $7.45^{+0.20}_{-0.19}$  \\
  Abell 860\_359  &  1.702  &  \citet{stark2014}  &  $13030^{+1050}_{-1150}$$^{\dagger}$  &  ---  &  $8.04^{+0.11}_{-0.10}$  \\
  SMACS J0304 ID 1.1  &  1.963  &  \citet{christensen2012}  &  $12300^{+380}_{-450}$$^{\dagger}$  &  ---  &  $8.15^{+0.05}_{-0.04}$  \\
  SL2SJ02176-0513  &  1.844  &  \citet{berg2018}  &  $15520^{+1990}_{-240}$$^{\dagger}$  &  ---  &  $7.56^{+0.03}_{-0.15}$  \\
  MACS 0451 ID 1.1b  &  2.060  &  \citet{stark2014}  &  $21220^{+4210}_{-2520}$$^{\dagger}$  &  ---  &  $7.36^{+0.11}_{-0.14}$  \\
  COSMOS-12805  &  2.159  &  \citet{kojima2017}  &  $11990^{+5070}_{-340}$$^{\dagger}$  &  ---  &  $8.29^{+0.03}_{-0.23}$  \\
  BX74  &  2.189  &  \citet{steidel2014}  &  $15010^{+1330}_{-1210}$$^{\dagger}$  &  ---  &  $7.98^{+0.07}_{-0.06}$  \\
  BX418  &  2.305  &  \citet{steidel2014}  &  $12880^{+1040}_{-260}$$^{\dagger}$  &  ---  &  $8.09^{+0.01}_{-0.08}$  \\
  BX660  &  2.174  &  \citet{steidel2014}  &  $12740^{+840}_{-430}$$^{\dagger}$  &  ---  &  $8.14^{+0.04}_{-0.07}$  \\
  Lynx Arc  &  3.357  &  \citet{villarmartin2004}  &  $17260^{+2350}_{-210}$$^{\dagger}$  &  ---  &  $7.84^{+0.03}_{-0.14}$  \\
  SMACS J2031 ID 1.1  &  3.507  &  \citet{christensen2012}  &  $16460^{+6290}_{-160}$$^{\dagger}$  &  ---  &  $7.74^{+0.01}_{-0.29}$  \\
  SGAS J105039.6+001730  &  3.625  &  \citet{bayliss2014}  &  $13880^{+950}_{-950}$$^{\dagger}$  &  ---  &  $8.08^{+0.08}_{-0.07}$  \\
  COSMOS-19985  &  2.188  &  \citet{sanders2023oii}  &  ---  &  $12350^{+1850}_{-1520}$  &  $7.99^{+0.23}_{-0.21}$  \\
  COSMOS-20062  &  2.185  &  \citet{sanders2023oii}  &  ---  &  $9880^{+1370}_{-1310}$  &  $8.37^{+0.32}_{-0.25}$  \\
  ERO 04590  &  8.496  &  \citet{curti2023}  &  $24340^{+2790}_{-2540}$  &  ---  &  $7.15^{+0.10}_{-0.10}$  \\
  ERO 06355  &  7.665  &  \citet{curti2023}  &  $12350^{+1020}_{-990}$  &  ---  &  $8.24^{+0.11}_{-0.10}$  \\
  ERO 10612  &  7.660  &  \citet{curti2023}  &  $18800^{+2240}_{-2020}$  &  ---  &  $7.67^{+0.11}_{-0.11}$  \\
  ERO 05144  &  6.378  &  \citet{nakajima2023}  &  $15390^{+1690}_{-1250}$  &  ---  &  $7.90^{+0.10}_{-0.11}$  \\
  GLASS 100003  &  7.877  &  \citet{nakajima2023}  &  $19680^{+4950}_{-3280}$  &  ---  &  $7.68^{+0.18}_{-0.19}$  \\
  GLASS 10021  &  7.286  &  \citet{nakajima2023}  &  $16560^{+2190}_{-1550}$  &  ---  &  $7.89^{+0.11}_{-0.13}$  \\
  GLASS 150029  &  4.584  &  \citet{nakajima2023}  &  $17240^{+1750}_{-1400}$  &  ---  &  $7.72^{+0.10}_{-0.10}$  \\
  GLASS 160133  &  4.015  &  \citet{nakajima2023}  &  $14530^{+680}_{-730}$  &  ---  &  $7.99^{+0.06}_{-0.05}$  \\
  GLASS 150008  &  6.229  &  \citet{jones2023}  &  $25280^{+13550}_{-6010}$$^{\dagger}$  &  ---  &  $7.36^{+0.25}_{-0.14}$  \\
  Q2343-D40  &  2.963  &  \citet{rogers2024}  &  $13190^{+610}_{-560}$  &  ---  &  $8.08^{+0.07}_{-0.07}$  \\
  SGAS1723+34  &  1.329  &  \citet{welch2024}  &  $12490^{+480}_{-480}$  &  $12210^{+1210}_{-1230}$  &  $8.13^{+0.06}_{-0.04}$  \\
  Sunburst Arc  &  2.371  &  \citet{welch2025}  &  $15390^{+670}_{-520}$  &  $22450^{+10610}_{-6500}$  &  $7.92^{+0.14}_{-0.09}$  \\
  MACS0647‑JD  &  10.165  &  \citet{hsiao2024}  &  $16680^{+2380}_{-2020}$  &  ---  &  $7.79^{+0.16}_{-0.13}$  \\
  A1703-zd5.2  &  6.429  &  \citet{topping2025}  &  $25580^{+2350}_{-2070}$$^{\dagger}$  &  ---  &  $7.26^{+0.07}_{-0.07}$  \\
  A1703-23  &  6.086  &  \citet{topping2025}  &  $28140^{+6390}_{-4590}$$^{\dagger}$  &  ---  &  $7.41^{+0.16}_{-0.06}$  \\
  A1703-zd6  &  7.043  &  \citet{topping2025}  &  $28110^{+2180}_{-1990}$  &  ---  &  $7.35^{+0.06}_{-0.05}$  \\
  GN-z11  &  10.603  &  \citet{alvarez2025}  &  $14040^{+2020}_{-1840}$  &  ---  &  $7.89^{+0.20}_{-0.14}$  \\
  ID60001  &  4.693  &  \citet{zhang2025}  &  $17100^{+310}_{-280}$  &  ---  &  $7.75^{+0.02}_{-0.02}$  \\
  40081  &  3.955  &  \citet{scholte2025}  &  $13810^{+600}_{-620}$  &  $18820^{+2900}_{-2300}$  &  $7.91^{+0.06}_{-0.05}$  \\
  45177  &  2.901  &  \citet{scholte2025}  &  $14090^{+1050}_{-1050}$  &  $12830^{+3390}_{-2700}$  &  $8.09^{+0.16}_{-0.07}$  \\
  47557  &  3.234  &  \citet{scholte2025}  &  $15200^{+1990}_{-1700}$  &  $16930^{+5350}_{-4140}$  &  $7.86^{+0.17}_{-0.10}$  \\
  48659  &  6.796  &  \citet{scholte2025}  &  $27910^{+6380}_{-5850}$  &  ---  &  $7.30^{+0.21}_{-0.07}$  \\
  52422  &  4.025  &  \citet{scholte2025}  &  $21130^{+1790}_{-1690}$  &  $19870^{+12080}_{-6170}$  &  $7.47^{+0.12}_{-0.06}$  \\
  56875  &  4.000  &  \citet{scholte2025}  &  $18050^{+3200}_{-2920}$  &  ---  &  $7.64^{+0.18}_{-0.15}$  \\
  57498  &  3.696  &  \citet{scholte2025}  &  $14830^{+580}_{-600}$  &  ---  &  $7.97^{+0.05}_{-0.04}$  \\
  59720  &  4.367  &  \citet{scholte2025}  &  $12470^{+1410}_{-1460}$  &  $<10470$  &  $8.12^{+0.16}_{-0.13}$  \\
  63962  &  4.359  &  \citet{scholte2025}  &  $16960^{+1900}_{-1980}$  &  $<16620$  &  $7.72^{+0.13}_{-0.10}$  \\
  69991  &  4.937  &  \citet{scholte2025}  &  $16210^{+2650}_{-2450}$  &  $<23330$  &  $7.92^{+0.19}_{-0.15}$  \\
  70864  &  5.255  &  \citet{scholte2025}  &  $12560^{+1170}_{-1170}$  &  ---  &  $8.19^{+0.13}_{-0.11}$  \\
  94335  &  1.812  &  \citet{scholte2025}  &  $11380^{+700}_{-870}$  &  $9590^{+1220}_{-1140}$  &  $8.33^{+0.17}_{-0.10}$  \\
  121806  &  5.225  &  \citet{scholte2025}  &  $14890^{+1010}_{-1010}$  &  ---  &  $7.96^{+0.08}_{-0.07}$  \\
  123597  &  3.798  &  \citet{scholte2025}  &  $11480^{+1320}_{-1370}$  &  $<14330$  &  $8.25^{+0.18}_{-0.14}$  \\
  45393  &  4.236  &  \citet{scholte2025}  &  $13950^{+1390}_{-1320}$  &  ---  &  $7.98^{+0.12}_{-0.10}$  \\
  59009  &  4.133  &  \citet{scholte2025}  &  $12850^{+590}_{-560}$  &  ---  &  $8.16^{+0.06}_{-0.06}$  \\
  73535  &  2.210  &  \citet{scholte2025}  &  $11070^{+1220}_{-1170}$  &  ---  &  $8.31^{+0.16}_{-0.13}$  \\
  93897  &  4.080  &  \citet{scholte2025}  &  $14200^{+790}_{-790}$  &  ---  &  $8.21^{+0.07}_{-0.07}$  \\
  95839  &  4.958  &  \citet{scholte2025}  &  $14510^{+690}_{-690}$  &  ---  &  $8.02^{+0.06}_{-0.05}$  \\
  104937  &  1.652  &  \citet{scholte2025}  &  $12740^{+1620}_{-1530}$  &  ---  &  $8.17^{+0.17}_{-0.14}$  \\
  119504  &  7.917  &  \citet{scholte2025}  &  $20540^{+2770}_{-2860}$  &  ---  &  $7.63^{+0.14}_{-0.11}$  \\
  123837  &  2.617  &  \citet{scholte2025}  &  $17190^{+1910}_{-1610}$  &  ---  &  $7.71^{+0.11}_{-0.10}$  \\
  RXCJ2248-ID  &  6.106  &  \citet{topping2024}  &  $26500^{+1120}_{-1330}$  &  ---  &  $7.42^{+0.06}_{-0.05}$  \\
  GS-NDG-9422  &  5.943  &  \citet{cameron2024}  &  $18200^{+1610}_{-1400}$  &  ---  &  $7.60^{+0.08}_{-0.08}$  \\
  J0217-0208  &  6.204  &  \citet{harikane2025}  &  $12290^{+1170}_{-1300}$  &  ---  &  $8.20^{+0.15}_{-0.12}$  \\
  SXDF-NB1006-2  &  7.212  &  \citet{harikane2025}  &  $14550^{+1430}_{-1400}$  &  ---  &  $7.94^{+0.14}_{-0.12}$  \\
  60001  &  4.694  &  \citet{stiavelli2025}  &  $17810^{+650}_{-600}$  &  ---  &  $7.69^{+0.04}_{-0.04}$  \\
  40149  &  5.403  &  \citet{stiavelli2025}  &  $22550^{+3310}_{-2920}$  &  ---  &  $7.53^{+0.13}_{-0.11}$  \\
  40004  &  5.682  &  \citet{stiavelli2025}  &  $26100^{+9710}_{-6320}$  &  ---  &  $7.48^{+0.25}_{-0.12}$  \\
  COS-3018 Main  &  6.850  &  \citet{scholtz2025}  &  $15200^{+910}_{-920}$  &  ---  &  $7.93^{+0.07}_{-0.06}$  \\
  COS-3018 North  &  6.850  &  \citet{scholtz2025}  &  $20110^{+2610}_{-2420}$  &  ---  &  $7.61^{+0.13}_{-0.10}$  \\
\enddata
\tablenotetext{a}{All \temotp\ vales were derived using [\ion{O}{3}]\W4364 except for the 14 sources marked with $\dagger$, for which \ion{O}{3}]\W1666 was used.}
\tablenotetext{b}{For the JADES sources reported in \citet{laseter2024}, we adopt the line flux measurements from the catalog of \citet{clarke2024}.}
\end{deluxetable*}

\section{Sulfur ion electron temperatures}\label{app:tems}

Electron temperatures of S$^{2+}$ and S$^+$ are reported in Table~\ref{tab:tems} for objects with detections of the [\ion{S}{3}]\W6314 and/or [\ion{S}{2}]\W4070 auroral line.

\begin{deluxetable*}{ l l l l }[b]
 \tablecaption{Electron temperatures derived from sulfur ion auroral lines.\label{tab:tems}}
 \tablehead{ \colhead{ID} & \colhead{$z_{\mathrm{spec}}$} & \colhead{\temstp} & \colhead{\temsp} \\ & & \colhead{\scriptsize{K}} & \colhead{\scriptsize{K}} }
 \startdata
\multicolumn{4}{c}{AURORA sample} \\
\hline
  GOODSN-17940  &  4.4115  &  $9120^{+1150}_{-1290}$  &  $14100^{+4370}_{-3280}$  \\
  GOODSN-21033  &  3.1120  &  $13260^{+1040}_{-1180}$  &  $<17830$  \\
  GOODSN-22235  &  2.4298  &  $12760^{+860}_{-890}$  &  $11730^{+1710}_{-1640}$  \\
  GOODSN-25004  &  2.0487  &  $14190^{+1410}_{-1250}$  &  $<8280$  \\
  GOODSN-28209  &  3.2325  &  $10760^{+1470}_{-1480}$  &  $<16280$  \\
  GOODSN-30053  &  2.2454  &  $<9780$  &  $10240^{+2840}_{-2100}$  \\
  COSMOS-4029  &  2.0765  &  $12920^{+1380}_{-1340}$  &  $<12390$  \\
  COSMOS-4156  &  2.1897  &  $13080^{+1080}_{-1060}$  &  $15480^{+3640}_{-3120}$  \\
  COSMOS-4740  &  3.1556  &  $11930^{+990}_{-890}$  &  $<14050$  \\
  COSMOS-5283  &  2.1742  &  $10740^{+590}_{-570}$  &  $12400^{+1700}_{-1580}$  \\
  COSMOS-6825  &  1.9748  &  $11300^{+1350}_{-1510}$  &  ---  \\
  COSMOS-8363  &  3.2475  &  $13060^{+1690}_{-1490}$  &  $<8450$  \\
\hline
\multicolumn{4}{c}{Literature sample} \\
\hline
  Q2343-D40  &  2.963  &  $16580^{+2050}_{-1610}$  &  ---  \\
  Sunburst Arc  &  2.371  &  $10470^{+660}_{-540}$  &  $11370^{+3770}_{-2340}$  \\
  ID60001  &  4.693  &  $16940^{+1860}_{-1670}$  &  ---  \\
\enddata
\end{deluxetable*}

\bibliography{aurora_calibrations}{}
\bibliographystyle{aasjournalv7}

\end{document}